\documentclass[a4paper,11pt]{article}

\pdfoutput=1 % if you are submitting a pdflatex (i.e., if you have
\usepackage{jcappub} % for details on the use of the package, please
\usepackage[toc,page]{appendix}
\usepackage{subcaption}
\usepackage{amsmath}
\usepackage{amssymb}
\usepackage{graphicx}                
\usepackage{multirow}
\usepackage[T1]{fontenc} % if needed
\usepackage[table]{xcolor}
\usepackage{float}

\usepackage{orcidlink}

\def\hank1{H_\nu^{(1)}}

\def\Mp{M_{\rm Pl}}

\def\cs{c_{\sigma}}
\def\sij{\sigma _{ij}}
\def\tg{\tilde{\gamma}}
\def\gij{\gamma _{ij}}
\def\hanka{H^{(1)} _{i\mu}}
\def\hankb{H^{(2)} _{-i\mu}}

\newcommand{\blueedit}[1]{{\color{blue}#1}}
\hypersetup{pdftitle={Cosmological Gravitational Wave Collider Physics}
}
\usepackage{bm}
\renewcommand{\arraystretch}{1.15}

\title{\boldmath 
 Cosmological Gravitational Wave Collider Physics}

\author[1]{Sukannya Bhattacharya,}
\author[1]{Matteo Fasiello,}
\affiliation[1]{Instituto de F\'{i}sica T\'{e}orica UAM-CSIC, c/ Nicol\'{a}s Cabrera 13-15, 28049, Madrid, Spain}
\author[2]{Anish Ghoshal,}
\affiliation[2]{Department of Physics and Astronomy, University of Sussex,
Brighton, BN1 9RH, United Kingdom}
\author[3,4]{Gianmassimo Tasinato}
\affiliation[3]{Department of Physics, Swansea University, Swansea SA2 8PP, United Kingdom,} \affiliation[4]{Dipartimento di Fisica e Astronomia, Universit\`a di Bologna, Via Irnerio 46, 40126 Bologna, Italy}
\emailAdd{sukannya.bhattacharya@ift.csic.es}
\emailAdd{matteo.fasiello@csic.es}
\emailAdd{a.ghoshal@sussex.ac.uk}
\emailAdd{g.tasinato22.at.gmail.com}

\abstract{
Primordial tensor non-Gaussianity can probe massive particles present during inflation. We investigate this possibility in an effective field theory where the massless graviton mixes with the helicity-two mode of an additional massive spin-two field $\sigma_{ij}$ with cubic self-interactions. For constant sound speed, we compute the leading tensor power-spectrum correction and the tree-level bispectrum, finding novel analytical expressions for these quantities. In the $m_\sigma>3H/2$ regime, we find a nonanalytic squeezed contribution that exhibits logarithmic oscillations with characteristic dependence $(k_L/k_S)^{3/2\pm i\mu}$, where $\mu=\sqrt{m_\sigma^2/H^2-9/4}$. The oscillation frequency is set by the mass $m_\sigma$, while the polarisation contractions determine the relative amplitudes of the different graviton helicity configurations. The same mixing generates corrections to the power spectrum and the bispectrum. We  derive the associated quadrupolar modulation of short-scale tensor power spectrum and relate its amplitude to the model parameters. Finally, we explore how a slowly varying sound speed modifies the squeezed signal.~These results connect the two-point, three-point, and quadrupolar tensor signatures of massive spin-two fields in the cosmological gravitational wave collider.}

\begin{document}
	\maketitle
	\flushbottom

\section{Introduction}
\label{sec:introduction}

Primordial non-Gaussianity provides a direct probe of interactions and additional degrees of freedom active during inflation. In cosmological collider physics, fields with masses of order the Hubble scale generate nonanalytic momentum dependence in late-time correlation functions, allowing information about their masses and spins to be extracted from squeezed configurations~\cite{Arkani-Hamed:2015bza,Chen:2009zp,Chen:2009we,Chen:2017ryl,Lee:2016vti,Kumar:2017ecc,Baumann:2017jvh,GoonTensorShapes2019,Noumi:2012vr,Bordin:2018pca,Hook:2019vcn}.

In exact four-dimensional de Sitter space characterized by a Hubble constant $H$,  fields are classified by unitary irreducible representations of the isometry group $SO(1,4)$. For a massive spin-two field $\sigma_{ij}$, the principal series corresponds to $m_\sigma>3H/2$. Writing $\nu=i\mu$, with $\mu=\sqrt{m_\sigma^2/H^2-9/4}$, the two late-time solutions scale as $(-\tau)^{\Delta_\pm}$ with $\Delta_\pm=3/2\pm i\mu$. Their imaginary part gives rise to oscillations in $\ln(-\tau)$ and, in the squeezed limit, to the characteristic dependence $(k_L/k_S)^{3/2\pm i\mu}$. In the complementary series, $\nu$ is real and $\Delta_\pm=3/2\pm\nu$, so the nonanalytic contribution is instead a noninteger power law without logarithmic oscillations. For an isolated unitary spin-two representation in exact de Sitter space, the Higuchi bound requires $m_\sigma^2\geq2H^2$, leaving $\sqrt{2}H<m_\sigma<3H/2$ as the complementary-series range, with the partially massless point at $m_\sigma^2=2H^2$~\cite{Higuchi:1986py}. Extensions of the Higuchi bound to cosmological backgrounds were studied in Refs.~\cite{Fasiello:2012rw,Fasiello:2013woa}.

During inflation, however, the preferred cosmological background breaks exact de Sitter invariance. So this classification is best viewed as a guide to the late-time behaviour of the field rather than as a complete description of the interacting EFT. For a principal-series field, the characteristic branches $(k_L/k_S)^{3/2\pm i\mu}$  combine into logarithmic oscillations. Their frequency is set by the mass in Hubble units, while the angular and polarisation dependence of the correlator carries information about spin.

In this work, we discuss a cosmological-collider
setup with an extra massive spin-two field during inflation. 
Most cosmological-collider studies have focused on scalar correlators, in particular the curvature bispectrum~\cite{Arkani-Hamed:2015bza,Chen:2010xka,Babich:2004gb,Senatore:2009gt,Creminelli:2005hu,Chen:2006nt,Ferguson:2008ra,Alvarez:2014vva,Weinberg:2005vy,Koyama:2010xj,PhysRevD.82.023502,Bartolo:2004if,Agullo:2010ws,Assassi:2015jqa}. Tensor correlators provide complementary information because the external graviton polarisations retain the tensor structure of the interaction. In particular, a squeezed tensor bispectrum can carry the mass dependence of an intermediate state through its logarithmic oscillations and its spin dependence through contractions of the graviton polarisation tensors~\cite{Baumann:2017jvh,Baumann:2020dch,Pajer:2013ana,Bordin:2020eui}.

There is by now a substantial literature on cosmological-collider signals from heavy fields and on methods for computing the corresponding correlators. Scalar exchange, turning trajectories, heavy-field self-interactions, Standard Model fields, gauge bosons, Higgs-sector physics, higher-spin states, parity violation, and related mechanisms have all been considered~\cite{ChenWangLargeMass2012,PiSasakiTurning2012,GongPiSasaki2013,ChenWangXianyuLoops2016,ChenWangXianyuSM2017,ChenWangXianyuMass2017,ChenWangXianyuNeutrino2018,CosmologicalHiggsCollider2020,AlexanderSUSY2019,LiuParityCollider2020,WangXianyuGauge2020,MaruOkawaGUT2021,CuiXianyuLeptogenesis2022,YouRHNeutrino2026}. The range of calculational tools has also expanded considerably, including bootstrap, cutting-rule, Mellin-Barnes, flow-equation, and loop-level methods~\cite{CosmologicalPolytopes2017,CosmologicalBootstrap2020,SnowmassBootstrap2024,PimentelWangBoostless2022,TongWangZhuCutting2022,CosmoFlow2023,XianyuZangMultiple2024,QinLoopCorrelators2025}. At the same time, dedicated CMB and large-scale-structure analyses have begun to search for these signatures directly~\cite{MeerburgProspects2017,MeerburgPNG2019,InflationTheoryObservations2022,CabassLSSCollider2024,SohnPlanckCollider2024,KumarPlanckHeavy2026,KumarScalarPlanck2026}.

For the tensor sector, several related constructions provide useful points of comparison. Additional massive spin-two fields can generate graviton-bispectrum shapes beyond the two leading graviton self-interaction shapes in de Sitter space~\cite{GoonTensorShapes2019}. Helicity-selective production from a chemical potential gives a different class of tensor and mixed correlators~\cite{TongSpinTwo2022,Wang:2020uic}, while massive-gravity models modify the observable graviton itself and can produce a blue tensor spectrum and a squeezed bispectrum controlled by the graviton mass~\cite{Ricciardone:2016lym,Ricciardone:2017kre,FujitaMassiveGravity2020}. Other inflationary sources can also generate squeezed tensor bispectra, for example through an excited scalar mediator~\cite{PengTensorBispectrum2024}
or through a brief phase of non-slowroll evolution
\cite{Mylova:2018yap,Ozsoy:2019slf}.
 The setup considered here instead keeps the observable graviton massless and couples it to the helicity-two mode $\sigma_{ij}$ of an additional massive spin-two field.

The interactions relevant for the tensor observables are a derivative bilinear mixing, controlled by the dimension-one coefficient $\rho$, and the cubic self-interaction $\kappa\,\mathrm{Tr}(\sigma^3)$. We assume that the bilinear mixing gives the leading coupling between the two tensor sectors, while mixed cubic operators are subleading in the parameter range of interest. The leading fractional correction to the tensor power spectrum then scales as $(\rho/H)^2$, whereas the nonanalytic bispectrum generated by the $\sigma^3$ interaction and three insertions of the mixing scales as $(\rho/H)^3(\kappa/H)$. The same mixing therefore enters both observables, so the tensor power spectrum constrains the parameter range in which the three-point signal can be enhanced.

We present the first explicit calculation of the primordial tensor bispectrum generated by this spin-two EFT in the cosmological-collider regime, $m_\sigma>3H/2$. For a general constant subluminal propagation speed $0<c_\sigma\leq1$, we compute the full tensor bispectrum and derive its squeezed nonanalytic contribution, retaining the complete dependence on the spin-two mass and propagation speed. We also derive the corresponding tensor power-spectrum correction in semi-analytical form. This allows the bispectrum signal and the modification of the tensor two-point function to be treated consistently within the same EFT and, in particular, determines how the power-spectrum constraint restricts the parameter space available for an observable cosmological-collider signal.

The power-spectrum correction we compute shows two useful regimes. Near the principal-series threshold, small $\mu$ gives the largest response. At larger $\mu$, the non-time-ordered contribution is exponentially suppressed, while the time-ordered interference term gives the leading inverse-mass contribution. The dependence on $c_\sigma$ is oscillatory because the principal-series mode functions contain powers $c_\sigma^{\pm i\mu}$. It is therefore not described by the simple inverse-power small-$c_\sigma$ enhancement familiar from complementary-series fields. This fact is important when assessing the parameter region with small propagation speed.

In the squeezed limit, with $r\equiv k_L/k_S\ll1$, the nonanalytic part of the tensor bispectrum has the schematic form
\begin{equation}
 B_{\gamma,\rm nonanalytic}^{\lambda_1\lambda_2\lambda_3}
 \propto P_\gamma(k_S)P_\gamma(k_L)
 \left(\frac{\rho}{H}\right)^3
 \left(\frac{\kappa}{H}\right)
 \mathcal A^{\lambda_1\lambda_2\lambda_3}
 W(\mu,c_\sigma)r^{3/2}
 \cos\!\left(\mu\ln r+\varphi\right).
 \label{eq:introduction_clock_summary}
\end{equation}
Here $\mathcal A^{\lambda_1\lambda_2\lambda_3}$ is the contraction of the external polarisation tensors, while $W(\mu,c_\sigma)$ contains the mass- and propagation-speed dependence of the nonanalytic amplitude. The $LLL/RRR$ and $LLR/RRL$ configurations are unsuppressed at leading order in the squeezed limit, whereas the $LRR/RLL$ amplitude carries an additional $r^2$ suppression. The logarithmic frequency is set by $\mu$, while the relative amplitudes of the helicity configurations provide an independent check of the tensor structure.

We additionally consider a slowly varying propagation speed through the parametrisation $c_\sigma(k)=c_{\rm ini}(k/k_\ast)^{s_2}$. Within the horizon-crossing approximation used here, a constant $s_2$ shifts the logarithmic frequency from $\mu$ to $\mu(1+s_2)$ and changes the amplitude because the long and short modes probe different values of $c_\sigma$. Reduced propagation speeds and broken boosts have also been studied in scalar cosmological-collider correlators~\cite{PimentelBoostless2022,JazayeriLowSpeed2023}; in the present case the reduced speed belongs to the intermediate helicity-two mode.

Finally, the squeezed tensor bispectrum induces a quadrupolar modulation of the short-scale tensor power. Such tensor fossils have been discussed more generally in Refs.~\cite{DimastrogiovanniTensorFossils2014,DimastrogiovanniGravityFossils2020}. Here the quadrupole is fixed by the same squeezed bispectrum rather than introduced as an independent phenomenological response. We compute its primordial amplitude and compare it with the tensor power-spectrum constraint. The quantities quoted here are primordial amplitudes; a forecast for a particular CMB experiment would additionally require transfer functions, noise, foregrounds, delensing, sky coverage, and a dedicated estimator. Also, a focussed
study of their consequences for interferometric detections would be welcome \cite{Bartolo:2018qqn}.

The paper is organised as follows. Section~\ref{sec:eft_framework} defines the tensor-sector EFT and the assumptions used in the calculation. Section~\ref{sec:cc_review} reviews the origin of the nonanalytic squeezed signal and introduces the mixed-propagator formalism. Section~\ref{sec:powspec} derives the tensor power correction, its limiting behaviour, scale dependence, and perturbative constraints. Section~\ref{subsec:tensor_collider_bispectrum} presents the full and squeezed tensor bispectra for constant and scale-dependent propagation speed. Section~\ref{sec:results} discusses the bispectrum amplitudes and the induced quadrupolar anisotropy. 
Section~\ref{sec:conclusion} contains our conclusions.
The Appendices collect the propagator derivation, the power-spectrum integrals, the polarisation contractions, and the squeezed-bispectrum coefficients.

\section{The Effective Field Theory Framework}
\label{sec:eft_framework}

The effective field theory of inflation (EFTI) organizes fluctuations around a background that spontaneously breaks time reparameterizations, with the Goldstone mode $\pi$ restoring the broken symmetry~\cite{Cheung:2007st}.  Extensions containing additional massive and spinning fields have been developed in Refs.~\cite{Bordin:2018pca,Dimastrogiovanni:2018gkl,Baumann:2020dch}.  Since the time dependence of the inflationary background breaks exact de Sitter isometries, the interactions of additional fields and their spectrum need not coincide with those of an isolated unitary representation in exact de Sitter space~\cite{Lee:2016vti,Arkani-Hamed:2015bza}.  This freedom allows for direct couplings between the inflaton (or the graviton) and additional fields: from these stems the non-analytic wavenumber behaviour in cosmological correlators.

Since our interest is in the gravitational-wave sector, we focus on the quadratic and cubic terms in the EFT Lagrangian involving the (GR) massless graviton $\gamma_{ij}$ and the helicity-two mode $\sigma_{ij}$ of an additional massive spin-two field. We refer the reader to Ref.~\cite{Bordin:2018pca} for the full EFT description; see also Refs.~\cite{Cheung:2007st,1dcc0a40-9483-382a-96d0-afd7171fa972}. Our starting point is $\mathcal{L}_{\gamma +\sigma}=\mathcal{L}_{\rm free}^{(2)}+\mathcal{L}_{\rm int}^{(2)} + \mathcal{L}_{\rm int}^{(3)},$ with
\begin{eqnarray}
%\mathcal{L}_{\gamma +\sigma}&=&\mathcal{L}_{\rm free}^{(2)}+\mathcal{L}_{\rm int}^{(2)} + \mathcal{L}_{\rm int}^{(3)},\\
%{\rm with,}\nonumber \\
\label{eq:full} \mathcal{L}_{\rm free}^{(2)}=\frac{a^3}{8}\bigg[\dot{\tg}_{ij}^2-\frac{1}{a^2}(\partial_i\tg_{jk})^2]&+&\frac{a^3}{4}\bigg[\dot{\sigma}_{ij}^2-\frac{\cs^2}{a^2}(\partial_i\sigma_{jk})^2-m_{\sigma}^2\sigma_{ij}^2],\label{eq:free}\\
\mathcal{L}_{\rm int}^{(2)}=\frac{a^3}{2}\rho\sij\dot{\tg}^{ij}, ~ ~ ~ && \mathcal{L}_{\rm int}^{(3)}=-a^3\kappa(\sij)^3.\label{eq:mix}
\end{eqnarray}
Throughout the manuscript, $\operatorname{Tr}(\sigma^3)\equiv\sigma_i{}^j\sigma_j{}^k\sigma_k{}^i$, so $\mathcal L_{\rm int}^{(3)}=-a^3\kappa\,\operatorname{Tr}(\sigma^3)$.
Here $\tilde\gamma_{ij}\equiv M_{\rm Pl}\gamma_{ij}$ is the canonically normalized metric tensor fluctuation. We restrict attention to the helicity-two mode $\sigma_{ij}$ of the massive spin-two field, whose propagation speed is $c_\sigma$.
%The truncated action in Eq.~\eqref{eq:free} is not, by itself, assumed to define a complete Lorentz-invariant five-polarization massive-spin-two theory. If such an embedding exists, the remaining helicities are assumed to decouple over the range of validity considered here or to be encoded in higher-dimensional operators suppressed by the EFT cutoff. 
The graviton sound speed is naturally set to unity, whereas $c_\sigma$ may differ from unity, leading
to interesting phenomenology. Constant $c_\sigma$ is used in the main calculation that follows; its slowly varying extension is introduced later.

We assume that the bilinear mixing is the leading interaction between the extra tensor sector and gravity. The leading non-Gaussian signal is generated by the intrinsic interaction $\kappa\,\mathrm{Tr}(\sigma^3)$ together with three insertions of the quadratic mixing\footnote{Note that the perturbative treatment of the quadratic interaction term requires it to be small with respect to the rest of the quadratic Lagrangian, which determines the wavefunction. Ref. \cite{Bordin:2018pca} includes the condition $\rho\ll H$, which we also employ here.}. Operators such as $\gamma\sigma^2$ and $\gamma^2\sigma$, are also present. We work in the regime in which these are parametrically suppressed. 
%After perturbatively diagonalizing the quadratic mixing, the $\sigma^3$ vertex generates mixed cubic interactions whose coefficients are fixed by powers of $\rho/H$, rather than by new independent couplings. This portal hierarchy is an assumption of the truncated EFT, not a consequence of the quadratic action alone. 
There is a direct analogy between the diagrams we  study here and the exchange diagrams in quasi-single-field inflation 
% further motivates isolating the intrinsic $\sigma^3$ contribution~\cite{Xianyu:2025lbk,Kumar:2026ogn}, 
 -
see also Ref.~\cite{GoonTensorShapes2019} for a discussion on the broader massive-spin-two tensor-bispectrum classification.
The tensor components of the spin-two fields $\gij$ and $\sij$ are decomposed as
\begin{equation}
T_{ij}({\bf x},\tau)=\int \frac{d^3k}{(2\pi)^3}e^{i{\bf k}\cdot{\bf x}}\sum_{\lambda}\epsilon_{ij}^{\lambda}(\hat{k})T_{\bf k}^{\lambda}(\tau). 
\end{equation}
With $\rho /H \ll 1$ and $\kappa/H \ll 1$, the quadratic and cubic terms in Eq.~\eqref{eq:mix} are perturbations on top of the free Lagrangian~\eqref{eq:free}. The free solutions of $\sigma$ and $\gamma$ are
\begin{eqnarray}
\gamma _k(\tau)&=&\frac{2iH}{\Mp\sqrt{2k^3}}(1+ik\tau)e^{-ik\tau},\label{gammamode}\\
\sigma _k(\tau)&=& -i\sqrt{\frac{\pi}{2}}e^{i\pi(\nu/2+1/4)}H(-\tau)^{3/2}\hank1(-\cs k\tau),\label{sigmode}
\end{eqnarray}
where $\hank1$ is the Hankel function of the first kind. The mass parameter is defined by $\nu^2\equiv 9/4-m_\sigma^2/H^2$.  Thus $\nu=i\mu$ in the heavy, principal-series regime $m_\sigma>3H/2$, with 
\begin{equation}
    \mu\equiv \sqrt{{m_\sigma^2\over H^2}-{9\over4}}, \label{eq:defmu}
\end{equation}
whereas $0<\nu<3/2$ for a lighter complementary-series field. The couplings $\rho$ and $\kappa$ have mass dimension one, and we work perturbatively in the dimensionless ratios $\rho/H$ and $\kappa/H$.

\begin{figure}[t]
\centering{
\includegraphics[width=0.85\textwidth]{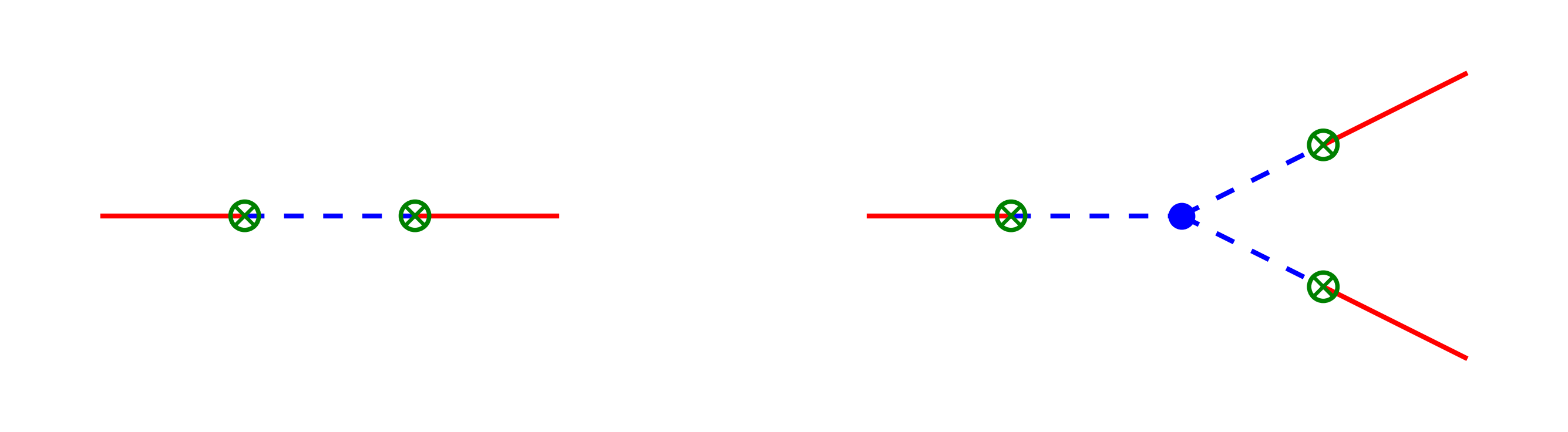}\\
\caption{\small Diagrammatic representation of key contribution to the tensor observables.  \textit{Left:} the leading correction to the graviton two-point function, generated by two insertions of the bilinear mixing vertex $\rho\,\sigma_{ij}\dot{\tilde\gamma}^{ij}$.  \textit{Right:} the contribution to the GW  bispectrum, due to three 2-vertices and one three-vertex from the $\kappa\,\operatorname{Tr}(\sigma^3)=\kappa\,\sigma_i{}^j\sigma_j{}^k\sigma_k{}^i$ interaction term.  %The figure makes clear why $\delta P_\gamma\propto(\rho/H)^2$, whereas $B_\gamma\propto(\rho/H)^3(\kappa/H)$. 
%\textit{Bottom panel:} Mixed graviton-spin-two propagator.  The black line denotes the conversion of a heavy helicity-two fluctuation $\sigma_{ij}$ into the observable graviton $\gamma_{ij}$ through the bilinear mixing interaction.  Two copies of this conversion kernel, one for each external graviton leg, enter the power-spectrum correction, whereas three copies enter the tensor bispectrum; consequently, the same mass and propagation-speed dependence controls both observables.
}\label{fig:sq_cubic}}
\end{figure}
Figure~\ref{fig:sq_cubic} aims at showing the perturbative counting used throughout the manuscript. Each bilinear conversion supplies one power of $\rho$: the tensor two-point correction contains two conversions, while the bispectrum contains three conversions and one cubic vertex.

\section{Cosmological collider signal}
\label{sec:cc_review}

In this section we review the origin of the oscillatory signal in the so-called cosmological-collider context.
There are several ways in which a massive field may affect the 2-point and 3-point correlators of scalar and tensor fluctuations.
\subsection{Massive fields in de Sitter and the origin of nonanalyticity}
\label{subsec:cc_de_sitter_modes}

We focus on the squeezed tensor bispectrum mediated by the helicity-two mode $\sigma_{ij}$ of the additional spin-2 field. In this limit, the nonanalytic signal is controlled by the late-time behaviour of this massive mode in quasi-de Sitter space~\cite{ChenWangLargeMass2012,PiSasakiTurning2012,GongPiSasaki2013,CosmologicalBootstrap2020}. In the quasi-de Sitter approximation used throughout this work, with $H$, $m_\sigma$, and $c_\sigma$ treated as approximately constant over the relevant time interval, the principal-series regime corresponds to $\nu=i\mu$. Denoting by $\sigma_k$ the mode function multiplying the corresponding polarization tensor, its late-time behaviour is
\begin{equation}
    \lim_{\tau\rightarrow 0}\sigma_k(\tau)
    \sim
    z^{3/2}
    \left[
    \alpha_k(\mu,c_\sigma)(c_\sigma z)^{i\mu}
    +
    \beta_k(\mu,c_\sigma)(c_\sigma z)^{-i\mu}
    \right],
\end{equation}
where $z\equiv-k\tau$, and $\alpha_k, \beta_k$
coefficients to be fixed by boundary conditions.

The two complex powers are the origin of the logarithmic oscillations appearing in the squeezed bispectrum. The factor $z^{3/2}$ describes the late-time decay of the massive mode, while $z^{\pm i\mu}$ carries the oscillatory phase. The propagation speed $c_\sigma$ affects both the coefficients $\alpha_k(\mu,c_\sigma)$ and $\beta_k(\mu,c_\sigma)$ and the explicit factors $(c_\sigma z)^{\pm i\mu}$, and therefore modifies both the amplitude and phase of the resulting logarithmic oscillations. For the spin-two field, the polarization tensors encode the helicity and angular dependence of the correlator.

A useful interpretation follows from relating a squeezed configuration to the separation between the characteristic times of the long and short modes. For $k_L\ll k_S$, this interval is of order
\begin{equation}
    \Delta N \simeq \log\left({k_S\over k_L}\right).
\end{equation}
Since the massive mode oscillates as $\exp[\pm i\mu\log(-\tau)]$, the phase accumulated over this interval is proportional to $\mu\Delta N$. Equivalently, the late-time complex powers of the long mode generate the characteristic nonanalytic dependence
\begin{equation}
\label{eq:cc_nonanalytic_scaling}
    \left({k_L\over k_S}\right)^{3/2\pm i\mu}
    =
    \left({k_L\over k_S}\right)^{3/2}
    \exp\left[
    \pm i\mu\log\left({k_L\over k_S}\right)
    \right].
\end{equation}
The two branches combine into the nonanalytic contribution with logarithmic oscillations schematically displayed in Eq.~\eqref{eq:introduction_clock_summary}. Its frequency in $\log(k_L/k_S)$ is set by $\mu$, while its amplitude and phase depend on the mixed-propagator coefficients calculated below.

For large $\mu$, the nonanalytic signal is suppressed by heavy-particle production. In the present calculation, the precise mass dependence is determined by the mixed-propagator integrals rather than imposed from a general estimate. The characteristic complex powers $(k_L/k_S)^{\pm i\mu}$ distinguish the nonanalytic contribution associated with propagation of the massive mode from the analytic terms that can be reproduced by local contact interactions after the heavy field is integrated out.

%\subsection{Power spectra with heavy exchange}
For complementary-series spin-2 fields, the helicity-two contribution to the tensor power spectrum is known to exhibit a power-law dependence on the propagation speed, with the sourced contribution scaling as $c_\sigma^{-2\nu}$ in the weak-mixing regime~\cite{Dimastrogiovanni:2018gkl,Iacconi:2020yxn}. In the principal-series regime considered here, $\nu=i\mu$, so the corresponding powers of $c_\sigma$ become oscillatory phases. The full dependence on $c_\sigma$ is determined by the mixed-propagator integrals derived in \blueedit{Sec.~\ref{sec:powspec}}.

Following the usual cosmological-collider decomposition~\cite{Arkani-Hamed:2015bza}, we separate the squeezed tensor bispectrum into analytic and nonanalytic contributions (the prime denotes the correlator without the momentum-conserving delta function):
\begin{align}
  \langle \gamma^{\lambda_1}_{{\bf k}_1}
          \gamma^{\lambda_2}_{{\bf k}_2}
          \gamma^{\lambda_3}_{{\bf k}_3}\rangle'
  &=
  \mathcal{A}^{\lambda_1\lambda_2\lambda_3}
  B_{\gamma}(k_1,k_2,k_3),
  \label{eq:schematic_exchange_bispectrum}\\
  B_{\gamma,\rm sq}(k_S,k_S,k_L)
  &=
  B_{\rm analytic}(k_S,k_S,k_L)
  +B_{\rm nonanalytic}(k_S,k_S,k_L),\\
 \frac{ B_{\rm analytic}(k_S,k_S,k_L)}{P_\gamma(k_L)P_\gamma(k_S)}
  &=
%  P_\gamma(k_L)P_\gamma(k_S)
  \sum_{n=0}^{\infty}
  a_n\left({k_L\over k_S}\right)^n,\\
  \frac{B_{\rm nonanalytic}(k_S,k_S,k_L)}{P_\gamma(k_L)P_\gamma(k_S)}
  &=
  \left({\rho\over H}\right)^3
  \left({\kappa\over H}\right)
  \left({k_L\over k_S}\right)^{3/2}
  \left[
    b_+(\mu,c_\sigma)
    \left({k_L\over k_S}\right)^{i\mu}
    +
    b_-(\mu,c_\sigma)
    \left({k_L\over k_S}\right)^{-i\mu}
  \right].
  \label{eq:gen_ana_nonana}
\end{align}
 The analytic contribution admits an expansion in integer powers of the squeezed momentum ratio, while the nonanalytic contribution contains the characteristic cosmological-collider signal from the propagation of the massive field~\cite{Arkani-Hamed:2015bza}. The distinctive squeezed-limit momentum dependence of cosmological-collider signals has motivated dedicated forecasts and template-based searches, including 21-cm forecasts~\cite{MeerburgProspects2017}, analyses of galaxy clustering~\cite{CabassLSSCollider2024}, and searches in CMB data~\cite{SohnPlanckCollider2024,KumarPlanckHeavy2026,KumarScalarPlanck2026}. For the spin-two interaction considered here, the nonanalytic amplitude scales as $(\rho/H)^3(\kappa/H)$, with each factor of $\rho$ associated with the conversion between the helicity-two mode and the graviton.

The nonanalytic contribution can equivalently be written in a manifestly real form. Writing $r\equiv k_L/k_S$,
\begin{equation}
    b_+ r^{3/2+i\mu}
    +b_- r^{3/2-i\mu}
    =
    2|b_+|r^{3/2}
    \cos\!\left(\mu\ln r+\arg b_+\right).
    \label{eq:complex_to_real_clock}
\end{equation}
Here $|b_+|$ and $\arg b_+$ determine the amplitude and phase of the oscillation. The period in $\ln r$ is $2\pi/\mu$.

For tensor correlators, the helicity structure is encoded in the polarization contractions~\cite{Dimastrogiovanni:2018gkl,Iacconi:2020yxn}. For the cubic interaction we considered,
\begin{equation}
    \mathcal A^{\lambda_1\lambda_2\lambda_3}
    =
    \epsilon^{\lambda_1}_{\alpha\beta}(-\hat k_1)
    \epsilon^{\lambda_2}_{\beta\gamma}(-\hat k_2)
    \epsilon^{\lambda_3}_{\gamma\alpha}(-\hat k_3),
\end{equation}
which we derive explicitly in Appendix~\ref{appendix:helicity}. The helicity dependence therefore carries the spin information in the tensor bispectrum.

\subsection{Diagrammatic interpretation and in-in structure}
\label{subsec:inin_review}
In primordial cosmology, observational quantities are late-time expectation values of quantum operators evaluated within a given initial state (`in' state). In the canonical in-in (or Schwinger-Keldysh) operator formalism, a late-time expectation value of an operator $Q(\tau)$ is expressed as:
\begin{equation}\label{schematic:inin}
\langle Q(\tau)\rangle
=
\left\langle \Omega \right|
\left[ \bar T \exp\left(i\int_{-\infty}^{\tau}d\tau'H_I(\tau')\right) \right]
Q_I(\tau)
\left[ T \exp\left(-i\int_{-\infty}^{\tau}d\tau'H_I(\tau')\right) \right]
\left| \Omega \right\rangle,
\end{equation}
where $T$ and $\bar T$ denote time-ordering and anti-time-ordering operators, $Q_I$ is the operator in the interaction picture, and $H_I$ is the interaction Hamiltonian.

This canonical formulation introduces two notable computational bottlenecks: (a) when derivative couplings are present, the equality $H_I=-\mathcal{L}_{\text{int}}$ does not hold in general. Instead, nontrivial canonical-momentum relations generate additional field-dependent contact terms in $H_I$ that are absent in the classical Lagrangian.  (b) evaluating $n$-point correlation functions involves multiple time integrals over nested domains for processes involving heavy particle exchange. In squeezed configurations, these integrals span the parametrically separated time scales associated with the long and short modes.  

To streamline the calculations we adopt the Schwinger-Keldysh Path Integral (SKPI) diagrammatic formulation developed in Ref.~\cite{Chen:2017ryl,Chen:2009zp}. Its mixed-propagator organization is  related to  cutting-rule, bootstrap, flow-equation, and Mellin-Barnes approaches~\cite{CosmologicalBootstrap2020,TongWangZhuCutting2022,PimentelWangBoostless2022,CosmoFlow2023,XianyuZangMultiple2024,QinLoopCorrelators2025}. This approach offers three key advantages. First, the Schwinger-Keldysh path integral avoids the explicit construction of an interaction Hamiltonian in terms of field-dependent canonical momenta. For the derivative bilinear mixing considered here, the interaction is incorporated through the doubled quadratic action and the corresponding mixed propagator. %This statement is specific to the consistently defined path-integral treatment used below and is not intended as a general claim for arbitrary higher-derivative theories.
  Second, by inserting a complete basis of states at late times ($\tau_f \to 0$), the time-ordered and anti-time-ordered branches are cast into a double-path integral over two copies of the field, $\phi_+$ (forward in time) and $\phi_-$ (backward in time), sewn together at the future boundary $\phi_+(\tau_f) = \phi_-(\tau_f)$.  Third, the diagrammatic organization of SKPI allows the mixing subdiagrams to be computed once and absorbed into a mixed propagator. By pre-computing the linear mixing between fields into a late-time bulk-to-boundary mixed propagator $\mathcal{G}_{\pm}(k;\tau)$, nested multi-vertex diagrams condense into significantly simpler single-vertex time integrations.  

For the system studied here, each conversion integral is evaluated once in terms of the mixed propagator $\mathcal G_\pm$. Accounting for one bilinear insertion, $\mathcal G_\pm$ includes one factor of $\rho$ -- see Appendix \ref{appendixA} for the evaluation of $\mathcal G_\pm$ and our conventions on propagators. Defining $\widehat{\mathcal G}_\pm\equiv\mathcal G_\pm/\rho$ then, 
\begin{align}
\delta P_\gamma&\propto\left(\frac{\rho}{H}\right)^2\widehat{\mathcal G}^{\,2},\nonumber\\
B_\gamma&\propto\left(\frac{\kappa}{H}\right)\left(\frac{\rho}{H}\right)^3\widehat{\mathcal G}^{\,3}.
\end{align}
Increasing $\rho$ therefore raises the bispectrum and the fractional power correction simultaneously. In the vacuum-dominated benchmarks we impose $|\delta P_\gamma|\ll1$ and $1+\delta P_\gamma>0$.

By employing the SKPI mixed-propagator technique, computing $B_\gamma$ reduces from evaluating 16 nested canonical field-contraction terms to a single, UV-convergent time integral over the product of three mixed propagators $\mathcal{G}_+(k_i;\tau)$:
\begin{equation}
B_\gamma(k_1,k_2,k_3) \propto \left({\kappa \over H}\right) \text{Im} \int_{-\infty}^{0} \frac{d\tau}{(-H\tau)^4} \mathcal{G}_+(k_1;\tau) \mathcal{G}_+(k_2;\tau)\mathcal{G}_+(k_3;\tau).
\end{equation}
Importantly, 
this approach yields fully analytic (or low-cost semi-analytic) expressions that preserve the explicit dependence on the heavy-field mass parameter $\mu \equiv \sqrt{m^2/H^2 - 9/4}$ and the sound speed $\cs$.

\section{Leading correction to the tensor power spectrum}\label{sec:powspec}
The $\sigma_{ij}$  field contributes to the tensor power spectrum  as shown in Fig.~\ref{fig:sq_cubic}.  The correction is  proportional to $\rho^2$ and depends on the mass parameter $\mu$ and propagation speed $c_\sigma$. The dimensionless tensor power spectrum is $\mathcal{P}_{\gamma}(k)\equiv \frac{k^3}{2\pi^2}P_{\gamma}(k)$, where 
\begin{align}
\left\langle\gamma^\lambda_{\mathbf k}\gamma^{\lambda'}_{\mathbf k'}\right\rangle
&=(2\pi)^3\delta^{(3)}(\mathbf k+\mathbf k')\delta_{\lambda\lambda'}P_\gamma^\lambda(k),
\nonumber\\
P_\gamma(k)\equiv\sum_{\lambda=L,R}P_\gamma^\lambda(k)
&=\frac{4H^2}{M_{\rm Pl}^2k^3}\left[1+\delta P_\gamma(\mu,c_\sigma)\right]\nonumber\\
&=\frac{4H^2}{M_{\rm Pl}^2k^3}\left[1+\frac{\rho^2}{H^2}\widetilde P(\mu,c_\sigma)\right].
\label{eq:tenspow}
\end{align}
Throughout the remainder of the paper, $P_\gamma(k)$ denotes the tensor power spectrum summed over the two graviton helicities, whereas $P_\gamma^\lambda(k)=P_\gamma(k)/2$ denotes the power in one helicity for a parity-symmetric state.  The symbol $\delta P_\gamma$ is reserved for the dimensionless fractional correction,
\begin{equation}
 \delta P_\gamma(\mu,c_\sigma)\equiv
 \left(\frac{\rho}{H}\right)^2\widetilde P(\mu,c_\sigma),
 \quad {\rm and} \quad
 P_\gamma=P_\gamma^{(0)}(1+\delta P_\gamma), \qquad P_\gamma^{(0)}\equiv  4H^2/(M_{\rm Pl}^2k^3)
\label{eq:fractional_power_correction}
\end{equation}
Since $P_\gamma$ is the sum over the two helicities, the total dimensionful sourced contribution implied by Eq.~\eqref{eq:tenspow} is
\begin{equation}
\Delta P_{\gamma,\rm tot}^{\rm(dim)}(k)
\equiv P_\gamma^{(0)}(k)\delta P_\gamma(k)
=\frac{4\rho^2}{M_{\rm Pl}^2k^3}\widetilde P(\mu,c_\sigma).
\label{eq:powercorr}
\end{equation}
For a parity-symmetric state the sourced contribution per helicity is half of this value.
%\begin{equation}
%\blueedit{\Delta P_\gamma^{\lambda,\rm(dim)}(k)
%=\frac{2\rho^2}{M_{\rm Pl}^2k^3}\widetilde P(\mu,c_\sigma),
%\qquad
%\Delta P_{\gamma,\rm tot}^{\rm(dim)}=\sum_{\lambda=L,R}\Delta P_\gamma^{\lambda,\rm(dim)}.}
%\label{eq:powercorr_per_helicity}
%\end{equation}
%\blueedit{This resolves a possible factor-of-four ambiguity: 
Hence, the vacuum spectrum is $2H^2/(M_{\rm Pl}^2k^3)$ per helicity and $4H^2/(M_{\rm Pl}^2k^3)$ after summing $L$ and $R$. The Schwinger--Keldysh integral defining $\widetilde P$ uses the same per-helicity convention as the external propagator, and the total result is obtained by the indicated helicity sum.

The expressions for the correction $\tilde{P}(\mu,\cs)$ are evaluated in Appendix~\ref{appendixA}. There are two contributions to $\widetilde P(\mu,c_\sigma)$ when we separate out the non-time-ordered contribution $\mathcal{C}_1$ from a time-ordered one $\mathcal{C}_2$.
\begin{eqnarray}
\widetilde P(\mu , \cs)&\equiv &\mathcal{C}_1(\mu,\cs)+\mathcal{C}_2(\mu,\cs),\label{eq:delp0} ~ ~ {\rm where,}\\
\mathcal{C}_1(\mu,\cs)&\equiv &\frac{\pi}{2}e^{-\pi\mu}\bigg\vert \int _0^{\infty} \frac{dx}{x^{1/2}}\hanka (\cs x) e^{ix} \bigg\vert ^2,\label{eq:def_c1} \\ 
\mathcal{C}_2(\mu,\cs)&\equiv &-\pi e^{-\pi\mu}{\rm Re}\bigg[\int _0^{\infty} \frac{dx}{x^{1/2}}\hanka (\cs x) e^{-ix} \int _x^{\infty} \frac{dy}{y^{1/2}}\hankb (\cs y) e^{-iy}\bigg].\label{eq:def_c2}
\end{eqnarray}
The complete results for the coefficients for arbitrary $\mu$ and $c_\sigma$ are
\begin{eqnarray}
\mathcal{C}_1(\mu,\cs)&= &\frac{\pi ^2}{\cs\cosh ^2(\pi\mu)}\bigg\vert{}_2F_1[1/2-i\mu,1/2+i\mu,1,\frac{\cs-1}{2\cs}]\bigg\vert^2, ~ ~ {\rm and}\label{eq:c1expr}\\
\mathcal{C}_2(\mu,\cs)&= &-\frac{2}{(1+\cs)\sinh(\pi\mu)}{\rm Re}\bigg[e^{\pi\mu}\bigg(\frac{2\cs}{1+\cs}\bigg)^{2i\mu}\mathcal{J}_+(\mu,\cs)-e^{-\pi\mu}\mathcal{J}_-(\mu,\cs)\bigg].\label{eq:c2expr}
\end{eqnarray}
The functions $\mathcal{J}_{\pm}(\mu,\cs)$ are convergent integrals, whose complete form can be found in the Appendix~\ref{appendixB}. Therefore, using the mixed propagator technique, we obtained  an analytical form of the correction to the tensor power spectrum as functions of the mass parameter $\mu$ and the propagation speed $\cs$ for the massive field, with the only numerical part being the exact evaluation of $\mathcal{J}_{\pm}(\mu,\cs)$. Equations~\eqref{eq:c1expr} and~\eqref{eq:c2expr} are exact for constant $0<c_\sigma\leq1$. For slowly varying $c_\sigma(k)$ they are used only through an adiabatic horizon-crossing prescription, evaluating the constant-$c_\sigma$ result at the value sampled by each mode. (But they are not an exact solution for arbitrary $c_\sigma(\tau)$.) To the best of our knowledge, this is the first instance of such an analytical calculation for the tensor power correction for subluminal $\cs$ in the context of cosmological colliders. In particular, in the limiting cases for $\mu$ and $\cs$, the result can be written fully analytically, which is discussed in the next subsection.

\subsection{Limits}
\label{sec:limits}
\begin{figure}[!]
\includegraphics[width=0.5\textwidth]{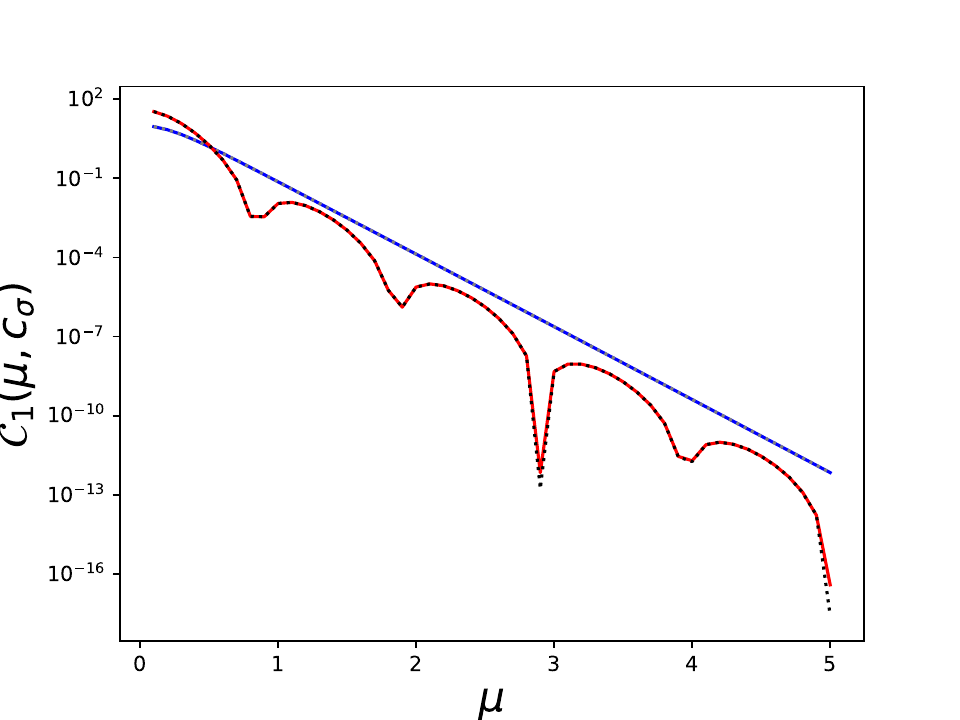}
\includegraphics[width=0.5\textwidth]{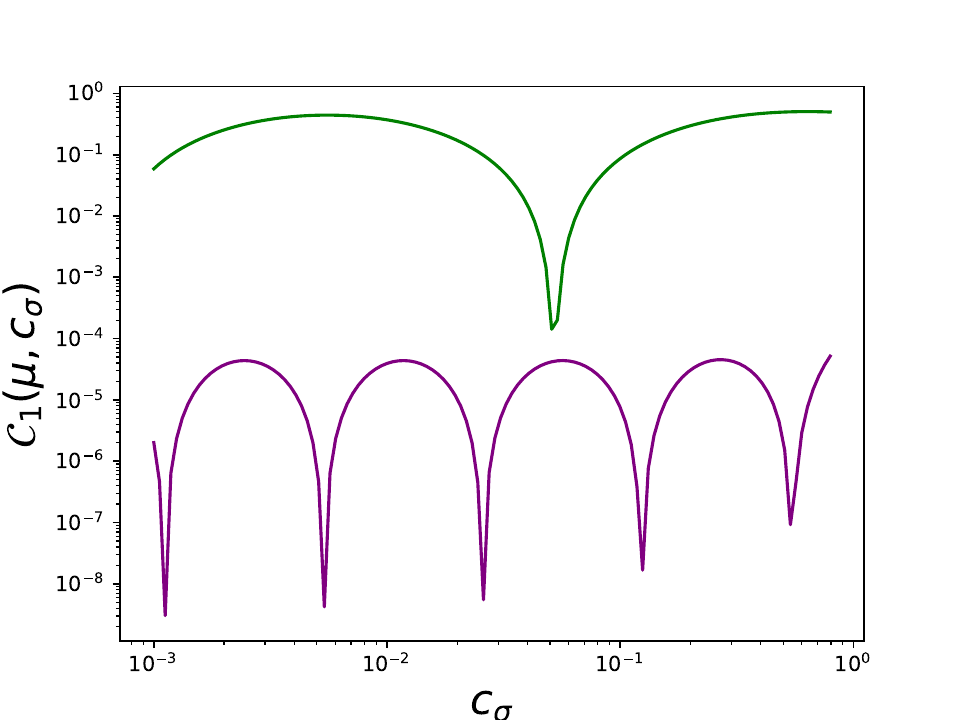}\\
\includegraphics[width=0.5\textwidth]{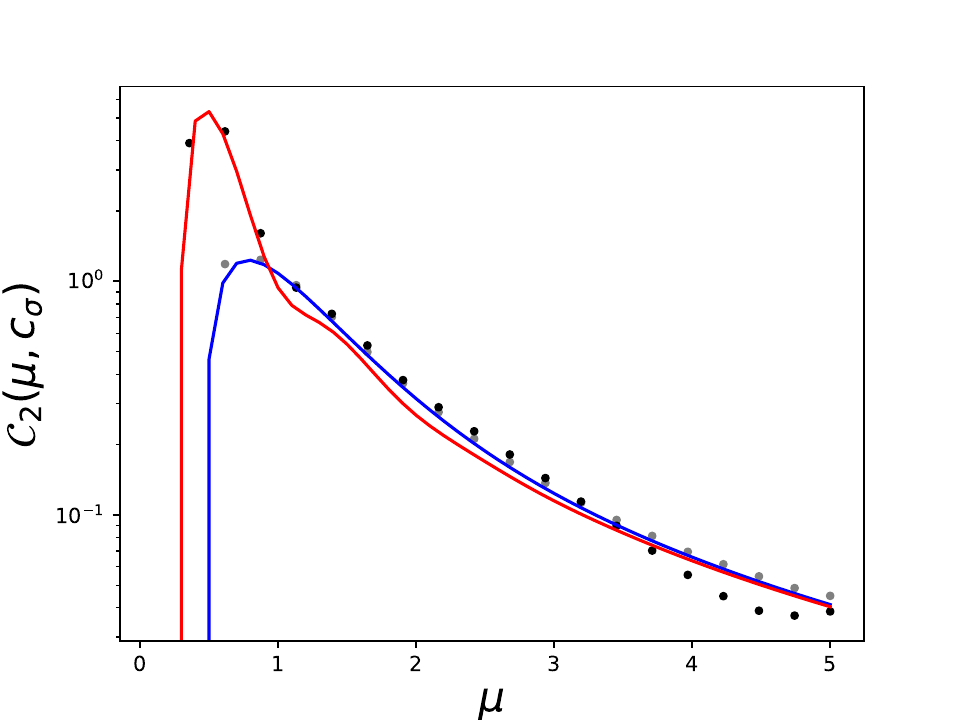}
\includegraphics[width=0.5\textwidth]{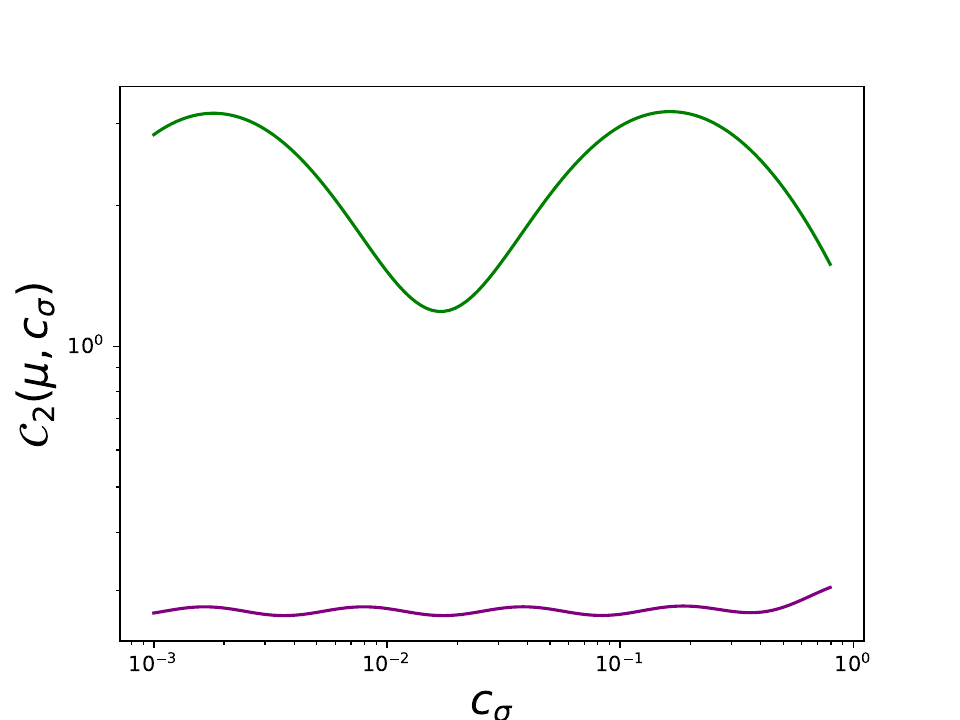}\\
\includegraphics[width=0.5\textwidth]{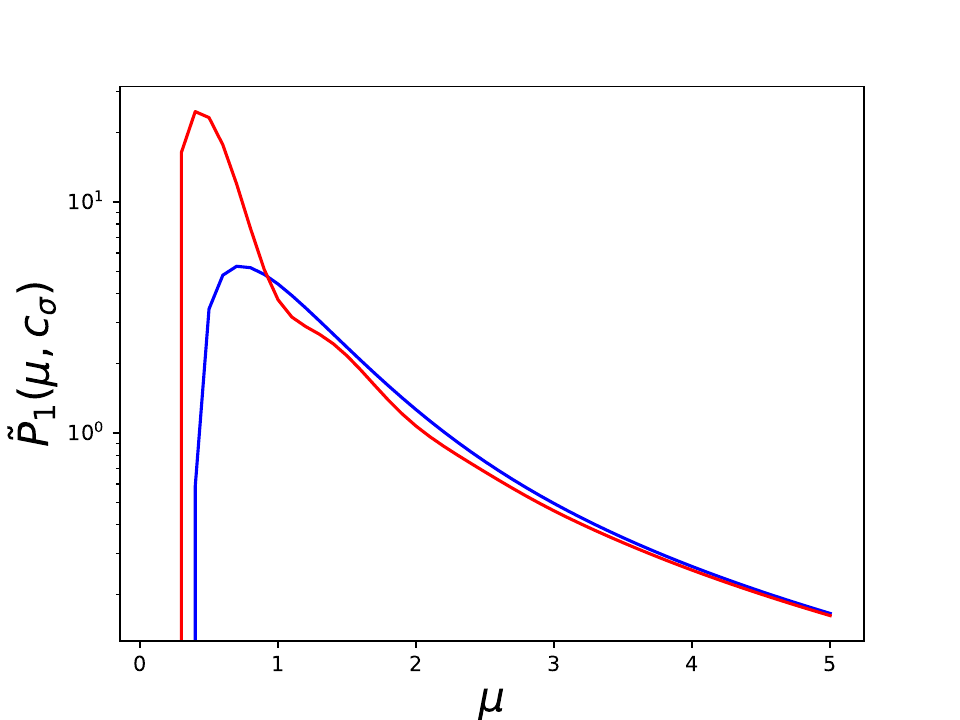}
\includegraphics[width=0.5\textwidth]{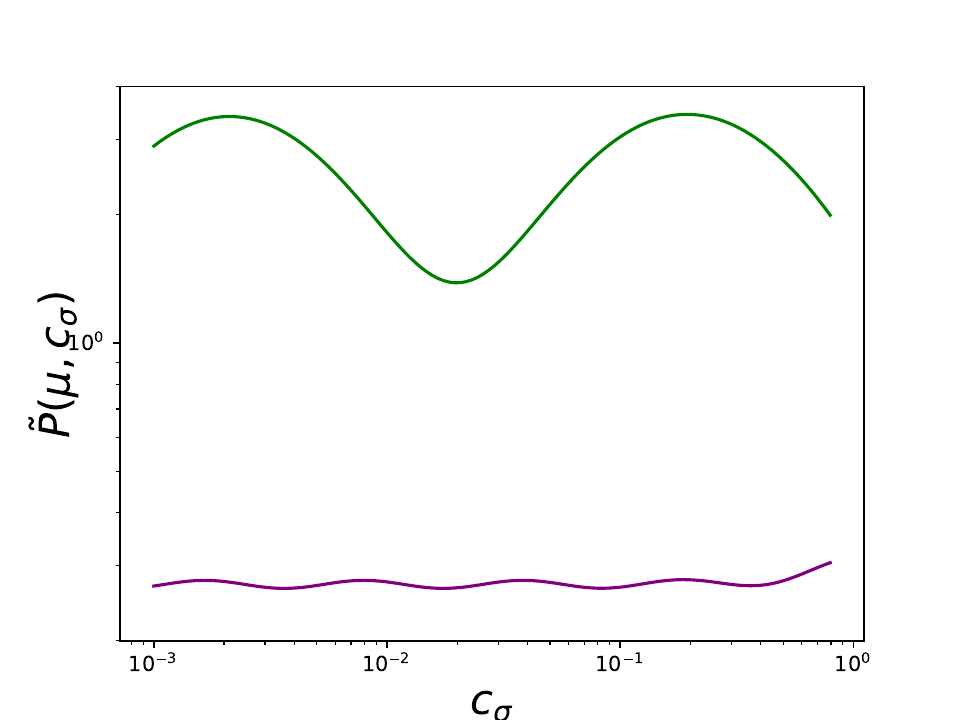}
\caption{\small Power-spectrum correction induced by the mixed propagator.  In the three rows $\mathcal C_1(\mu,c_\sigma)$, $\mathcal C_2(\mu,c_\sigma)$ and $\tilde P(\mu,c_\sigma)=\mathcal C_1+\mathcal C_2$ are plotted using analytical forms in~\eqref{eq:c1expr} and~\eqref{eq:c2expr}. The plots in the {\it left panel} show them as functions of the mass parameter $\mu$, for representative constant sound speeds $c_\sigma=1$ (blue) and $0.1$ (red). For $\mathcal C_{1,2}$, their numerical values are shown in dotted lines for $\mathcal C _1$, and scattered points for $\mathcal C _2$(gray and black for $c_\sigma=1,0.1$ respectively). The plots in the {\it right panel} show the same quantities as functions of $c_\sigma$, for representative masses $\mu=2$ (purple) and $0.7$ (green).  }\label{fig:pow}
\end{figure}
Several features of Fig.~\ref{fig:pow} are physically informative.  First, $\mathcal C_1\geq0$ because it is a squared non-time-ordered amplitude, while $\mathcal C_2$ is an interference contribution and may change sign.  The total correction $\widetilde P$ therefore need not be monotonic in either $\mu$ or $c_\sigma$.  Second, increasing $\mu$ suppresses the response because heavy-particle production becomes inefficient and the time integrals become more oscillatory.  Third, changing $c_\sigma$ alters the relative phase between the graviton and heavy-mode wavefunctions; the visible oscillations therefore arise from interference between contributions with different phases, rather than from a power-law enhancement at small $c_\sigma$.

\begin{table}[htbp]
\centering
\small
\begin{tabular}{|c|c|l|l|}
\hline
$\mu$ & $\cs$ & \multicolumn{1}{c|}{$\mathcal{C}_{1}\simeq$} & \multicolumn{1}{c|}{$\mathcal{C}_{2}\simeq$} \\[4pt]
\hline
$\mu \rightarrow \infty$ & $\cs < 1$ 
& $\displaystyle \frac{8\pi}{\mu} e^{-2\pi\mu}\frac{\cos^2(\mu \theta -\pi/4)}{\sqrt{1-\cs^2}}$
& $\displaystyle \frac{2}{\mu(1+\cs)} \sin\left[ 2\mu \ln\left(\frac{2\cs}{1+\cs}\right) \right] + \mathcal{O}\left(\frac{1}{\mu^2}\right)$ \\[10pt]
& & {\small where $\theta \equiv \cosh^{-1}(1/\cs)$} & \\[6pt]
\hline
Real $\mu$ & $\cs \ll 1$ & Oscillation in $\mu \log (2\cs)$ & Oscillations $\sim e^{\pi\mu}\cos[2\mu\log(2\cs)]$ \\[6pt]
\hline
Real $\mu$ & $\cs \rightarrow 1$ 
& $\displaystyle \frac{\pi^2}{\cs\cosh^2(\pi\mu)}\left(1+\left(\frac{1}{4}+\mu^2\right)(\cs -1)\right)$ 
& $\displaystyle -\frac{1}{2}\mathrm{Re}\left[\psi^{(1)}\left(\frac{1}{4}+\frac{i\mu}{2}\right)-\psi^{(1)}\left(\frac{3}{4}+\frac{i\mu}{2}\right)\right]$ \\[12pt]
\hline
$\mu \ll 1$ & $\cs \leq 1$ & Finite constant & $\displaystyle \mathcal{C}_2(\mu\ll 1, \cs)\simeq -\frac{4\mathcal{J}_{\pm}(\mu \ll 1, \cs)}{(1+\cs)}$ \\[10pt]
\hline
\end{tabular}
\caption{\small Analytical forms of $\mathcal{C}_{1,2}$ at different limits of $\mu , \cs$.}
\label{tab:2}
\end{table}

Asymptotes for $\mathcal{C}_{1,2}$ attained for different limits of $(\mu,\cs)$ are shown in Table~\ref{tab:2}. Details of the analytical forms used for arriving at these limits can be found in Appendices~~\ref{subsec:lim1}-\ref{subsec:C1C2limits}.  
For small $\mu$, $\mathcal{C}_1$ approaches a finite constant, while $\mathcal{C}_2(\mu\ll 1, \cs)\simeq -4\mathcal{J}_{\pm}(\mu \ll 1, \cs)/(1+\cs)$ is real and finite. %Therefore, for a given $\cs$, $\mathcal{C}_2$ attains a maximum at $\mu _{\rm max }(\cs) \simeq  \pi /\bigg(  \left\vert  \ln\left(\frac{2\cs}{1+\cs}\right) \right\vert \bigg)$ with the maximum amplitude $\mathcal{C}_{2, \rm max}\simeq 8\left\vert  \ln\left(\frac{2\cs}{1+\cs}\right) \right\vert/(\pi (1+\cs))$.  
These asymptotic estimations are useful to understand the contributions, particularly re-establishing the expected oscillatory dependence associated with $\cs ^{\pm i\mu}$ in the small $\cs$ limit.  
In the remainder of the analysis, we use the full expressions for $\mathcal{C}_{1,2}$ in Eqs.~\eqref{eq:c1expr},~\eqref{eq:c2expr}, which -- along with the full contribution -- are shown in the Fig.~\ref{fig:pow} as functions of $\mu,\cs$.

The limiting formulae explain the numerical hierarchy in Fig.~\ref{fig:pow}.  At fixed $0<c_\sigma<1$ and large $\mu$, the non-time-ordered term obeys
\begin{equation}
 \mathcal C_1=\mathcal O\!\left(\frac{e^{-2\pi\mu}}{\mu}\right),
 \qquad
 \mathcal C_2=\mathcal O\!\left(\frac{1}{\mu}\right)
 \times\hbox{oscillatory phase},
 \label{eq:C1C2_large_mu_scaling}
\end{equation}
so $\mathcal C_2$ dominates parametrically even though both terms may be comparable near the principal-series threshold.  In the small-$c_\sigma$ expansion, the explicit factor $1/c_\sigma$ in Eq.~\eqref{eq:c1expr} is canceled by the leading $|{}_2F_1|^2\propto c_\sigma$ behavior.  Consequently, $\mathcal C_1$ remains bounded and oscillatory in $\ln c_\sigma$ rather than developing a simple inverse-power enhancement.  These conclusions agree with the qualitative numerical trends: the total correction is largest near small $\mu$, while at larger mass the oscillatory $\mathcal C_2$ contribution controls the residual signal.  The comparison is asymptotic rather than pointwise, since possible cancellations between $\mathcal C_1$ and $\mathcal C_2$ might move the position of local zeros of $\widetilde P$.

This behavior also distinguishes the helicity-two mode of the additional massive spin-two field considered here.  In the latter, the observable tensor mode itself can acquire a mass, a blue spectrum, and a squeezed bispectrum whose scaling is controlled by that mass~\cite{FujitaMassiveGravity2020}.  Here the observable graviton remains massless, and $\widetilde P$ is a mixing correction whose principal-series sound-speed dependence is oscillatory.

\subsection{Tensor spectral index}
\label{sec:ntsec}

The scale dependence of the corrected tensor power spectrum is characterized by the tensor spectral index
\begin{equation}
n_{T}\equiv \frac{d\ln \mathcal{P}_{\gamma}}{d\ln k}
=\frac{d\ln \mathcal{P}_{\gamma}^{0}}{d\ln k}
+\frac{d }{d\ln k}\ln (1+\delta P_{\gamma})
\equiv n_{T,0}+n_{T,\sigma},
\end{equation}
where $\mathcal{P}_{\gamma}^{0}\equiv 2H^2/\pi^2M_{\rm Pl}^2$ is the uncorrected slow-roll tensor power spectrum. The second term contains the scale dependence induced by the mixed-propagator correction in Eq.~\eqref{eq:tenspow}. Using the explicit dependence of $\tilde{P}$ on $\cs$ from Eqs.~\eqref{eq:c1expr} and~\eqref{eq:c2expr}, we obtain
\begin{equation}
n_{T}=-2\epsilon \bigg[1- \frac{\delta P_{\gamma}}{1+\delta P_{\gamma}}
\bigg(1+{1 \over 2\epsilon }\frac{d\ln \cs}{d\ln k}
\frac{d\ln \tilde{P}}{d\ln \cs}\bigg)\bigg].
\label{eq:nT}
\end{equation}
Here $\epsilon\equiv-\dot H/H^2$ is the first slow-roll parameter. Equation~\eqref{eq:nT} assumes that the dimensionful mixing coefficient $\rho$ is approximately time independent over the range of scales considered, and that the slow variation of the mass parameter $\mu$ can be neglected. Consequently,
$\delta P_\gamma=(\rho^2/H^2)\tilde P$ receives a contribution
$d\ln(\rho^2/H^2)/d\ln k\simeq2\epsilon$.
If instead the dimensionless ratio $\rho/H$ is taken to be constant, this contribution is absent. For constant $\cs$, the two prescriptions give
$n_T=-2\epsilon/(1+\delta P_\gamma)$ and $n_T=-2\epsilon$, respectively. Thus the mixed correction does not generate a blue tensor spectrum for constant propagation speed within the perturbative regime.

To parameterize a slowly varying propagation speed, we take
\begin{equation}
\cs(k)\equiv c_{\rm ini}\bigg(\frac{k}{k_*}\bigg)^{s_2},
\label{eq:csdef}
\end{equation}
where $k_*=k_{\rm CMB}=0.05\,{\rm Mpc}^{-1}$ and
$d\ln\cs/d\ln k=s_2$. The scale dependence arises because a slowly varying
$\cs(\tau)$ is evaluated at the characteristic time associated with each
mode. We assume $|s_2|\ll1$ and use the constant-$\cs$ mixed-propagator
kernel with $\cs$ evaluated at the corresponding scale. It is useful to characterize the response of the power correction directly by
\begin{equation}
 \mathcal R_c(\mu,c_\sigma)
 \equiv\frac{d\widetilde P(\mu,c_\sigma)}{d\ln c_\sigma}.
 \label{eq:sound_speed_response}
\end{equation}
In terms of this quantity, the condition for a blue tensor tilt following from Eq.~\eqref{eq:nT} is
\begin{equation}
\left(\frac{\rho}{H}\right)^2 s_2\,\mathcal R_c(\mu,c_\sigma)>2\epsilon .
\label{eq:s2forntrun}
\end{equation}
This form remains regular at the zeros of $\widetilde P$. Since $\widetilde P$ is oscillatory in $\cs$, its derivative may change sign and generate local excursions in $n_T$, but a sustained blue spectrum requires Eq.~\eqref{eq:s2forntrun} to hold over the complete range of scales under consideration.

The functional forms of $\mathcal{C}_{1,2}$ in Eqs.~\eqref{eq:c1expr} and~\eqref{eq:c2expr} show that the principal-series correction is oscillatory in $\cs$, with only a mild variation of its envelope. In the parameter region explored here we do not find a sustained blue tensor spectrum while maintaining perturbative mixing and slowly varying propagation. The resulting tensor tilt therefore remains red and of slow-roll size. A wider numerical scan over $0.1\leq\mu\leq5$ and $-0.5\leq s_2\leq0.01$ gives the same qualitative result, although values with $|s_2|$ approaching unity should only be regarded as diagnostic since they lie outside the strict adiabatic regime.

This behaviour has a direct consequence for the power-spectrum constraint. For constant $\cs$, $\widetilde P(\mu,\cs)$ is independent of scale, so that any residual scale dependence of $\delta P_\gamma$ comes only from the chosen evolution of $\rho/H$. For $s_2<0$, on the other hand, increasing $k$ decreases $\cs(k)$ and moves the system through the oscillatory small-$\cs$ region of the power correction. The perturbativity condition must then be imposed over the complete range of momenta rather than at a single reference scale.

\begin{figure}[!]
\centering
\includegraphics[width=0.49\textwidth]{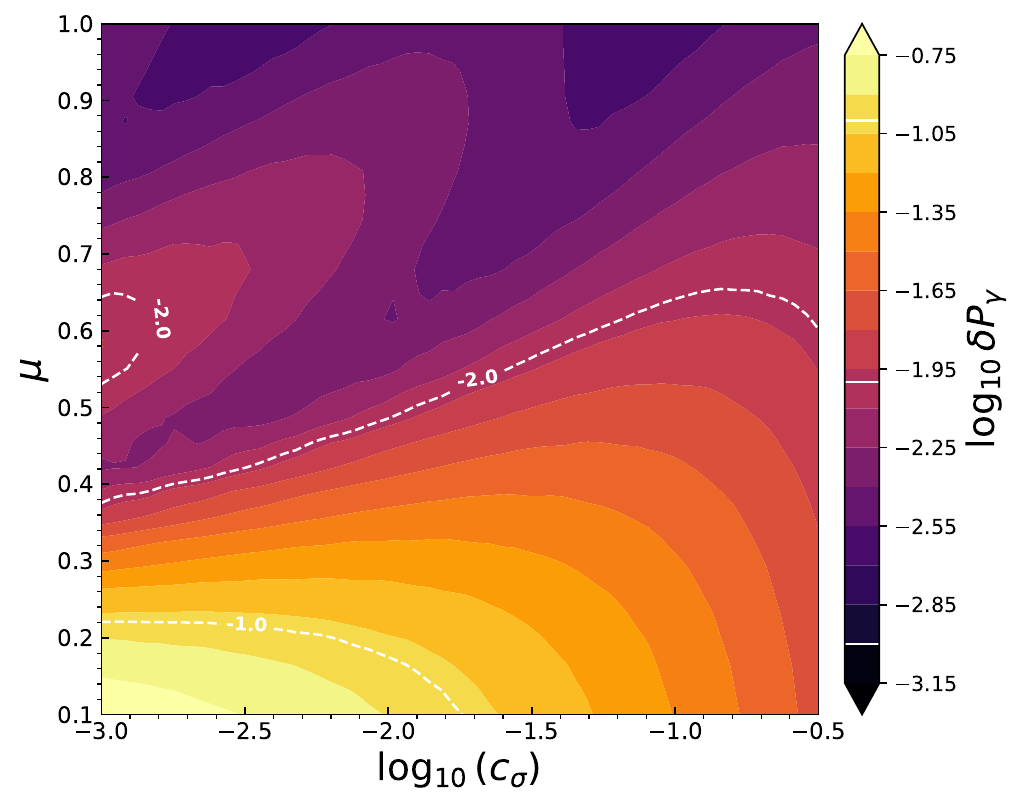}
\includegraphics[width=0.49\textwidth]{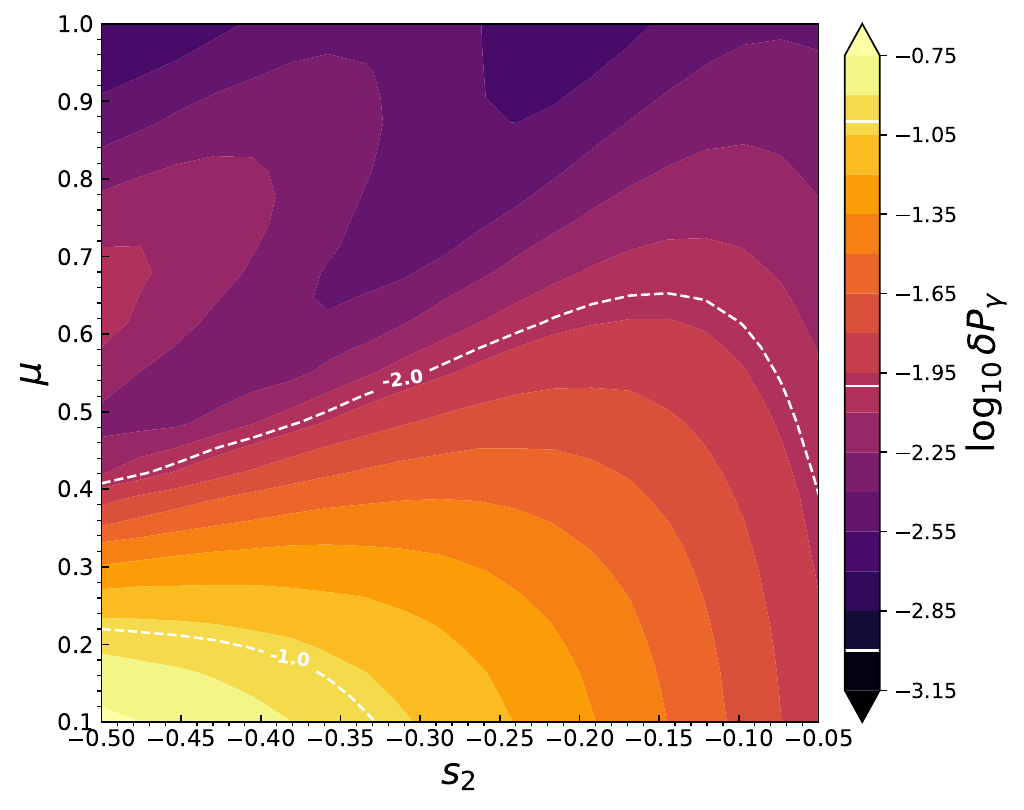}
\caption{\small Contours for $\delta P_{\gamma}$ with $\mu$ and $\cs$ (\textit{left panel}) and with $\mu$ and $s_2$ at $k=10^4$ Mpc$^{-1}$ (\textit{right panel}) for $\rho/H=0.05$.}
\label{fig:contourpower}
\end{figure}

The left panel of Fig.~\ref{fig:contourpower} shows that the largest fractional power corrections are concentrated near the small-$\mu$, small-$c_\sigma$ corner, where the massive field approaches the principal-series threshold and the sound horizon is most separated from the graviton horizon. In the right panel, negative $s_2$ lowers $c_\sigma(k)$ toward high wavenumber and therefore moves the contours toward stronger corrections at fixed $k=10^4\,{\rm Mpc}^{-1}$. These are pointwise constraints from the tensor power spectrum rather than independent strong-coupling bounds. For a finite momentum interval, perturbative control must hold at every $k$ in that interval.

It is useful to contrast this behaviour with the complementary-series spin-2 field studied in Ref.~\cite{Iacconi:2020yxn}. For real $\nu$, the sourced tensor power contains the power-law dependence $\cs^{-2\nu}$, so that the ansatz in Eq.~\eqref{eq:csdef} generates a corresponding power-law scale dependence. In the principal-series regime considered here, $\nu=i\mu$, and the analogous sound-speed dependence becomes oscillatory in $\ln\cs$ rather than producing a monotonic inverse-power enhancement. This is why a slowly varying propagation speed can generate a sizeable tensor tilt in the light-field case while the heavy-field correction considered here remains oscillatory and comparatively mild.

The present-day gravitational-wave energy density associated with the tensor spectrum is
\begin{equation}
    \Omega_{\rm GW}(k,\eta_0)
    = {1\over12}
    \left({k\over a_0H_0}\right)^2
    T_\gamma^2(k,\eta_0)
    \mathcal P_\gamma(k) .
    \label{eq:OmegaGW_transfer_results}
\end{equation}
For the benchmark normalization considered here, the scale dependence generated by the massive field  does not appreciably enhance the high-frequency stochastic background, which remains of order $\mathcal O(10^{-16})$. The  mechanism considered in this work therefore does not by itself bring the primordial tensor background into the sensitivity range of upcoming interferometers. This motivates our focus on the tensor bispectrum and its associated quadrupolar signal at CMB scales.

Additional mechanisms that enhance the tensor power spectrum at small scales could change this conclusion. In that case the total tensor tilt may satisfy
$n_{T,0}+n_{T,\sigma}+n_{T,\mathrm{additional}}>0$, making smaller scales phenomenologically interesting, including those relevant to interferometers such as LISA -- see e.g. \cite{Bartolo:2016ami}. Such an additional tensor-amplification sector is not included here. If present, the oscillatory correction generated by the scale-dependent $\cs$ considered above would be superimposed on the resulting enhanced tensor spectrum.

\subsection{Perturbativity limits}
\label{sec:pertval}
Our main phenomenological goal is to obtain an observable oscillatory signal in the tensor bispectrum from the heavy spin-2 sector. Therefore, it is essential to identify the region of parameter space in which the calculation remains under perturbative control and the effective description is internally consistent. In particular, we can expect that in a cosmological-collider setup  the very same interactions that generate the non-Gaussian signal also modify the two-point function -- while we wish that loop corrections and strong-coupling effects  remain under control. The EFT treatment of additional spinning fields during inflation provides a natural language in which to organize these requirements \cite{Cheung:2007st,Bordin:2018pca,Dimastrogiovanni:2018gkl,Baumann:2020dch}.

In our setup the leading couplings relevant for the tensor observables are the bilinear mixing between the graviton and the extra spin-2 field, schematically controlled by $\rho$, and the cubic self-interaction of the heavy field, controlled by $\kappa$. A minimal perturbativity requirement is therefore
\begin{equation}
\frac{\rho}{H} \ll 1,
\qquad
\frac{\kappa}{H} \ll 1,
\label{eq:basic_perturbativity}
\end{equation}
so that the interaction terms can be treated as perturbations on top of the free quadratic dynamics. In practice, however, Eq.~\eqref{eq:basic_perturbativity} is necessary but not sufficient. Loop corrections and higher-order interaction terms must remain subdominant relative to the contributions retained in the analysis. This requires the strong-coupling scale associated with the EFT operators to remain above the inflationary Hubble scale, and higher-dimensional operators omitted from the truncated action not to compete with the interactions used to generate the signal \cite{Cheung:2007st,Bordin:2018pca,Dimastrogiovanni:2018gkl}.

We also impose a conservative condition on the tensor power spectrum, requiring the sourced contribution to remain parametrically smaller than the vacuum tensor spectrum on all scales where the EFT is applied:
\begin{equation}
\left|\delta P_\gamma(k)\right| \ll 1 .
\label{eq:small_deltaP}
\end{equation}
This is stronger than the requirement of weak mixing itself. In particular, Ref.~\cite{Bordin:2018pca} showed that, for sufficiently small propagation speed, the tensor power spectrum can be dominated by the contribution sourced by the additional spin-2 field while the underlying bilinear mixing remains weak. Their non-perturbative treatment of the coupled graviton--spin-2 system similarly shows that a large sourced contribution does not by itself imply a breakdown of the EFT. We nevertheless impose Eq.~\eqref{eq:small_deltaP} as a conservative working criterion, since it keeps the calculation in the vacuum-dominated regime where the leading mixed-propagator correction to the tensor two-point function is manifestly small.

In the principal-series regime, $\nu=i\mu$, the correction to the tensor power spectrum $\mathcal{P}_{\gamma}(k)$ is oscillatory in the sound speed, as discussed in Section~\ref{sec:ntsec}. For the scale-dependent sound speed in Eq.~\eqref{eq:csdef}, the power-spectrum condition must therefore be checked over the complete range of scales rather than being imposed only at the smallest value of $c_\sigma$. With $\widetilde P=\mathcal C_1+\mathcal C_2$, the corresponding constraint on $\rho/H$ at fixed $\mu$ is
\begin{equation}
\rho/H < (\rho/H)_{\rm max} \equiv
\sqrt{
\frac{\Delta _P}
{\displaystyle \max_{k\in[k_L^{\rm min},k_S]}
\left|\tilde{P}\big(\mu,c_\sigma(k)\big)\right|}
},
\label{eq:rhomax}
\end{equation}
where $\left|\delta P_{\gamma}\right|<\Delta _P$ defines the chosen power-spectrum criterion. For the range of parameter space of our interest, $0.1\leq \mu \leq 5$ and $10^{-3}<\cs<1$, we find a most restrictive value $(\rho/H)_{\rm max} \simeq 0.033$ for $\Delta _P=0.1$, while the less restrictive choice $\Delta _P=1$ gives $(\rho/H)_{\rm max} \simeq 0.1$.

The bound in Eq.~\eqref{eq:rhomax} should therefore be understood as the bound implied by our chosen power-spectrum criterion, rather than as the absolute boundary of weak coupling. A sourced contribution with $|\delta P_\gamma|\gtrsim1$ is not necessarily strongly coupled, although in that regime the leading-order treatment of the quadratic mixing used here should be checked against a treatment of the coupled quadratic system beyond the leading mixing insertion, as in Ref.~\cite{Bordin:2018pca}. Additional restrictions may arise from loop corrections involving $\kappa$, radiatively generated mixed operators, and the small-$c_\sigma$ strong-coupling scale. These requirements must be imposed together with the EFT consistency conditions of Sec.~\ref{sec:pertval} for every $k\in[k_L^{\rm min},k_S]$. For a scale-dependent propagation speed, the local requirements are $0<c_\sigma(k)\leq1$ and $|s_2|\ll1$ throughout the complete momentum interval.
Thus a small sound speed is treated as a controlled enhancement only if it does not lower the strong-coupling scale below $H$ or drive the mixed contribution outside the chosen power-spectrum domain.
In particular, numerical examples with $c_\sigma\sim10^{-3}$ should be interpreted conditionally: they belong to the EFT domain only if the cutoff $\Lambda_\sigma(c_\sigma,\rho,\kappa)$ remains above $H$ and additional operators do not compete with those retained in Eq.~\eqref{eq:mix}. The power-spectrum condition alone does not establish this cutoff hierarchy.

The power-spectrum condition places a bound on the cubic power of the mixing parameter entering the nonanalytic bispectrum. At fixed $(\mu,c_\sigma)$,
\begin{equation}
 \left(\frac{\rho}{H}\right)^3
 <\frac{\Delta_P^{3/2}}{|\widetilde P(\mu,c_\sigma)|^{3/2}}.
 \label{eq:rho_cube_power_bound}
\end{equation}
This limit will be useful below for estimating the maximum bispectrum compatible with the chosen power-spectrum criterion; it should not be interpreted as an absolute upper bound imposed by weak coupling.

\section{Cosmological gravitational-wave collider bispectrum}
\label{subsec:tensor_collider_bispectrum}

\subsection{Full bispectrum and shape}
\label{sec:fullB}

We now compute the primordial tensor bispectrum mediated by the massive $\sigma$ field. The calculation follows directly from the interaction structure in Eq.~\eqref{eq:mix}, with the leading contribution containing three insertions of the bilinear mixing and one insertion of the cubic heavy-sector interaction, shown in the top right panel of Fig.~\ref{fig:sq_cubic}. 
The schematic parametric scaling and nonanalytic structure follow from Eqs.~\eqref{eq:schematic_exchange_bispectrum} and~\eqref{eq:gen_ana_nonana}, respectively, while the detailed momentum dependence is determined by the mixed-propagator time integrals.

Explicitly writing the momentum-conserving delta function in Eq.~\eqref{eq:schematic_exchange_bispectrum}, we define the bispectrum as
\begin{align}
\label{eq:bisp0}
\left\langle
\gamma^{\lambda_1}_{{\bf k}_1}
\gamma^{\lambda_2}_{{\bf k}_2}
\gamma^{\lambda_3}_{{\bf k}_3}
\right\rangle_\sigma
&=(2\pi)^3\delta^{(3)}({\bf k}_1+{\bf k}_2+{\bf k}_3)
\mathcal A^{\lambda_1\lambda_2\lambda_3}
B_\sigma(k_1,k_2,k_3),\\
\label{def:Apol}
\mathcal A^{\lambda_1\lambda_2\lambda_3}
&\equiv
\epsilon^{\lambda_1}_{\alpha\beta}(-\hat k_1)
\epsilon^{\lambda_2}_{\beta\gamma}(-\hat k_2)
\epsilon^{\lambda_3}_{\gamma\alpha}(-\hat k_3) .
\end{align}
The factor $\mathcal A^{\lambda_1\lambda_2\lambda_3}$ contains the spin and helicity information, which can be evaluated for given bispectrum shapes (see Appendix~\ref{appendix:helicity}). The scalar function $B_\sigma$ denotes the contribution of the $\sigma^3$ diagram after the polarization factor $\mathcal A^{\lambda_1\lambda_2\lambda_3}$ has been factored out. In the squeezed limit it can be separated into analytic and nonanalytic contributions,
\begin{equation}
 B_\sigma=B_{\sigma,\rm analytic}+B_{\sigma,\rm nonanalytic}.
 \label{eq:bispectrum_analytic_clock_split}
\end{equation}
The nonanalytic contribution $B_{\sigma,\rm nonanalytic}$ is obtained by isolating the $z^{3/2\pm i\mu}$ branches of the long-wavelength propagator.

A direct in-in evaluation of $B_\sigma(k_1,k_2,k_3)$ contains four time integrations: three associated with the bilinear mixing vertices and one associated with the cubic $\sigma^3$ vertex. The mixed propagators incorporate the three bilinear mixing integrals, leaving a single integration over the time of the cubic vertex. The upper and lower signs on $\mathcal G_\pm$ label the two Schwinger-Keldysh contour branches. Using the mixed propagator derived in Appendix~\ref{appendixA},
\begin{equation}
 \mathcal G_{\pm}(k,\tau)=\frac{\pi\rho H}{4M_{\rm Pl}k^3}I_{\pm}(\mu,c_\sigma;-k\tau).
 \label{eq:mixed_propagator_physical_main}
\end{equation}
The late-time tree-level bispectrum sourced by $\sigma$ can then be written as
\begin{align}
\label{eq:bisp1}
B_\sigma(k_1,k_2,k_3)
&=2\kappa ~{\rm Im}\int _{-\infty}^0 \frac{d\tau}{(-H\tau)^4}\mathcal{G}_{+}(k_1,\tau)\mathcal{G}_{+}(k_2,\tau)\mathcal{G}_{+}(k_3,\tau)\nonumber \\
&=
{\pi^3\over32}
\left({\rho\over M_{\rm Pl}}\right)^3
{\kappa \over H}
{1\over k_2^3k_3^3}
\,\mathcal I_{\nu,c_\sigma}(k_1,k_2,k_3),\\
\label{def:calI}
{\rm where} ~ ~ ~\mathcal I_{\nu,c_\sigma}(k_1,k_2,k_3)
&\equiv
{\rm Im}\int_0^\infty {dz\over z^4}
I_+(\mu,c_\sigma;z)
I_+\!\left(\mu,c_\sigma;{k_2\over k_1}z\right)
I_+\!\left(\mu,c_\sigma;{k_3\over k_1}z\right).
\end{align}

The integration variable is $z=-k_1\tau$. In the first line of Equation~\eqref{eq:bisp1}, the factor $2~{\rm Im}$ is included to account for the fact that two different diagrams with $\sigma _{+}$ and $\sigma_{-}$ at the cubic self-interaction vertices are complex conjugates of each other. Equation~\eqref{def:calI} is written with $k_1$ as the reference momentum; cyclically equivalent forms are obtained by choosing a different reference leg. The apparent asymmetry in $k_2$ and $k_3$ is therefore only a consequence of the chosen parametrization of the time integral \footnote{
It is useful to make this symmetry explicit, since the cubic vertex $\kappa(\sigma_{ij})^3$ does not distinguish the three external heavy legs.  Let
\[
{\cal N}_{\kappa\rho}\equiv
{\pi^3\over32H}
\left({\rho\over M_{\rm Pl}}\right)^3\kappa,
\qquad
(a,b,c)\in\{(1,2,3),(2,3,1),(3,1,2)\} .
\]
A manifestly cyclic representation of the same reduced scalar bispectrum is then
\begin{equation}
\label{eq:bisp1_cyclic_average}
B_\sigma(k_1,k_2,k_3)=
{\cal N}_{\kappa\rho}\,{1\over3}
\sum_{(a,b,c)_{\rm cyc}}
{1\over k_b^3 k_c^3}
\mathcal I_{\nu,c_\sigma}(k_a,k_b,k_c) .
\end{equation}
The factor $1/3$ should not be interpreted as a new combinatorial factor in the in-in diagram.  It only averages three exactly equivalent choices of the variable $z=-k_a\tau$.  In practical numerical work Eq.~\eqref{eq:bisp1} and Eq.~\eqref{eq:bisp1_cyclic_average} agree when the time integrals are evaluated with the same $i\epsilon$ prescription and integration accuracy. In the squeezed calculation we use the ordering
\[
 k_1\simeq k_2\equiv k_S,\qquad k_3\equiv k_L,
\]
so that the nonanalytic part comes from the mixed propagator whose argument is proportional to $k_L/k_S$.  The two short legs must then be kept unexpanded, while the long leg is expanded with Eq.~\eqref{def:calI}.  If instead one chooses $k_1=k_L$ as the reference leg, the same nonanalytic term is recovered after the corresponding cyclic relabelling of Eq.~\eqref{eq:bisp1_cyclic_average}.}.

For comparison with standard non-Gaussian templates, we introduce a dimensionless shape through
\begin{equation}
\label{eq:shape_def_sec44}
\left\langle
\gamma^{\lambda_1}_{{\bf k}_1}
\gamma^{\lambda_2}_{{\bf k}_2}
\gamma^{\lambda_3}_{{\bf k}_3}
\right\rangle_\sigma
=(2\pi)^4\mathcal P_\zeta^2
{S^{\lambda_1\lambda_2\lambda_3}(k_1,k_2,k_3)
\over k_1^2k_2^2k_3^2}
\delta^{(3)}({\bf k}_1+{\bf k}_2+{\bf k}_3).
\end{equation}
With this convention,
\begin{align}
\label{eq:shapepol}
S^{\lambda_1\lambda_2\lambda_3}(k_1,k_2,k_3)
&=
\mathcal C_\kappa\,
\mathcal A^{\lambda_1\lambda_2\lambda_3}
\left({k_1\over k_2}\right)
\left({k_1\over k_3}\right)
\mathcal I_{\nu,c_\sigma}(k_1,k_2,k_3),\\
\mathcal C_\kappa
&\equiv
{\pi^3\over32}
\left({\rho\over M_{\rm Pl}}\right)^3
{\kappa \over H}
{1\over2\pi\mathcal P_\zeta^2} .
\end{align}
The normalization $\mathcal C_\kappa$ contains all dependence on the interaction strengths. The remaining factors determine the momentum dependence and the helicity dependence of the shape.

The nonlinearity parameter, with the polarization factor stripped off, is defined as
\begin{equation}
F_{\rm NL}^{\rm sym}(k_1,k_2,k_3)
    \equiv
\frac{B_\sigma(k_1,k_2,k_3)}
{P_\gamma(k_1)P_\gamma(k_2)+P_\gamma(k_2)P_\gamma(k_3)+P_\gamma(k_3)P_\gamma(k_1)}\,.
\label{eq:FNLgen}
\end{equation}
For a nearly scale-invariant tensor spectrum, $P_\gamma(k)\propto k^{-3}$, using Eqs.~\eqref{eq:bisp1} and~\eqref{eq:FNLgen} gives
\begin{equation}
 F_{\rm NL}^{\rm sym}(k_1,k_2,k_3)
 =
 \frac{\mathcal C_\kappa}{2\pi^3 r_{\rm tens}^2}
 \frac{k_1^3}{k_1^3+k_2^3+k_3^3}
 \mathcal I_{\nu,c_\sigma}(k_1,k_2,k_3),
 \label{eq:FNLgencalI}
\end{equation}
where $r_{\rm tens}\equiv\mathcal P_\gamma/\mathcal P_\zeta$. This expression is dimensionless.

Scale invariance provides a further analytical check on the full three-point function. Under a common rescaling $k_i\rightarrow\lambda k_i$, the reduced time integral $\mathcal I_{\nu,c_\sigma}$ is dimensionless, while the prefactor in Eq.~\eqref{eq:bisp1} scales as $\lambda^{-6}$. Hence
\begin{equation}
 B_\sigma(\lambda k_1,\lambda k_2,\lambda k_3)
 =\lambda^{-6}B_\sigma(k_1,k_2,k_3)
 \label{eq:bispectrum_homogeneity}
\end{equation}
up to slow-roll running and an explicitly scale-dependent $c_\sigma$. Correspondingly, the shape $S^{\lambda_1\lambda_2\lambda_3}$ is invariant under a common rescaling and depends only on two independent momentum ratios. This homogeneity provides a useful numerical check: changing the overall triangle size at fixed momentum ratios must leave the constant-$c_\sigma$ reduced shape unchanged.

We pause here to briefly comment on the breaking of  consistency relations. In single-field inflation, where there is only one adiabatic mode, it turns out the  leading term in the squeezed bispectrum limit is a gauge artifact \cite{Maldacena:2002vr} and is therefore not observable\footnote{See also Ref. \cite{Acquaviva:2002ud} for similar result discussion.}. In this case the leading squeezed-limit bispectrum can be thought of as a result of applying a gauge transformation to the power spectrum. The analysis is similar for scalar, mixed, and tensor bispectra.
One ought to therefore check whether a consistency relation is in place that relates the power spectrum and the bispectrum contribution described by the diagram on the RHS of Fig.~\ref{fig:sq_cubic}. Fortunately, this question has already been addressed in the literature,  specifically for the diagram in hand. The work in \cite{Assassi:2012zq} studied consistency relations in the scalar sector of quasi-single-field (QSF) inflation. As previously mentioned, ours is a Lagrangian that is the exact tensor analogous of the QSF. It was shown in \cite{Iacconi:2020yxn} that the diagram in Fig.~\ref{fig:sq_cubic} breaks consistency relations so that its contribution is directly physical, already at leading order. The quickest route to the result is to notice that the parameter $\kappa$ is entirely absent from the quadratic Lagrangian, yet the squeezed bispectrum is linearly proportional to it. It is straightforward to see then that parametrically the key bispectrum contribution may not be  obtained from the power spectrum.

% colo2

\subsection{Squeezed limit and non-analytic signal}
\label{subsubsec:squeezed_limit_sec44}
In the squeezed limit
\begin{equation}
    k_1\simeq k_2\equiv k_S,
    \qquad
    k_3\equiv k_L\ll k_S,
    \qquad
    r\equiv {k_L\over k_S} ,
\end{equation}
The long leg is evaluated at late mixed-propagator time, so that $I_+(z\ll1)$ controls the nonanalytic part of the answer.  In the massive regime $\nu=i\mu$, the late-time expansion is
\begin{align}
I_+(c_\sigma,z\ll1)
&\simeq
s_+(\mu,c_\sigma)
\left({c_\sigma\over2}\right)^{i\mu}
 z^{3/2+i\mu}
+s_-(\mu,c_\sigma)
\left({c_\sigma\over2}\right)^{-i\mu}
 z^{3/2-i\mu},
\label{eq:Izsmall}
\end{align}
with the branch coefficients defined through Eq.~\eqref{eq:I_small_z_audited}. 
The two powers $z^{3/2\pm i\mu}$ are the late-time de Sitter falloffs of the heavy mode.
Since the long mixed propagator is evaluated at an argument proportional to $k_L/k_S$,
these give rise to $r^{3/2\pm i\mu}$ in the bispectrum. The factor $r^{3/2}$
describes the dilution of the massive mode, while $r^{\pm i\mu}$ produces the
logarithmic oscillations characteristic of the cosmological-collider signal.

\subsection{Squeezed bispectrum for constant sound speed}
\label{subsec:constantcs}

Substituting Eq.~\eqref{eq:Izsmall} into Eq.~\eqref{def:calI}, while keeping the two short-mode mixed propagators unexpanded because they are evaluated near their own sound horizon, gives
\begin{align}
\label{eq:Shapesq}
S^{\lambda_1\lambda_2\lambda_3}(k_S,k_S,k_L)
&\simeq
\mathcal C_\kappa
\mathcal A^{\lambda_1\lambda_2\lambda_3}
 r^{1/2}
{\rm Im}\left[
 w_+(\mu,c_\sigma)
 \left({c_\sigma r\over2}\right)^{i\mu}
+w_-(\mu,c_\sigma)
 \left({c_\sigma r\over2}\right)^{-i\mu}
\right],\\
w_\pm(\mu,c_\sigma)
&\equiv
s_\pm(\mu,c_\sigma)
\int_0^\infty dz\,
I_+^2(\mu,c_\sigma;z)
 z^{-5/2\pm i\mu} .
\label{eq:wpm}
\end{align}
The coefficients $w_\pm$ are evaluated numerically using the contour prescription described in Appendix~\ref{AppendixA2}~\footnote{There is a relative factor $r$ between the definition of the shape $S^{\lambda _1\lambda _2\lambda _3}(k_S,k_S,k_L)\sim r^{1/2}$ and the bispectrum $B_{\sigma}(k_S,k_S,k_L)\sim r^{3/2}$ due to convention in Eqs.~\eqref{eq:shape_def_sec44} and~\eqref{eq:bisp1}. On the other hand, the non-Gaussianity parameter $F_{\rm NL}$ for the nonanalytic collider component scales as $r^{3/2}$, as in Eq.~\eqref{eq:squeezed_signal_schematic}.}.
It is useful to rewrite Eq.~\eqref{eq:Shapesq} in terms of real coefficients,
\begin{align}
\label{eq:shapesqosc}
S^{\lambda_1\lambda_2\lambda_3}(k_S,k_S,k_L)
&\simeq
\mathcal C_\kappa
\mathcal A^{\lambda_1\lambda_2\lambda_3}
 r^{1/2}
\left[
 w_1\sin\left(\mu\log r+\phi\right)
+w_2\cos\left(\mu\log r+\phi\right)
\right],\\
 w_1&\equiv {\rm Re}(w_+-w_-),
\qquad
 w_2\equiv {\rm Im}(w_++w_-),
\qquad
\phi\equiv \mu\log(c_\sigma/2).
\end{align}
The oscillation frequency in $\log r$ is set by $\mu$, while both the amplitude and phase depend on $c_\sigma$. Equation~\eqref{eq:shapesqosc} can also be written as
\begin{align}
\label{eq:shapesqosc2}
S^{\lambda_1\lambda_2\lambda_3}(k_S,k_S,k_L)
&\simeq
\mathcal C_\kappa
\mathcal A^{\lambda_1\lambda_2\lambda_3}
 r^{1/2}
W\cos\left(\mu\log r+\theta\right),
\\
W&\equiv \sqrt{w_1^2+w_2^2},
\qquad
e^{i\Omega}\equiv\frac{w_2-iw_1}{W},
\qquad
\theta\equiv \phi+\Omega .
\end{align}

\begin{figure}[t]
\centering
\includegraphics[width=0.495\textwidth]{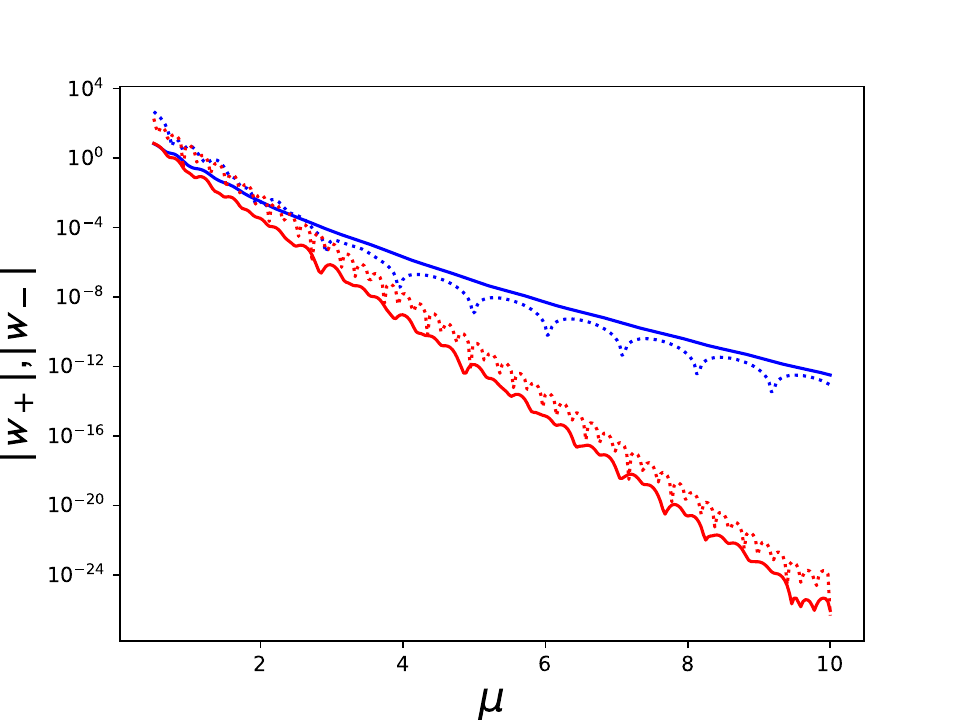}
\includegraphics[width=0.495\textwidth]{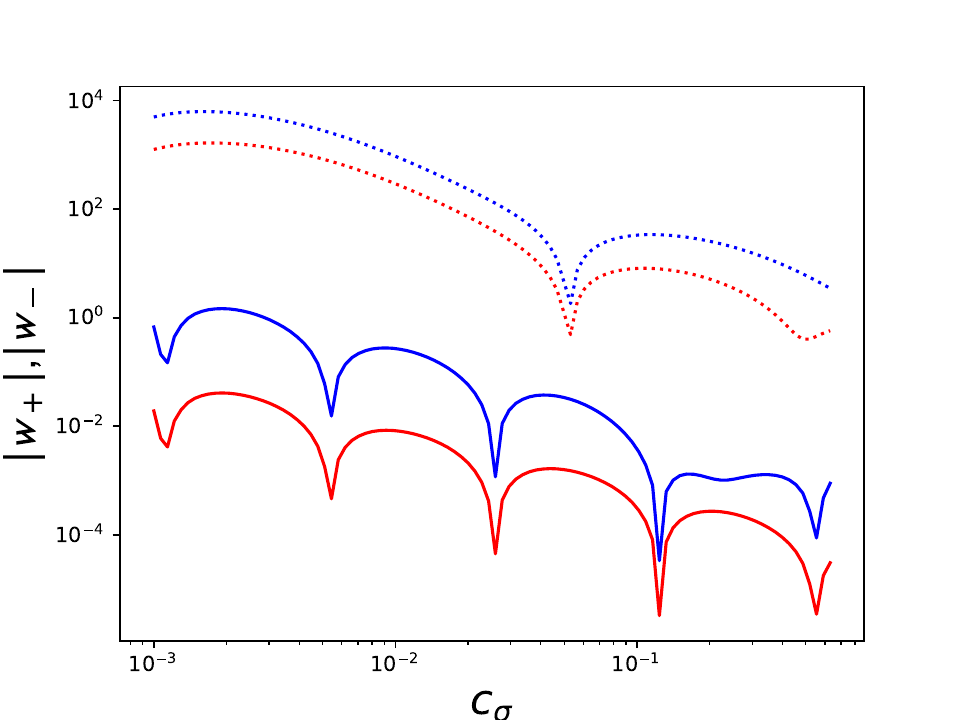}\\
\includegraphics[width=0.495\textwidth]{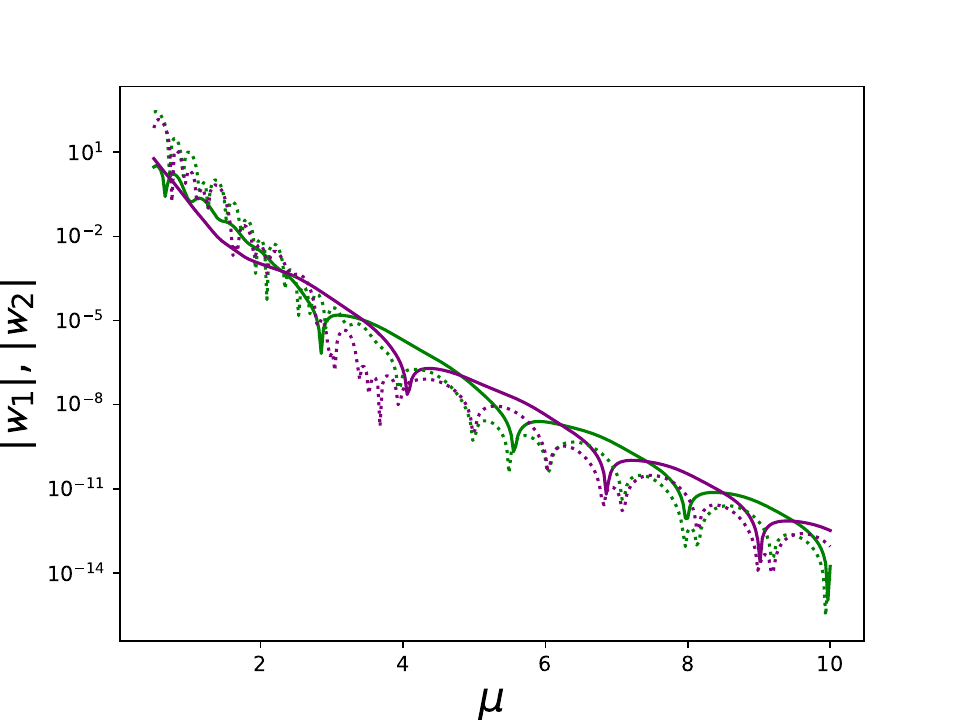}
\includegraphics[width=0.495\textwidth]{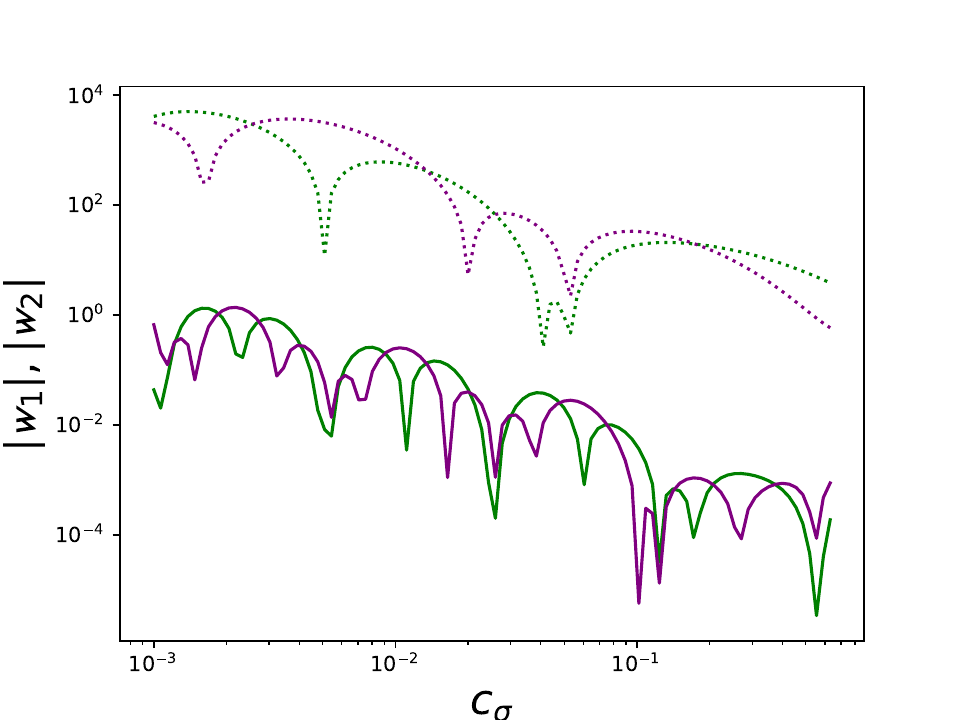}\\
\includegraphics[width=0.47\textwidth]{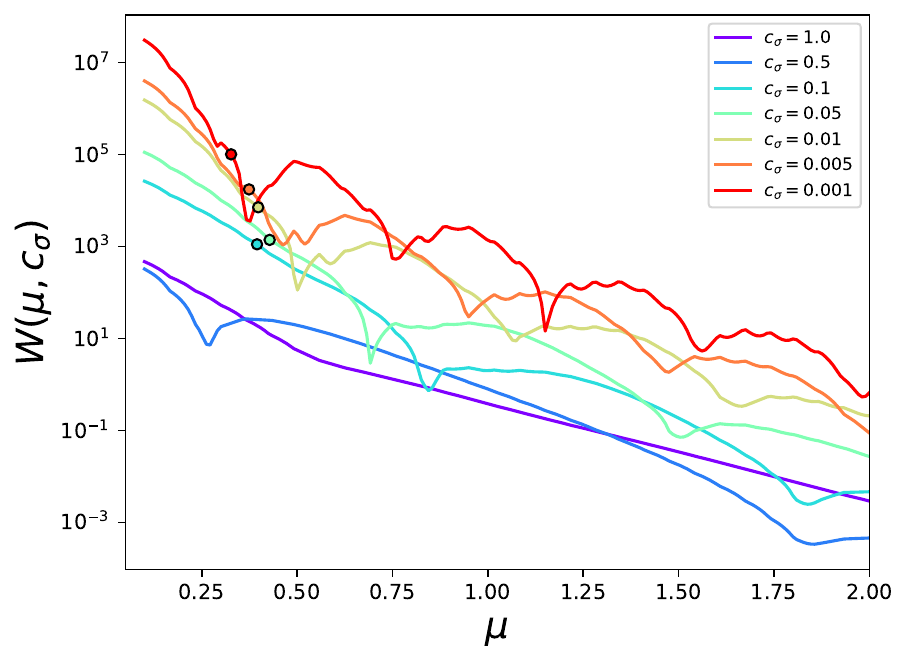}
\includegraphics[width=0.46\textwidth]{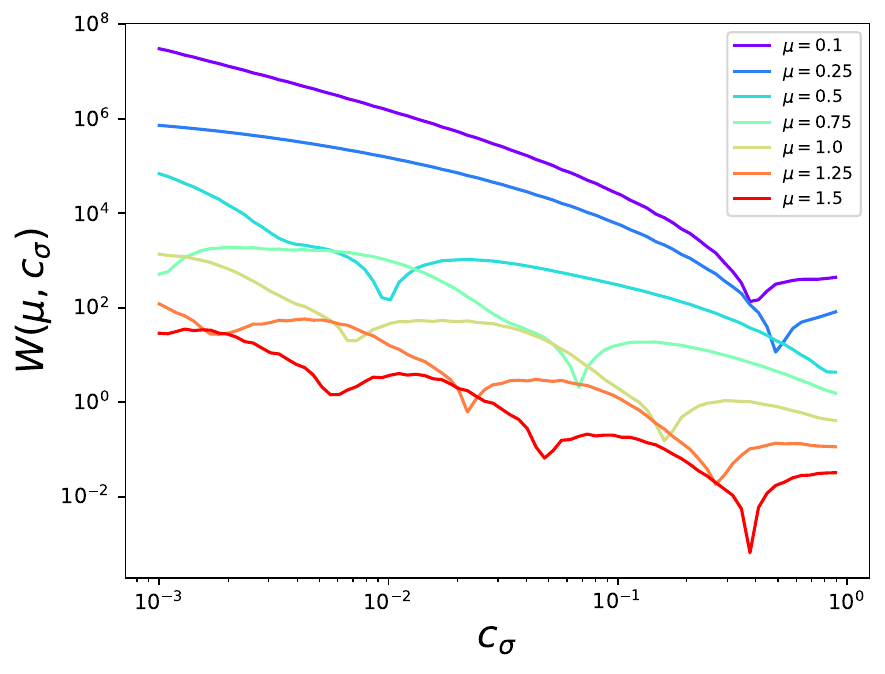}
\caption{\small Mass and sound-speed dependence of the squeezed-limit coefficients. \textit{Top panel:} $w_+(\mu,c_\sigma)$ (blue) and $w_-(\mu,c_\sigma)$ (red). \textit{Middle panel:} $w_1$ (green) and $w_2$ (purple). The {\it lhs} plots show the variation with $\mu$ for $\cs=1$ (solid lines) and $\cs=0.1$ (dotted lines), while the {\it rhs} plots show the variation with $\cs$ for $\mu=2$ (solid lines) and $\mu=0.7$ (dotted lines). \textit{Bottom panel:} $W(\mu,\cs)$ as a function of $\mu$ (left) and $\cs$ (right). The amplitude increases towards small $\mu$ and small $\cs$. In the left panel, the black circles indicate, for each value of $\cs$, the minimum value of $\mu$ below which $\widetilde P(\mu,\cs)>10$.}
\label{fig:wpm}
\end{figure}

Figure~\ref{fig:wpm} shows the dependence of the coefficients on $\mu$ and $c_\sigma$. At large $\mu$, $w_-$ is more strongly suppressed than $w_+$, while the total amplitude is determined by $W=\sqrt{w_1^2+w_2^2}$. A zero crossing of either $w_1$ or $w_2$ therefore changes the phase but does not in general imply a vanishing bispectrum. The coefficients become larger towards small $\mu$, close to the principal-series threshold, and small $c_\sigma$. The correction $\widetilde P(\mu,c_\sigma)$ to the tensor power spectrum can also become large in this region. The black circles indicate the condition specified in the caption; they are not a strong-coupling boundary.

The helicity dependence enters through $\mathcal A^{\lambda_1\lambda_2\lambda_3}$. In the squeezed limit, the $LRR/RLL$ configurations are suppressed because
$\mathcal A^{\lambda_1\lambda_2\lambda_3}\propto r^2$
(see Appendix~\ref{appendix:helicity}). For the $LLL/RRR$ and $LLR/RRL$ configurations, the leading polarization factor instead approaches $1/4$. Thus the logarithmic oscillations and the relative amplitudes of the different helicity configurations contain independent information about the mass and tensor structure of the exchanged field.

\begin{figure}[t]
\centering
\includegraphics[width=0.45\textwidth]{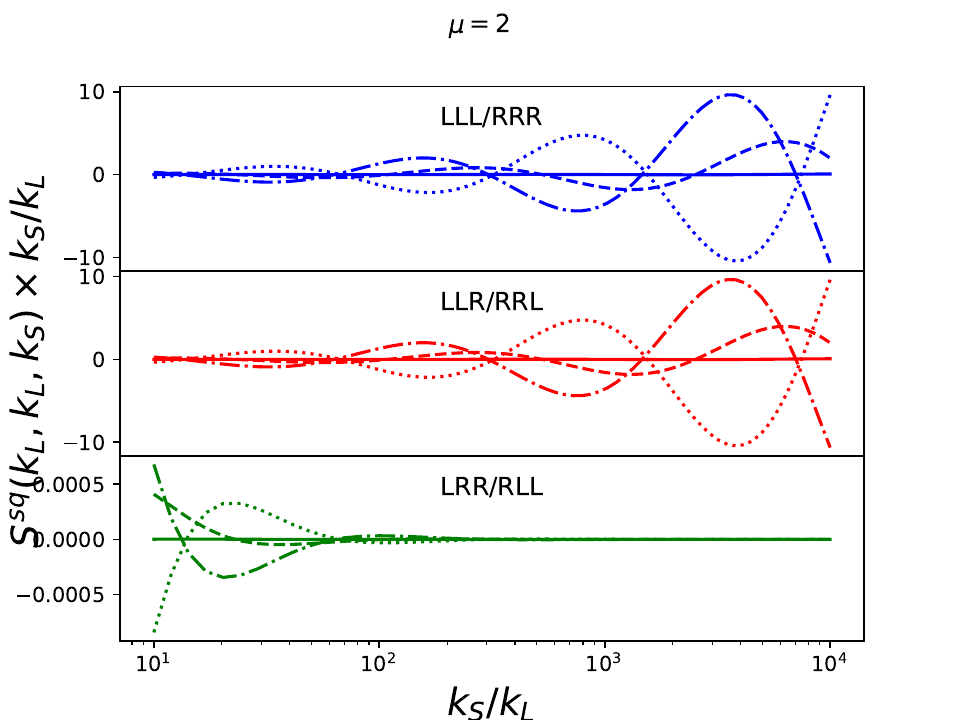}
\includegraphics[width=0.45\textwidth]{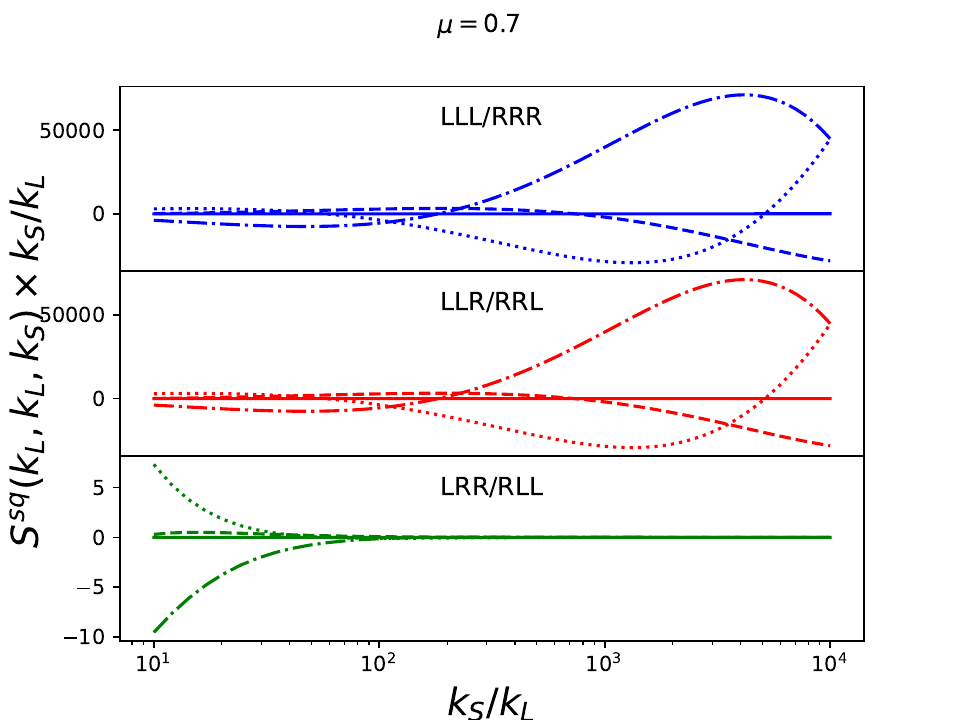}
\caption{\small Squeezed tensor-bispectrum shapes (up to the normalization $\mathcal{C}_{\kappa}$), amplified by $k_S/k_L=1/r$ to make the logarithmic oscillations visible. Left and right panels correspond to different values of the heavy-mass parameter $\mu$. Different colors denote different helicity combinations. Solid, dashed, dotted and dot-dashed curves correspond to constant values $c_\sigma=1,0.01,0.003,0.001$, respectively. Increasing $\mu$ increases the oscillation frequency in $\log(k_L/k_S)$, while varying $c_\sigma$ changes the phase and amplitude.}
\label{fig:5}
\end{figure}

Figure~\ref{fig:5} shows the corresponding squeezed shapes. To make the oscillations visible, the plotted quantity is amplified by $1/r$. Increasing $\mu$ increases the oscillation frequency in $\log r$, while changing $c_\sigma$ modifies the phase and amplitude. The $LRR/RLL$ configurations are suppressed relative to $LLL/RRR$ and $LLR/RRL$.

For constant $c_\sigma$, adjacent maxima and minima are separated by approximately
$\Delta\ln r=\pi/\mu$, while successive maxima are separated by $2\pi/\mu$. Varying $c_\sigma$ changes the phase and amplitude but not this logarithmic frequency. The additional $r^2$ suppression of the $LRR/RLL$ configurations follows from the polarization factor. Since Fig.~\ref{fig:5} plots $S/r$, its visual amplitude differs by one power of $r$ from that of the unrescaled shape and should not be interpreted directly as the squeezed scaling of the bispectrum.

Using the power-spectrum condition in Eq.~\eqref{eq:rho_cube_power_bound}, an envelope amplitude of the form
$F_{\rm NL,sq,env}\propto(\rho/H)^3(\kappa/H)W r^{3/2}$ satisfies
\begin{equation}
 F_{\rm NL,sq,env}^{\rm max}
 \propto
 \left(\frac{\kappa}{H}\right)
 \Delta_P^{3/2}
 \frac{W(\mu,c_\sigma)}{|\widetilde P(\mu,c_\sigma)|^{3/2}}
 r^{3/2},
 \label{eq:FNL_power_limited_scaling}
\end{equation}
up to the stated polarization and normalization factors. When this power-spectrum condition provides the relevant bound on $\rho/H$, the quantity controlling the largest allowed amplitude is therefore $W/|\widetilde P|^{3/2}$ rather than $W$ alone. 
%Near a zero of $\widetilde P$, however, the power-spectrum condition ceases to constrain the mixing, and the weak-mixing and EFT conditions discussed in Sec.~\ref{sec:pertval} must instead be imposed.

\subsection{Squeezed bispectrum for scale-dependent sound speed}
\label{subsubsec:scale_dependent_cs_sec44}

If the propagation speed of the helicity-two mode varies slowly during inflation, we parameterize the value sampled by a mode of wavenumber $k$ using Eq.~\eqref{eq:csdef}. At leading order in this slow variation, the coefficient associated with the long mode is evaluated at $c_\sigma(k_L)$, while the two short-mode kernels are evaluated at $c_\sigma(k_S)$. 

% For the power-law parametrization in Eq.~\eqref{eq:csdef},
% \[
% \frac{c_\sigma(k_L)}{c_\sigma(k_S)}
% =
% \left(\frac{k_L}{k_S}\right)^{s_2}
% =
% r^{s_2}.
% \]
% Hence the phase associated with the long-mode branch satisfies
% \[
% \left(\frac{c_\sigma(k_L)r}{2}\right)^{\pm i\mu}
% =
% \left(\frac{c_\sigma(k_S)}{2}\right)^{\pm i\mu}
% r^{\pm i\mu(1+s_2)} .
% \]
% Keeping this leading scale dependence gives

% Since
% \begin{equation}
%  \frac{c_\sigma(k_L)}{c_\sigma(k_S)}=r^{s_2},
%  \label{eq:cs_long_short_ratio}
% \end{equation}
% the ratio of the explicit $c_\sigma^{-1/2}$ factors contributes $r^{-s_2/2}$. 

\begin{align}
\label{eq:shape-cskdep}
S^{\lambda_1\lambda_2\lambda_3}(k_S,k_S,k_L)
&\simeq
\mathcal C_\kappa
\mathcal A^{\lambda_1\lambda_2\lambda_3}
 r^{1/2}
\Big[
 y_1(\mu,c_{\rm ini},s_2,k_S,k_L)
 \sin\big(\mu(1+s_2)\log r +\phi _S\big)\nonumber \\
&\hspace{2.5cm}
+y_2(\mu,c_{\rm ini},s_2,k_S,k_L)
 \cos\big(\mu(1+s_2)\log r +\phi _S\big)
\Big],
\\
\label{eq:ypm_cskdep}
y_\pm(\mu,c_{\rm ini},s_2,k_S,k_L)
&\equiv
s_\pm(\mu,c_\sigma(k_L))
\int_0^\infty dz\,
I_+^2\!\left(\mu,c_\sigma(k_S);z\right)
 z^{-5/2\pm i\mu},
\\
 y_1&\equiv {\rm Re}(y_+-y_-),
\qquad
 y_2\equiv {\rm Im}(y_++y_-),
\qquad
 \phi_S\equiv \mu \log\!\left[c_\sigma(k_S)/2\right].
\end{align}
For $s_2=0$ and $c_\sigma=c_{\rm ini}$, these expressions reduce to the constant-sound-speed coefficients $w_{1,2}$ defined in Sec.~\ref{subsec:constantcs}. For constant $s_2$, the explicit logarithmic oscillation therefore has frequency $\mu(1+s_2)$ in $\log r$. A scale-dependent logarithmic frequency would instead require a running $s_2$ or a more general evolution of $c_\sigma$.

The dependence of $s_\pm(\mu,c_\sigma)$ on the sound speed can be separated further. From the expression for the coefficients in Appendix~\ref{AppendixA2}, both branches contain the same factor $c_\sigma^{-1/2}$ and the same hypergeometric dependence on $c_\sigma$. 
For the power-law parametrization in Eq.~\eqref{eq:csdef},
\begin{equation}
\frac{c_\sigma(k_L)}{c_\sigma(k_S)}
=
\left(\frac{k_L}{k_S}\right)^{s_2}
=
r^{s_2}.\label{eq:cs_long_short_ratio}
\end{equation}
Hence the phase associated with the long-mode branch satisfies
\begin{equation}
\left(\frac{c_\sigma(k_L)r}{2}\right)^{\pm i\mu}
=
\left(\frac{c_\sigma(k_S)}{2}\right)^{\pm i\mu}
r^{\pm i\mu(1+s_2)}. \nonumber
\end{equation}
Keeping this leading scale dependence and defining the remaining ratio of hypergeometric functions by
\begin{align}
v(\mu,s_2,k_S,k_L)
&\equiv
\frac{_2F_1\!\left({1\over2}-i\mu,{1\over2}+i\mu;1;
{c_\sigma(k_L)-1\over2c_\sigma(k_L)}\right)}
{_2F_1\!\left({1\over2}-i\mu,{1\over2}+i\mu;1;
{c_\sigma(k_S)-1\over2c_\sigma(k_S)}\right)},
\end{align}
the shape can be written as
\begin{align}
\label{eq:ytildepm}
S^{\lambda_1\lambda_2\lambda_3}(k_S,k_S,k_L)
&\simeq
\mathcal C_\kappa
\mathcal A^{\lambda_1\lambda_2\lambda_3}
 r^{1/2-s_2/2}
\Big[
 \tilde{y}_1(\mu,c_{\rm ini},s_2,k_S,k_L)
 \sin\big(\mu(1+s_2)\log r +\phi _S\big)\nonumber \\
&\hspace{2.5cm}
+\tilde{y}_2(\mu,c_{\rm ini},s_2,k_S,k_L)
 \cos\big(\mu(1+s_2)\log r +\phi _S\big)
\Big],
\end{align}
where
\begin{align}
\tilde{y}_\pm(\mu,c_{\rm ini},s_2,k_S,k_L)
&\equiv
s_\pm(\mu,c_\sigma(k_S))\,
v(\mu,s_2,k_S,k_L)
\int_0^\infty dz\,
I_+^2\!\left(\mu,c_\sigma(k_S);z\right)
 z^{-5/2\pm i\mu},
\\
 \tilde{y}_1&\equiv {\rm Re}(\tilde{y}_+-\tilde{y}_-),
\qquad
 \tilde{y}_2\equiv {\rm Im}(\tilde{y}_++\tilde{y}_-).
\end{align}

For $s_2<0$ and $r<1$, Eq.~\eqref{eq:cs_long_short_ratio} implies
$c_\sigma(k_L)>c_\sigma(k_S)$. The change in the bispectrum amplitude is determined by the complete coefficients above: the explicit factor $r^{-s_2/2}$ is only one contribution, while the hypergeometric ratio $v$ and the short-mode integrals also depend on the sound speed.

\begin{figure}[!]
\centering
\includegraphics[width=0.48\textwidth]{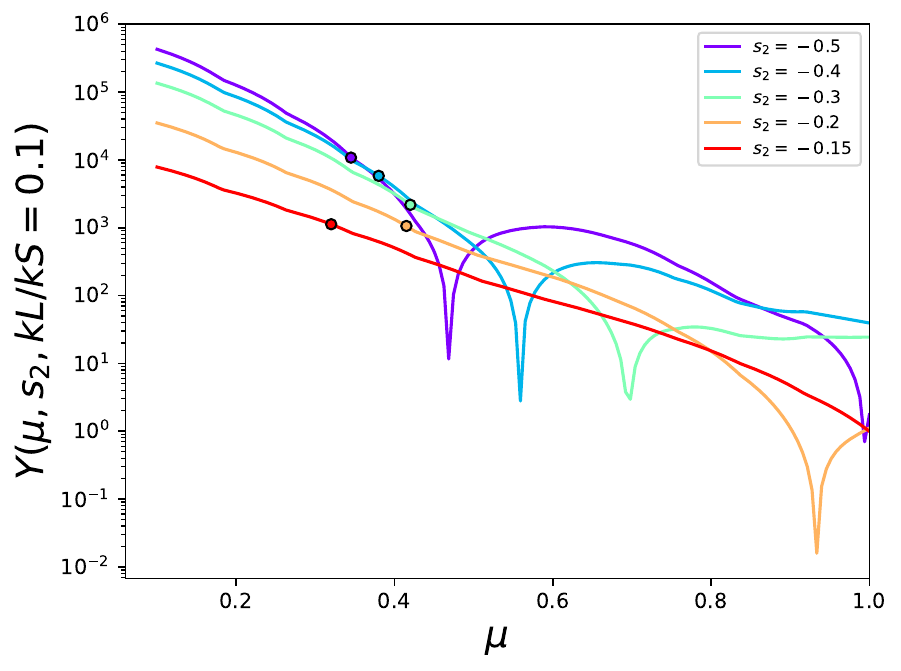}
\includegraphics[width=0.48\textwidth]{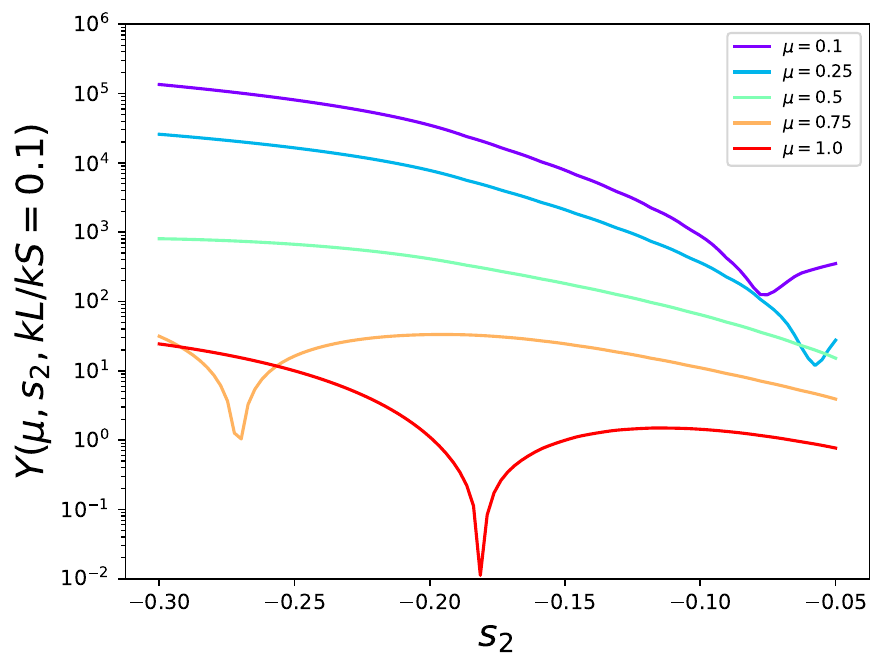}
\caption{\small The coefficient $Y(\mu,s_2)=\sqrt{y_1^2+y_2^2}$ as a function of $\mu$ (left panel) and $s_2$ (right panel) for $k_S=10^4\,{\rm Mpc}^{-1}$ and $k_L/k_S=0.1$. For the parameters shown, $Y$ increases towards small $\mu$ and more negative $s_2$. In the left panel, the black circles indicate, for each value of $s_2$, the minimum value of $\mu$ below which $\widetilde P(\mu,s_2,k=10^4\,{\rm Mpc}^{-1})>10$.}
\label{fig:Ymu}
\end{figure}

Figure~\ref{fig:Ymu} shows
\[
Y(\mu,s_2)\equiv\sqrt{y_1^2+y_2^2}
\]
for $k_S=10^4\,{\rm Mpc}^{-1}$ and $k_L=10^3\,{\rm Mpc}^{-1}$. The left panel shows the dependence on $\mu$ for several values of $s_2$, while the right panel shows the dependence on $s_2$ at fixed $\mu$. For the parameters displayed, the amplitude increases towards the principal-series threshold and towards more negative values of $s_2$. The black circles in the left panel indicate, for each $s_2$, the value of $\mu$ below which $\widetilde P(\mu,s_2,k=10^4\,{\rm Mpc}^{-1})>10$; they do not represent a perturbativity or strong-coupling boundary.

The scale dependence introduced through $s_2$ must be restricted to momentum ranges for which $0<c_\sigma(k)\leq1$ and the variation of $c_\sigma$ is sufficiently slow for the approximation used in Eqs.~\eqref{eq:shape-cskdep}--\eqref{eq:ytildepm}. The usual weak-mixing and EFT conditions must also hold throughout the same range of scales. For quantitative use we require $|s_2|\ll1$ across the complete interval and regard $|s_2|\lesssim0.1$ as the controlled range of the leading adiabatic template. Curves with $|s_2|=0.15$ or $0.2$ are retained only as diagnostic illustrations and are not used for precision forecasts.

\begin{figure}[!]
\includegraphics[width=0.49\textwidth]{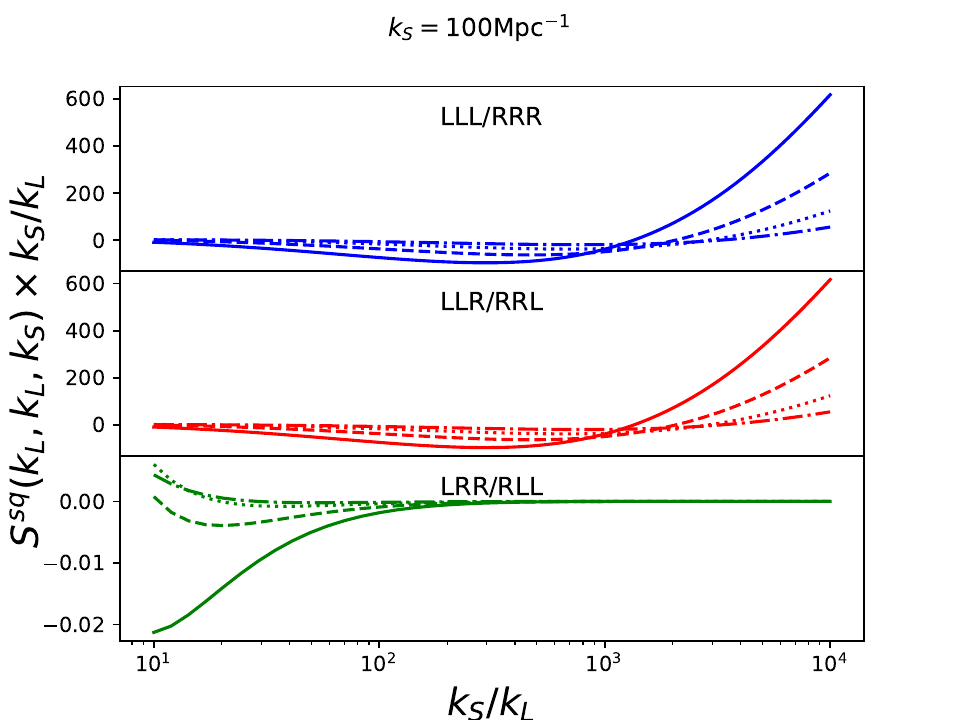}
\includegraphics[width=0.49\textwidth]{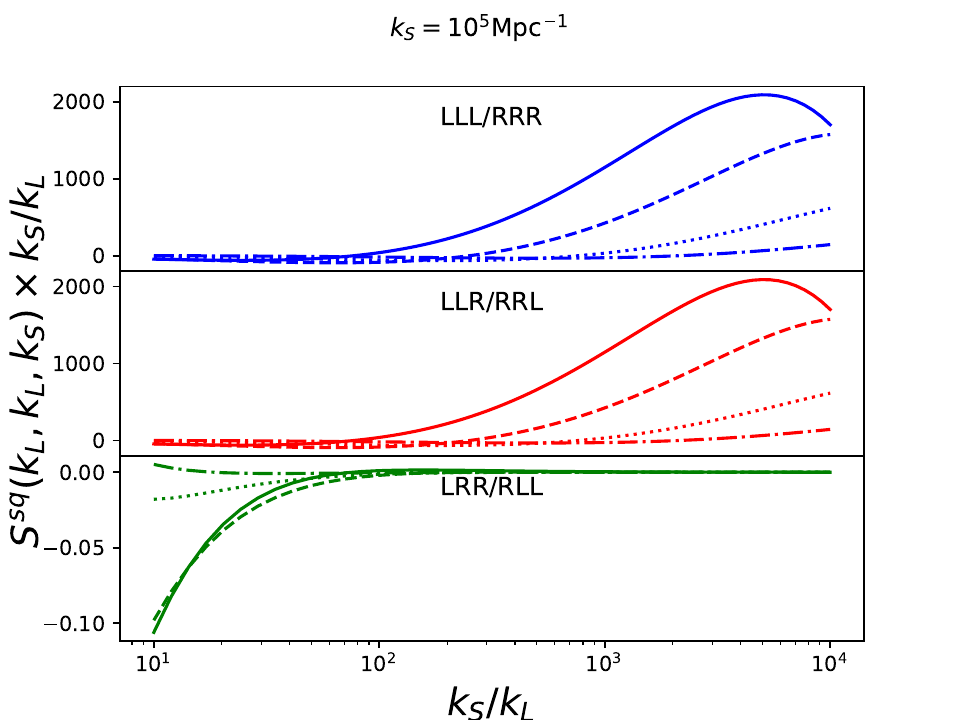}
\caption{\small Squeezed tensor-bispectrum shapes (up to the normalization $\mathcal{C}_{\kappa}$) for the scale-dependent parametrization $c_\sigma(k)=c_{\rm ini}(k/k_*)^{s_2}$, amplified by $k_{\rm S}/k_{\rm L}$. The solid, dashed, dotted, and dot-dashed curves correspond to $s_2=-0.2,-0.15,-0.1,-0.05$, respectively, with $c_{\rm ini}=1$ and $\mu=0.7$. The left and right panels use $k_S=100\,{\rm Mpc}^{-1}$ and $k_S=10^{5}\,{\rm Mpc}^{-1}$, respectively. Changing $s_2$ shifts the logarithmic oscillation frequency from $\mu$ to $\mu(1+s_2)$ and modifies the amplitude through the different values of $c_\sigma$ sampled by the long and short modes.}
\label{fig:shapes2}
\end{figure}

Figure~\ref{fig:shapes2} shows the squeezed shapes for representative values of $s_2$ and two choices of $k_S$. Relative to the constant-$c_\sigma$ result, changing $s_2$ modifies both the logarithmic oscillation frequency and the amplitude. For constant $s_2$, the explicit oscillatory factor has frequency $\mu(1+s_2)$ in $\log(k_L/k_S)$.

For the power-law parametrization with constant $s_2$, the explicit oscillation frequency in $\log r$ is constant. A running $s_2(k)$ or a more general evolution of $c_\sigma$ would instead produce a scale-dependent frequency. The panels at $k_S=100$ and $10^5\,{\rm Mpc}^{-1}$ are theoretical illustrations outside the CMB range used for the final phenomenological analysis and should not be interpreted as direct detector forecasts. As discussed in Sec.~\ref{sec:ntsec}, the present setup does not by itself guarantee an enhanced tensor power spectrum at interferometer scales, so we restrict the phenomenological discussion below to CMB scales.

For completeness, Eq.~\eqref{eq:ytildepm} may be written in terms of a single amplitude and phase. Defining
\[
\widetilde Y\equiv
\sqrt{\tilde y_1^2+\tilde y_2^2},
\qquad
e^{i\delta_Y}\equiv
\frac{\tilde y_2-i\tilde y_1}{\widetilde Y},
\]
one obtains
\begin{equation}
 S^{\lambda_1\lambda_2\lambda_3}(k_S,k_S,k_L)
 \simeq
 \mathcal C_\kappa
 \mathcal A^{\lambda_1\lambda_2\lambda_3}
 r^{(1-s_2)/2}
 \widetilde Y\,
 \cos\!\left[
 \mu(1+s_2)\ln r+\phi_S+\delta_Y
 \right].
 \label{eq:scale_dependent_shape_scaling}
\end{equation}
This form makes explicit the modified logarithmic frequency and the additional $r^{-s_2/2}$ dependence associated with the variation of the propagation speed.

\section{Results}
\label{sec:results}

This section summarizes the primordial observables obtained from the tensor bispectrum derived above. We first define the reduced squeezed bispectrum and its normalization, and then relate it to the quadrupolar modulation induced by a long-wavelength tensor mode. The quantities presented below are primordial amplitudes; a forecast for a specific CMB experiment would additionally require radiation transfer, instrumental noise, delensing, foreground modelling, sky coverage, and an explicit estimator.

For the numerical examples below we focus mainly on constant $c_\sigma$ and use
$r_{\max}=k_L/k_S=0.1$. This avoids introducing additional scale dependence from the propagation speed and allows us to isolate the dependence on the mass, couplings, and helicity structure. The scale-dependent case discussed in Sec.~\ref{subsubsec:scale_dependent_cs_sec44} can be treated in the same way provided $c_\sigma(k)$ remains slowly varying and within the EFT regime over the relevant momentum interval.  
%{\gt{GT: this notation part looks out of place here, maybe it goes at the end of the intro?}}
%\textbf{Notation.}
The dimensionful tensor power spectrum is denoted by $P_\gamma(k)$, while
$\mathcal P_\gamma(k)$ is the corresponding dimensionless spectrum,
as defined in Eq.~\eqref{eq:tenspow} and in the paragraphs following it.
%Throughout the paper,
%$P_\gamma(k)$ and $\mathcal P_\gamma(k)$ denote the sums over the two graviton
%helicities, while
%$P_\gamma^\lambda(k)$ and $\mathcal P_\gamma^\lambda(k)$ denote the
%corresponding quantities for a single helicity. For a parity-symmetric state,
%$P_\gamma^\lambda=P_\gamma/2$.

The tensor bispectrum for specified external helicities is denoted by
$B_\gamma^{\lambda_1\lambda_2\lambda_3}(k_1,k_2,k_3)$ and has momentum
dimension $k^{-6}$ under a common rescaling of the external momenta.
The reduced shape
$S^{\lambda_1\lambda_2\lambda_3}(k_1,k_2,k_3)$ is dimensionless.
The quantities $F_{\rm NL}$ and $F_{\rm NL,sq}$ defined below characterize
the bispectrum amplitude, while $\overline{\mathcal Q^2}$ denotes the
rotationally invariant variance of the induced tensor quadrupole.
%the factor $\mathcal N_\lambda$ where it enters.

\subsection{Reduced squeezed bispectrum}
\label{subsec:reduced_tensor_template}

For specified external graviton helicities, we define the reduced squeezed bispectrum using the tensor power spectrum summed over the two graviton helicities,
\begin{align}
 F_{\rm NL,sq}^{\lambda_S\lambda_S\lambda_L}(k_S,k_L)
 &\equiv
 \frac{B_{\gamma,\rm nonanalytic}^{\lambda_S\lambda_S\lambda_L}(k_S,k_S,k_L)}
 {P_\gamma(k_S)P_\gamma(k_L)},
 \label{eq:FNL}\\
 B_{\gamma,\rm nonanalytic}^{\lambda_1\lambda_2\lambda_3}
 &\equiv
 \mathcal A^{\lambda_1\lambda_2\lambda_3}
 B_{\sigma,\rm nonanalytic}.
 \label{eq:helicity_bispectrum_normalization}
\end{align}
Throughout this subsection we retain only the nonanalytic squeezed contribution; smooth analytic and coordinate-response contributions are not included.

The symmetric denominator in Eq.~\eqref{eq:FNLgen}, evaluated for
$k_1=k_2=k_S$ and $k_3=k_L$, is
$2P_\gamma(k_S)P_\gamma(k_L)+P_\gamma(k_S)^2$.
Neglecting the slow-roll tilt of the tensor spectrum, this gives
\begin{equation}
 F_{\rm NL,sq}^{\lambda_S\lambda_S\lambda_L}
 =2\mathcal{A}^{\lambda_S\lambda_S\lambda_L}F_{\rm NL,sym}
 \left[
 1+{1\over 2}
 \left({k_L\over k_S}\right)^3
 \frac{1+\delta P_{\gamma}(k_S)}
 {1+\delta P_{\gamma}(k_L)}
 \right],
 \label{eq:FNL_squeezed_v_symmetric}
\end{equation}
where $F_{\rm NL,sym}$ here denotes the nonanalytic contribution to the quantity defined in Eq.~\eqref{eq:FNLgen}.

For comparison with scalar-normalized amplitudes, we also define
\begin{equation}
f_{\rm NL}^{\lambda_1\lambda_2\lambda_3}(k_S,k_L)
 \equiv
 \frac{B_{\gamma,\rm nonanalytic}^{\lambda_1\lambda_2\lambda_3}(k_S,k_S,k_L)}
 {(3/5)P_\zeta(k_S)P_\zeta(k_L)}
 =\frac{5r_{\rm tens}^2}{3}
 F_{\rm NL,sq}^{\lambda_1\lambda_2\lambda_3}(k_S,k_L),
\label{eq:fNL}
\end{equation}
where
$r_{\rm tens}\equiv\mathcal P_\gamma/\mathcal P_\zeta$.
Equation~\eqref{eq:fNL} is a definition of the scalar-normalized tensor amplitude used below; it should not be confused with the usual local scalar $f_{\rm NL}$.

Using the late-time behavior of the mixed propagator, the bispectrum defined in Sec.~\ref{sec:fullB}, and the squeezed result of Sec.~\ref{subsubsec:squeezed_limit_sec44}, the nonanalytic contribution is
\begin{align}
B_{\gamma,\rm nonanalytic}^{\lambda_S\lambda_S\lambda_L}(k_S,k_S,k_L)
&=
{\mathcal A}^{\lambda_S\lambda_S\lambda_L}
{\pi ^3 \kappa \over 32H}
\left({\rho \over M_{\rm Pl}}\right)^3
\left({k_L\over k_S}\right)^{3/2}
{1\over k_L^3k_S^3}
\nonumber\\
&\quad\times
W(\mu,c_\sigma)
\cos\!\left[
\mu\ln\!\left({k_L\over k_S}\right)
+\varphi(\mu,c_\sigma)
\right]
\nonumber\\
&=
b^{\lambda_S\lambda_S\lambda_L}(\mu,c_\sigma)
P_\gamma(k_L)P_\gamma(k_S)
\left({\rho\over H}\right)^3
\left({\kappa\over H}\right)
\nonumber\\
&\quad\times
W(\mu,c_\sigma)
\left({k_L\over k_S}\right)^{3/2}
\cos\!\left[
\mu\ln\!\left({k_L\over k_S}\right)
+\varphi(\mu,c_\sigma)
\right],
\label{eq:squeezed_signal_schematic}\\
b^{\lambda_1\lambda_2\lambda_3}(\mu,c_\sigma)
&\equiv
\mathcal A^{\lambda_1\lambda_2\lambda_3}
{\pi ^3\over 2^{9}}
{M_{\rm Pl}\over H}
\left[1+\delta P_\gamma(k_L)\right]^{-1}
\left[1+\delta P_\gamma(k_S)\right]^{-1}.
\end{align}
Here $\varphi(\mu,c_\sigma)\equiv\theta=\phi+\Omega$, with $\theta$ defined in Eq.~\eqref{eq:shapesqosc2}.

For constant $c_\sigma$, the scale dependence of $b$ arises only through the factors
$1+\delta P_\gamma(k_{L,S})$ and through the polarization factor.
In the vacuum-dominated limit $|\delta P_\gamma|\ll1$, the unsuppressed configurations satisfy
\begin{equation}
b^{LLL/RRR}\simeq b^{LLR/RRL}\simeq 9.0\times10^2\nonumber
\end{equation}
for $H\simeq4.1\times10^{13}\,{\rm GeV}$, while
\begin{equation}
b^{LRR/RLL}\simeq
2.26\times10^2\left({k_L\over k_S}\right)^2.\nonumber
\end{equation}

\begin{figure}[!]
\centering
\includegraphics[width=0.68\textwidth]{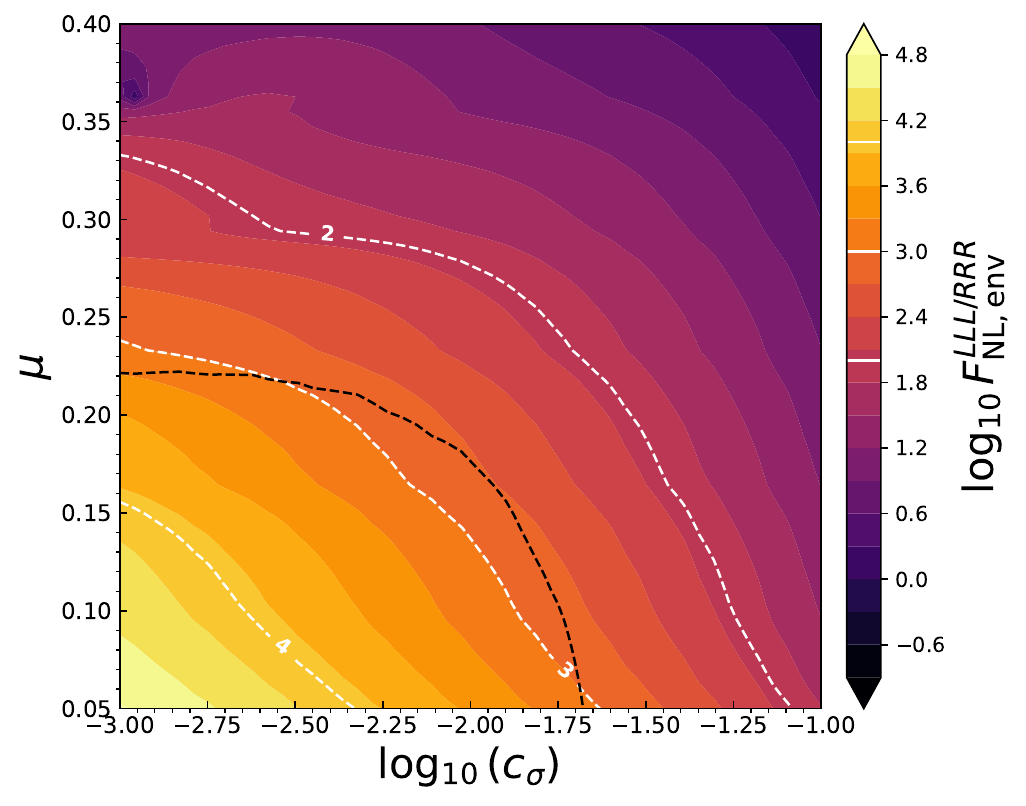}
\caption{\small Amplitude $F_{\rm NL,sq,env}^{LLL/RRR}$ at $k_L/k_S=0.1$ in the $(\mu,c_\sigma)$ plane for $H=4.1\times10^{13}\,{\rm GeV}$, $\rho/H=0.05$, and $\kappa/H=0.5$. The black dashed contour marks the chosen power-spectrum criterion $|\delta P_\gamma|=0.1$. It does not represent a strong-coupling boundary and does not include independent loop or cutoff constraints from the heavy sector.}
\label{fig:FNLcontour}
\end{figure}

Combining Eqs.~\eqref{eq:FNL} and~\eqref{eq:squeezed_signal_schematic}, the reduced squeezed amplitude becomes
\begin{align}
F_{\rm NL,sq}^{\lambda_S\lambda_S\lambda_L}(k_S,k_L)
&\simeq
b^{\lambda_S\lambda_S\lambda_L}(\mu,c_\sigma)
\left({\rho\over H}\right)^3
\left({\kappa\over H}\right)
W(\mu,c_\sigma)
r^{3/2}
\cos\!\left[\mu\ln r+\varphi\right]
\label{eq:FNL_template_constant_cs}\\
&\equiv
F_{\rm NL,sq,env}(\mu,c_\sigma,r)
\cos\!\left[\mu\ln r+\varphi\right].
\end{align}
For the unsuppressed helicity configurations, the factor $r^{3/2}$ gives the leading suppression as the squeezed limit $r\rightarrow0$ is approached. The amplitude
$F_{\rm NL,sq,env}$ contains the coupling dependence, the polarization factor, the normalization by the tensor power spectrum, and the dependence of $W$ on $\mu$ and $c_\sigma$.

Figure~\ref{fig:FNLcontour} shows this amplitude for the $LLL/RRR$ configuration. It is largest towards the principal-series threshold and small $c_\sigma$. The correction to the tensor power spectrum can also become large in this region, and the dashed contour shows the part excluded by the chosen criterion $|\delta P_\gamma|<0.1$. Points beyond this contour are not thereby excluded by weak coupling; they simply lie outside the vacuum-dominated region adopted for the benchmark analysis.\footnote{For comparison, the less restrictive criterion $|\delta P_\gamma|<1$ does not exclude any point in the parameter range shown.}

The helicity dependence follows directly from the polarization factors. In the squeezed limit,
\begin{equation}
 \frac{\mathcal A^{LRR/RLL}}{\mathcal A^{LLL/RRR}}
 =\frac{r^2/16}{1/4}
 =\frac{r^2}{4}.
 \label{eq:helicity_amplitude_ratio}
\end{equation}
The $LRR/RLL$ bispectrum is therefore suppressed by $r^2/4$ relative to the unsuppressed configurations. A quantity quadratic in this response is correspondingly suppressed by $r^4/16$ before summing over helicities. At $r=0.1$, these factors are $2.5\times10^{-3}$ and $6.25\times10^{-6}$, respectively.

When the power-spectrum criterion of Eq.~\eqref{eq:rho_cube_power_bound} provides the active upper bound on $\rho/H$, the maximum allowed amplitude at fixed $\kappa/H$ and $r$ scales as
$W/|\widetilde P|^{3/2}$. The largest value of $W$ alone therefore need not correspond to the largest value compatible with the chosen power-spectrum condition. 
Near zeros of $\widetilde P$, however, this criterion ceases to constrain the mixing, and the weak-mixing and EFT conditions discussed in Sec.~\ref{sec:pertval} must instead be imposed.

For the benchmark shown in Fig.~\ref{fig:FNLcontour}, values
$F_{\rm NL,sq,env}\gtrsim10^3$ remain possible while imposing
$|\delta P_\gamma|<0.1$. For $r_{\rm tens}\simeq0.03$, this corresponds through Eq.~\eqref{eq:fNL} to a scalar-normalized amplitude of order unity. We use Eq.~\eqref{eq:FNL_template_constant_cs} below to compute the quadrupolar modulation induced by long-wavelength tensor modes.

\subsection{From squeezed tensor non-Gaussianity to GW quadrupolar anisotropy}
\label{subsec:quadrupolar_anisotropy_results}

The squeezed tensor bispectrum modulates the short-scale tensor power in the presence of a long tensor mode. In position space, the local tensor power around a point ${\bf x}_c$ may be written as
\begin{equation}
    P_\gamma({\bf k}_S,{\bf x}_c)
    =
    P_\gamma(k_S)
    \left[
    1+\mathcal Q_{ij}({\bf k}_S,{\bf x}_c)
    \hat k_S^i\hat k_S^j+\cdots
    \right].
    \label{eq:local_power_modulation_results}
\end{equation}

To determine the response, consider first a fixed short-mode helicity $\lambda_S$. The fractional modulation produced by a long mode of helicity $\lambda_L$ is proportional to
\[
\frac{
B_{\gamma,\rm nonanalytic}^{\lambda_S\lambda_S\lambda_L}
(k_S,k_S,k_L)}
{
P_\gamma^{\lambda_S}(k_S)
P_\gamma^{\lambda_L}(k_L)
}.
\]
For the parity-symmetric state considered here,
$P_\gamma^\lambda=P_\gamma/2$. It is therefore convenient to define the polarization-independent squeezed amplitude
\[
\mathcal F_{\rm sq}(k_S,k_L)
\equiv
\frac{
B_{\sigma,\rm nonanalytic}(k_S,k_S,k_L)
}{
P_\gamma(k_S)P_\gamma(k_L)
},
\]
so that
\[
F_{\rm NL,sq}^{\lambda_S\lambda_S\lambda_L}
=
\mathcal A^{\lambda_S\lambda_S\lambda_L}
\mathcal F_{\rm sq}.
\]

At leading order in the squeezed limit, the contraction of the two equal short-mode polarization tensors gives
\[
\mathcal A^{\lambda_S\lambda_S\lambda_L}
=
-\frac12
\epsilon^{\lambda_L}_{ij}(-\hat{\bf k}_L)
\hat k_S^i\hat k_S^j
+\mathcal O(k_L/k_S).
\]
The fractional response is therefore the same for the two short-mode helicities at this order, and the quadrupole tensor of the total tensor power is
\begin{equation}
\mathcal Q_{ij}({\bf k}_S,{\bf x}_c)
=
-2
\sum_{\lambda_L}
\int {d^3k_L\over(2\pi)^3}
e^{i{\bf k}_L\cdot{\bf x}_c}
\mathcal F_{\rm sq}(k_S,k_L)
\epsilon^{\lambda_L}_{ij}(-\hat{\bf k}_L)
\gamma^{\lambda_L}_{-{\bf k}_L}.
\label{eq:Qij_results}
\end{equation}
The overall sign depends on the polarization convention and drops out of the variance.

The rotationally invariant quadrupole variance is
\begin{equation}
    \overline{{\mathcal Q}^2}(k_S)
    \equiv
    \left\langle\sum_{m=-2}^{2}|\mathcal Q_{2m}|^2\right\rangle
    =
    {8\pi\over15}
    \left\langle
    \mathcal Q_{ij}\mathcal Q^{\ast ij}
    \right\rangle .
    \label{eq:Qbar_definition_results}
\end{equation}
Using the tensor two-point function for the long mode gives
\begin{equation}
\overline{{\mathcal Q}^2}(k_S)
=
{32\pi\over15}
\int {d^2\hat k_L\over4\pi}
\int_{k_L^{\rm min}}^{k_L^{\rm max}}
{dk_L\over k_L}
\left|
\mathcal F_{\rm sq}(k_S,k_L)
\right|^2
\mathcal P_\gamma(k_L).
\label{eq:Qbar_general_results}
\end{equation}
The sum over the long-mode helicities is already contained in the total tensor spectrum
$\mathcal P_\gamma$, and no additional helicity multiplicity is required.

We expand the quadrupolar modulation as
\begin{equation}
 \mathcal Q_{ij}\hat k^i\hat k^j
 =
 \sum_{m=-2}^{2}\mathcal Q_{2m}Y_{2m}(\hat{\boldsymbol k}),
 \qquad
 \int d^2\hat k\,
 Y_{2m}(\hat k)Y^{\ast}_{2m'}(\hat k)
 =
 \delta_{mm'}.
 \label{eq:Q_spherical_expansion}
\end{equation}
For a symmetric trace-free tensor,
\begin{equation}
 \int d^2\hat k\,
 (\mathcal Q_{ij}\hat k^i\hat k^j)
 (\mathcal Q^{\ast}_{mn}\hat k^m\hat k^n)
 =
 \frac{8\pi}{15}
 \mathcal Q_{ij}\mathcal Q^{\ast ij},
 \label{eq:Q_angular_identity}
\end{equation}
which gives Eq.~\eqref{eq:Qbar_definition_results}. With the polarization normalization in Eq.~\eqref{eq:polnorm}, the transverse-traceless completeness relation is
\begin{equation}
 \sum_{\lambda=L,R}
 \epsilon^{\lambda}_{ij}(\hat{\boldsymbol k})
 \epsilon^{\lambda\ast}_{mn}(\hat{\boldsymbol k})
 =
 \frac12
 \left(
 \Pi_{im}\Pi_{jn}
 +\Pi_{in}\Pi_{jm}
 -\Pi_{ij}\Pi_{mn}
 \right),
 \qquad
 \Pi_{ij}\equiv\delta_{ij}-\hat k_i\hat k_j,
 \label{eq:polarization_completeness_Q}
\end{equation}
and, for the parity-symmetric state,
\begin{equation}
\sum_{\lambda_L}
P_\gamma^{\lambda_L}(k_L)
=
P_\gamma(k_L).
\end{equation}

The lower limit $k_L^{\rm min}$ is set by the largest coherent scale included in the analysis, while $k_L^{\rm max}$ is restricted by the squeezed approximation, for which we take
$k_L/k_S\lesssim0.1$. Motivated by CMB scales, we consider
$10^{-4}\lesssim k_S/{\rm Mpc}^{-1}\lesssim0.2$.

For a scale-invariant long-mode tensor spectrum and constant sound speed, Eq.~\eqref{eq:FNL_template_constant_cs} gives
\[
\mathcal F_{\rm sq}(k_S,k_L)
=
{\pi^3\over2^9}
{M_{\rm Pl}\over H}
\left({\rho\over H}\right)^3
\left({\kappa\over H}\right)
\frac{
W(\mu,c_\sigma)\,
r^{3/2}
\cos[\mu\ln r+\varphi]
}{
[1+\delta P_\gamma(k_S)]
[1+\delta P_\gamma(k_L)]
},
\qquad
r\equiv{k_L\over k_S}.
\]
Using
\[
\mathcal P_\gamma(k)
=
{2H^2\over\pi^2M_{\rm Pl}^2}
[1+\delta P_\gamma(k)],
\]
we obtain
\begin{equation}
\sqrt{\overline{{\mathcal Q}^2}(k_S)}
=
\mathcal A_Q
\left({\rho\over H}\right)^3
\left({\kappa\over H}\right)
W(\mu,c_\sigma)
\sqrt{\mathcal M^2(k_S)},
\label{eq:Q2M2}
\end{equation}
where
\[
\mathcal A_Q
\equiv
{\pi^{5/2}\over2^6\sqrt{15}},
\]
and
\begin{align}
\mathcal M^2(k_S)
 &=
\int_{r_{\rm min}}^{r_{\rm max}}
{dr\over r}\,
r^3
\frac{
\cos^2(\mu\log r+\varphi)
}{
[1+\delta P_\gamma(k_S)]^2
[1+\delta P_\gamma(k_L)]
}.
\label{eq:M2}
\end{align}

For constant $c_\sigma$, constant $\rho/H$, and neglecting slow-roll scale dependence,
$\delta P_\gamma(k_L)=\delta P_\gamma(k_S)\equiv\delta P_\gamma$.
The integral can then be evaluated analytically:
\begin{align}
 \mathcal M^2
 &=
 {1\over(1+\delta P_\gamma)^3}
 \Bigg[
 \frac{r_{\max}^3-r_{\min}^3}{6}
 +\frac{1}{2(9+4\mu^2)}
 \Bigg\{
 r_{\max}^3
 \left[
 3\cos\Theta_{\max}+2\mu\sin\Theta_{\max}
 \right]
 \nonumber\\
 &\hspace{4.7cm}
 -r_{\min}^3
 \left[
 3\cos\Theta_{\min}+2\mu\sin\Theta_{\min}
 \right]
 \Bigg\}
 \Bigg],
 \label{eq:M2_analytic_results}\\
 \Theta_a&\equiv2\mu\ln r_a+2\varphi.
\end{align}
The first term is the phase-averaged contribution, while the second term,
$\Delta I_{Q,\Theta}$, contains the dependence on the phase and the integration limits.
In the vacuum-dominated regime $|\delta P_\gamma|\ll1$, the prefactor
$(1+\delta P_\gamma)^{-3}$ may be approximated by unity.

For $r_{\rm max}\gg r_{\rm min}$,
\begin{align}
 \mathcal M^2
 &\simeq
 {1\over(1+\delta P_\gamma)^3}
 \left[
 \frac{r_{\rm max}^3}{6}
 +\Delta I_{Q,\Theta_{\rm max}}
 \right],
\\
 \sqrt{\overline{\mathcal Q^2}}
 &\simeq
 {\mathcal A_Q\over(1+\delta P_\gamma)^{3/2}}
 \left(\frac{\rho}{H}\right)^3
 \left(\frac{\kappa}{H}\right)
 W(\mu,c_\sigma)
 \sqrt{
 \frac{r_{\rm max}^3}{6}
 +\Delta I_{Q,\Theta_{\rm max}}
 }.
 \label{eq:Q_rms_exact_boundary}
\end{align}
If $|\Delta I_{Q,\Theta}|\ll r_{\rm max}^3/6$, the expansion is
\begin{equation}
 \sqrt{
 \frac{r_{\rm max}^3}{6}
 +\Delta I_{Q,\Theta}}
 =
 \frac{r_{\rm max}^{3/2}}{\sqrt6}
 \left[
 1+
 \frac{3\,\Delta I_{Q,\Theta}}{r_{\rm max}^3}
 +\mathcal O\!\left(
 \frac{\Delta I_{Q,\Theta}^2}{r_{\rm max}^6}
 \right)
 \right].
 \label{eq:Q_rms_phase_expansion}
\end{equation}

The phase-dependent term obeys
\begin{equation}
 \left|
 3\cos\Theta+2\mu\sin\Theta
 \right|
 \leq
 \sqrt{9+4\mu^2},
 \label{eq:phase_vector_bound}
\end{equation}
and therefore
\begin{equation}
 |\Delta I_{Q,\Theta}|
 \leq
 \frac{
 r_{\max}^3+r_{\min}^3
 }{
 2\sqrt{9+4\mu^2}
 }.
 \label{eq:quadrupole_boundary_bound}
\end{equation}
For $r_{\min}\ll r_{\max}$, the ratio of this bound to the phase-averaged contribution is
$3/\sqrt{9+4\mu^2}$. The phase dependence can therefore be order unity near the principal-series threshold and becomes suppressed at large $\mu$. Increasing the logarithmic integration range alone does not suppress this contribution because the integral is weighted towards its upper limit.

If $k_L^{\rm min}$ is fixed, then
$r_{\rm min}=k_L^{\rm min}/k_S$ and a residual $k_S$ dependence remains. This dependence is negligible when $r_{\rm min}\ll r_{\rm max}$. In this limit, the phase-averaged rms anisotropy scales as
\begin{equation}
 \sqrt{\overline{\mathcal Q^2}}
 \propto
 \left(\frac{\rho}{H}\right)^3
 \left(\frac{\kappa}{H}\right)
 \frac{W(\mu,c_\sigma)}
 {[1+\delta P_\gamma]^{3/2}}
 r_{\rm max}^{3/2}.
 \label{eq:quadrupole_scaling_summary}
\end{equation}
When the power-spectrum criterion in Eq.~\eqref{eq:rho_cube_power_bound} provides the active upper bound on $\rho/H$, the corresponding amplitude scales as
$W/|\widetilde P|^{3/2}$, up to the factor
$[1+\delta P_\gamma]^{-3/2}$. %Near zeros of $\widetilde P$, the power-spectrum criterion ceases to constrain the mixing and the weak-mixing and EFT conditions must instead be imposed.

\subsection{Numerical Results and Interpretation}
\label{subsec:numerical_trends_results}

The analytic expressions in the last subsection explain the qualitative behavior of the quadrupolar anisotropy. At fixed couplings, the signal is largest when the mixed propagator is large, which typically occurs for smaller sound speed and for masses close to the Hubble scale. It is suppressed for large $\mu$ because the nonanalytic contribution is Boltzmann suppressed and the oscillatory phase varies more rapidly across the long-mode integral. It is also restricted at very small $c_\sigma$ because the tensor power-spectrum correction grows and constrains the allowed value of $\rho/H$.

\begin{figure}[htbp!]
\centering
\includegraphics[width=0.48\textwidth]{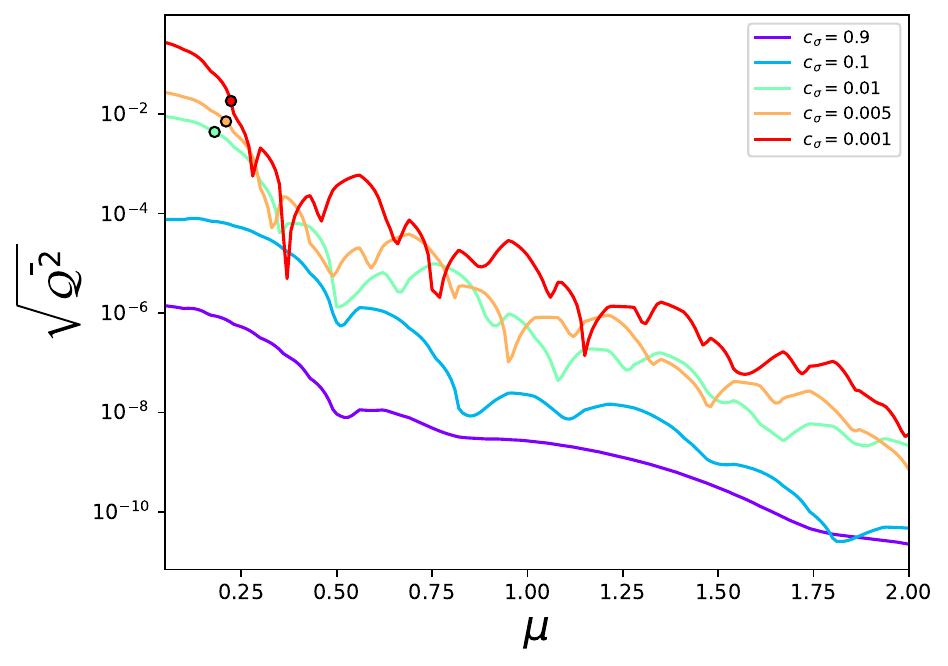}
\includegraphics[width=0.48\textwidth]{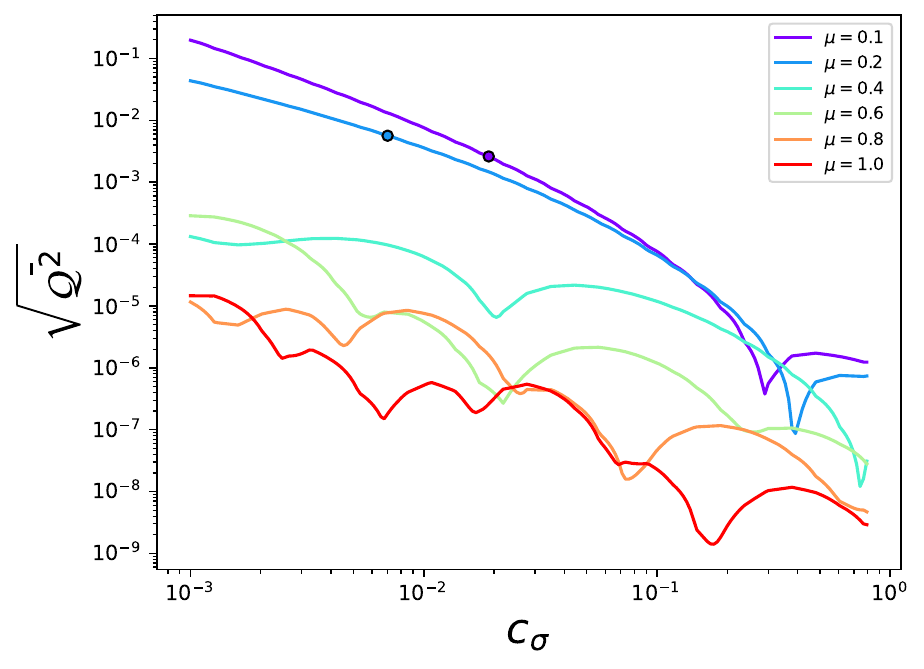}
\caption{\small Primordial rms quadrupolar anisotropy from Eq.~\eqref{eq:Q2M2} for $\rho/H=0.05$ and $\kappa/H=0.5$. The left panel varies $\mu$ at representative values of $c_\sigma$, and the right panel varies $c_\sigma$ at representative values of $\mu$. The markers denote the boundary $\delta P_\gamma(\mu,c_\sigma)=0.1$ for each one-dimensional scan.}
%\blueedit{The numerical values in the text and table use the corrected quadrupole normalization. The external figure file must be regenerated with this normalization.}
\label{fig:Qmu_Qcs}
\end{figure}

The two panels of Fig.~\ref{fig:Qmu_Qcs} display this competition directly. The anisotropy decreases with increasing $\mu$, while decreasing $c_\sigma$ can enhance the response. The markers indicate the points at which $\delta P_\gamma(\mu,c_\sigma)=0.1$ along each one-dimensional scan. They therefore mark the conservative power-spectrum criterion adopted here, rather than a strong-coupling or EFT boundary. The plotted quantity is the total primordial rms quadrupolar anisotropy for the specified parameter choices; it does not include an experiment-specific window function, reconstruction noise, or foreground covariance.

\begin{table}[h]
\begin{center}
\centering
\begin{tabular}{| c | c | c | c | c | c | c | }
\hline
 $\mu$ & $m_{\sigma}/H$ & $\cs$ & $\delta P_{\gamma}$ &
 $F_{\rm NL, env}^{LLL/RRR}$ &
 $f_{\rm NL, env}^{LLL/RRR}$ &
 $\sqrt{\overline{{\mathcal Q}^2}}$ \\
\hline
0.450 & 1.566 & 0.200 & 0.017 & 0.244 & $3.15\times10^{-4}$ & $1.95\times10^{-6}$\\
\hline
0.400 & 1.552 & 0.056 & 0.028 & 2.44 & $3.22\times10^{-3}$ & $1.76\times10^{-5}$\\
\hline
0.337 & 1.537 & 0.010 & 0.042 & 24.4 & $3.30\times10^{-2}$ & $1.23\times10^{-4}$\\
\hline
0.261 & 1.523 & 0.005 & 0.071 & 244 & 0.349 & $2.02\times10^{-3}$\\
\hline
0.1 & 1.503 & 0.002 & 0.200 & $1.22\times10^4$ & 21.9 & $5.66\times10^{-2}$\\
\hline
\end{tabular}
\end{center}
\caption{\small Non-linearity parameters and GW anisotropy for representative values of $\mu$ and $\cs$ with $\rho/H=0.05$, $\kappa/H=0.5$, $H=4.1\times10^{13}\,{\rm GeV}$, and $r_{\max}=0.1$. Corresponding corrections to the tensor power spectrum are quoted in the fourth column. The scalar-normalized amplitudes are evaluated using the tensor-to-scalar ratio implied by the same tensor spectrum.}
\label{tab:1}
\end{table}

For the benchmark choice $\rho/H=0.05$, $\kappa/H=0.5$, $H=4.1\times10^{13}\,{\rm GeV}$, and $r_{\max}=0.1$, the numerical results in Table~\ref{tab:1} exhibit a clear progression as the mass approaches the principal-series threshold and the propagation speed decreases. The non-linearity parameters are evaluated for the unsuppressed $LLL/RRR$ helicity channel, whereas the quadrupolar anisotropy is the total value obtained from Eq.~\eqref{eq:Q2M2}. The first four benchmarks satisfy the conservative condition $\delta P_\gamma\leq0.1$, while the additional illustrative point in the last row has $\delta P_\gamma=0.2$ and therefore lies outside this conservative region, although it remains below $|\delta P_\gamma|=1$.

For $\mathcal P_\zeta=2.1\times10^{-9}$, the chosen value of $H$ corresponds to a vacuum tensor-to-scalar ratio $r_{\rm tens}^{(0)}\simeq0.0274$. Including the sourced tensor-power correction gives $r_{\rm tens}\simeq0.0278$, $0.0281$, $0.0285$, $0.0293$, and $0.0328$ for the five rows of Table~\ref{tab:1}, respectively. The scalar-normalized amplitudes are evaluated consistently using $f_{\rm NL}=(5/3)r_{\rm tens}^2F_{\rm NL,sq}$ with the total tensor-power convention.

The sequence in Table~\ref{tab:1} illustrates the correlated scaling of the observables. Towards smaller $\mu$ and $c_\sigma$, the nonanalytic bispectrum amplitude grows, the rms quadrupole increases, and the tensor power correction simultaneously approaches and eventually exceeds the conservative power-spectrum criterion. With the normalization derived in Sec.~\ref{subsec:quadrupolar_anisotropy_results}, the first four benchmark points give rms anisotropies ranging from $1.95\times10^{-6}$ to $2.02\times10^{-3}$, while the final illustrative point reaches $5.66\times10^{-2}$.

\begin{figure}[b]
\centering
\includegraphics[width=0.55\textwidth]{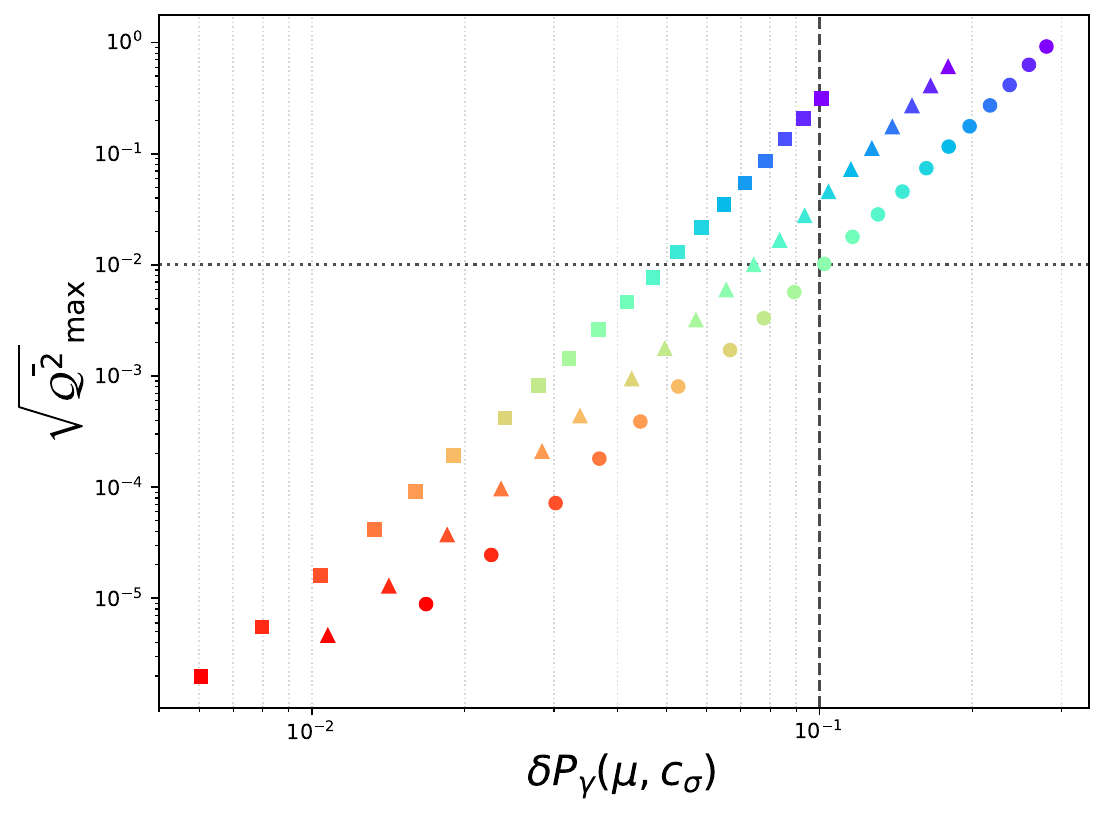}
\caption{\small The rms quadrupolar anisotropy versus the tensor power correction $\delta P_{\gamma}$ as the parameters $\mu$ and $10^{-3}\leq\cs\leq0.3$ are varied, for $\rho/H=0.05$ (circles), $\rho/H=0.04$ (triangles), and $\rho/H=0.03$ (squares). The coupling $\kappa/H=0.5$ is fixed, and colors red to violet signify approaching the bottom-left region of the $(\mu,\cs)$ parameter space.}
%\blueedit{The numerical discussion uses the corrected quadrupole normalization; the external plot must be regenerated consistently.}
\label{fig:Qmu_max}
\end{figure}

Figure~\ref{fig:Qmu_max} shows the correlation between the tensor power correction and the rms quadrupolar response as $\mu$ and $\cs$ are varied. At fixed $\kappa/H$, increasing $\rho/H$ moves points to larger $\delta P_\gamma\propto\rho^2$ and larger rms anisotropy $\sqrt{\overline{\mathcal Q^2}}\propto\rho^3$. Consequently, the three symbol families should not be compared at equal horizontal position, since points with the same $(\mu,c_\sigma)$ have the same color. The color progression tracks motion through parameter space, while departures from a single power law reflect the independent $(\mu,c_\sigma)$ dependence of $W$ and $\widetilde P$.

The benchmark points in Table~\ref{tab:1} at very small sound speed, especially $c_\sigma\lesssim10^{-2}$, remain conditional primordial amplitudes subject to the independent cutoff requirement $\Lambda_\sigma>H$.
Figure~\ref{fig:Qmu_Qcs} should therefore be read as a diagnostic of the primordial response, not as an unconstrained forecast.

\begin{figure}[!]
\centering
\includegraphics[width=0.49\textwidth]{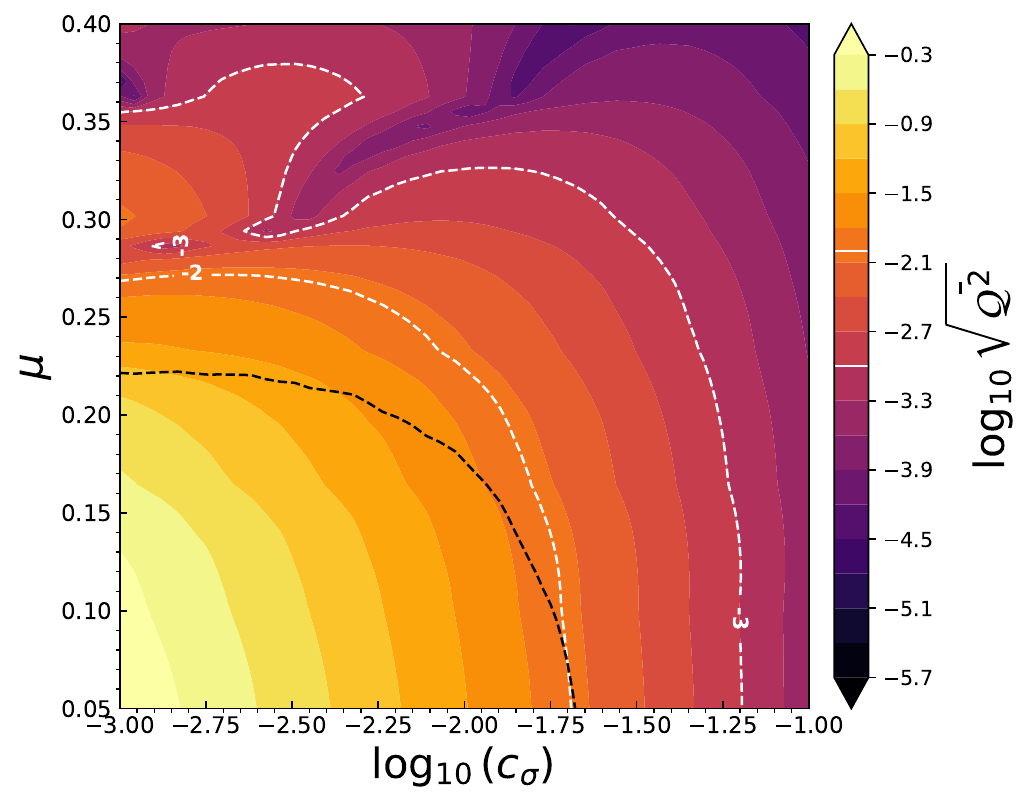}
\includegraphics[width=0.49\textwidth]{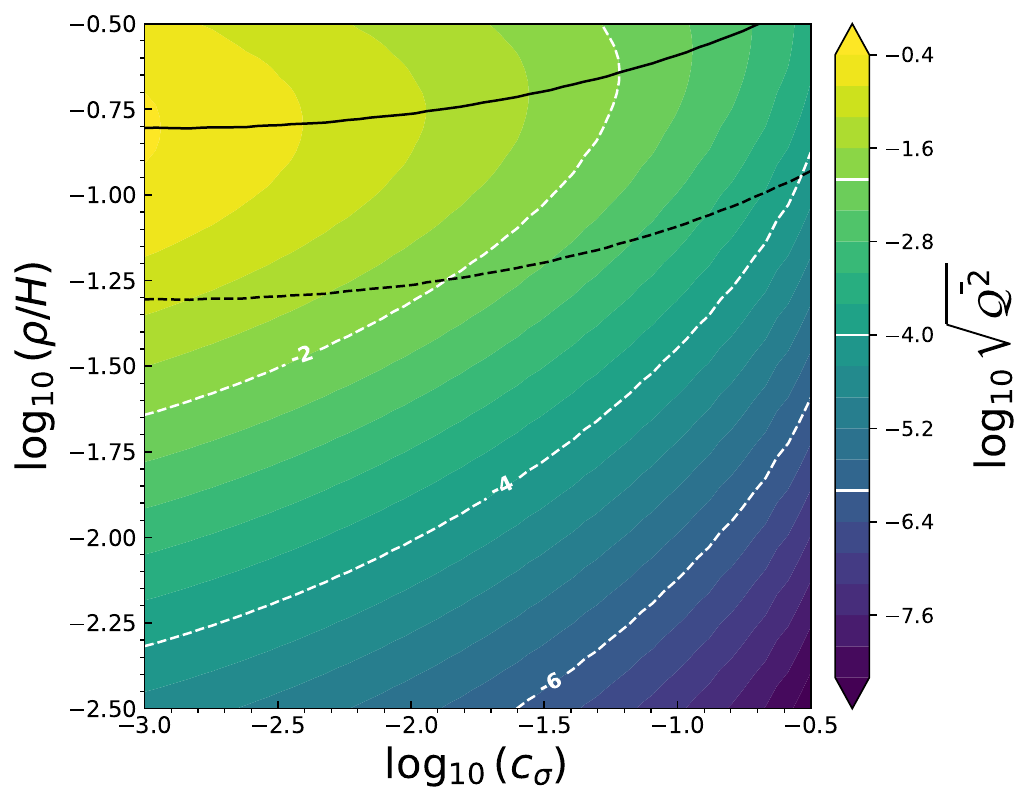}
\caption{\small Contours of the primordial rms quadrupolar tensor anisotropy. \textit{Left:} variation in the $(\mu,c_\sigma)$ plane for $\rho/H=0.05$ and $\kappa/H=0.5$. \textit{Right:} variation in the $(\rho/H,c_\sigma)$ plane for $\mu=0.22$ and $\kappa/H=0.5$. The black dashed curve marks the conservative power-spectrum boundary $\delta P_\gamma=0.1$, while the black solid curve in the right panel marks the relaxed boundary $\delta P_\gamma=1$. These curves constrain the tensor power correction but do not include independent loop or strong-coupling bounds.}
%\blueedit{The external contour file must be regenerated using the corrected quadrupole normalization; the power-spectrum boundaries are unchanged.}
\label{fig:Qcontout1}
\end{figure}

Figure~\ref{fig:Qcontout1} summarizes the primordial quadrupolar response and its power-spectrum restriction. The left panel varies $(\mu,c_\sigma)$ at fixed couplings, while the right panel varies $(\rho/H,c_\sigma)$ at fixed mass. The black dashed curve implements the conservative condition $\delta P_\gamma=0.1$, and the black solid curve in the right panel implements the relaxed condition $\delta P_\gamma=1$. The similarity between the left-panel contours and Fig.~\ref{fig:FNLcontour} follows from their common dependence on $W(\mu,c_\sigma)$, whereas their different powers of $\rho$ distinguish the bispectrum amplitude from the quadrupole variance. These curves constrain the tensor power correction only and do not include independent loop or small-$c_\sigma$ cutoff bounds.

Moving to larger $\rho/H$ increases the anisotropy as $\rho^3$ at the amplitude level and as $\rho^6$ in the variance. Similarly, decreasing $c_\sigma$ can enhance the mixed propagator, but this enhancement is viable only if the EFT remains weakly coupled and the tensor power correction remains within the chosen criterion, together with the independent loop and cutoff requirements.

\section{Discussion and Conclusion}
\label{sec:conclusion}

We have studied the tensor signatures generated by the helicity-two mode of an additional massive spin-two field during inflation. The same set of interactions produces a correction to the tensor power spectrum, a nonanalytic squeezed tensor bispectrum, and a quadrupolar modulation of the short-scale tensor power. Since these observables arise from the same mixed graviton--spin-two propagator, they depend on the same mass, propagation speed, and interaction strengths and cannot in general be varied independently.

The mixed propagator provides a convenient way of organizing the calculation. Using the mixed propagator technique in this work, we arrive at analytical (or moderately semi-analytical) results for the correlation functions. The fractional tensor power correction can be written as
\begin{equation}
 \delta P_\gamma(\mu,c_\sigma)
 =\left(\frac{\rho}{H}\right)^2\widetilde P(\mu,c_\sigma),
 \qquad
 \widetilde P=\mathcal C_1+\mathcal C_2.
 \label{eq:conclusion_power_summary}
\end{equation}
The coefficient $\mathcal C_1$ is the non-time-ordered contribution, while $\mathcal C_2$ arises from the time-ordered part of the in-in integral and can have either sign. At large $\mu$, $\mathcal C_1$ is exponentially suppressed and $\mathcal C_2$ gives the leading inverse-mass contribution with oscillatory dependence on the parameters. Near the principal-series threshold both terms can be important, including regions in which cancellations make the numerical evaluation more delicate.

The corresponding nonanalytic three-point function is, in the squeezed limit,
\begin{equation}
 B_{\gamma,\rm nonanalytic}^{\lambda_1\lambda_2\lambda_3}
 =
 b^{\lambda_1\lambda_2\lambda_3}(\mu,c_\sigma)\,
 P_\gamma(k_L)P_\gamma(k_S)
 \left(\frac{\rho}{H}\right)^3
 \left(\frac{\kappa}{H}\right)
 W(\mu,c_\sigma)\,
 r^{3/2}
 \cos\!\left(\mu\ln r+\varphi\right),
 \label{eq:conclusion_clock_summary}
\end{equation}
with $r\equiv k_L/k_S$ and
$b^{\lambda_1\lambda_2\lambda_3}\propto
\mathcal A^{\lambda_1\lambda_2\lambda_3}(M_{\rm Pl}/H)
[1+\delta P_\gamma(k_L)]^{-1}
[1+\delta P_\gamma(k_S)]^{-1}$, as given explicitly in Eq.~\eqref{eq:FNL_template_constant_cs}. The factor $r^{3/2}$ is the characteristic nonanalytic squeezed scaling of the massive intermediate state, while the logarithmic frequency $\mu$ determines its mass in Hubble units. The polarization contraction provides an independent dependence on the external helicities. In particular, the $LLL/RRR$ and $LLR/RRL$ configurations remain unsuppressed at leading order, whereas the $LRR/RLL$ amplitude is smaller by $r^2/4$ relative to an unsuppressed configuration.

The tensor bispectrum and the tensor power correction are correlated. If the adopted power-spectrum criterion
$|\delta P_\gamma|<\Delta_P$ supplies the active upper bound on $\rho/H$, then
\begin{equation}
 F_{\rm NL,sq,env}^{\rm max}
 \propto
 \left(\frac{\kappa}{H}\right)
 \Delta_P^{3/2}
 \frac{W(\mu,c_\sigma)}{|\widetilde P(\mu,c_\sigma)|^{3/2}}
 r^{3/2},
 \label{eq:conclusion_optimized_signal}
\end{equation}
up to the normalization and polarization factors given in Sec.~\ref{sec:results}. In this regime the relevant combination is therefore $W/|\widetilde P|^{3/2}$ rather than $W$ alone. This statement does not apply near zeros of $\widetilde P$, where the power-spectrum criterion ceases to constrain the mixing and the weak-mixing, loop, and EFT conditions must instead be imposed.

For the scale-dependent ansatz $c_\sigma(k)=c_{\rm ini}(k/k_\ast)^{s_2}$ with constant $s_2$, the ratio of long- and short-mode sound speeds is $c_\sigma(k_L)/c_\sigma(k_S)=r^{s_2}$. Within the horizon-crossing approximation used here, the logarithmic oscillation frequency becomes $\mu(1+s_2)$ and the amplitude acquires additional scale dependence, including the explicit factor $r^{-s_2/2}$. A direct treatment of $c_\sigma(\tau)$ would be required to determine more precisely the range of validity of this approximation.

The squeezed bispectrum also induces a quadrupolar modulation of the short-scale tensor power. For constant $c_\sigma$, a scale-invariant long-mode spectrum, $r_{\min}\ll r_{\max}$, and after phase averaging, the rms quadrupole scales as
\begin{equation}
 \sqrt{\overline{\mathcal Q^2}}
 \propto
 \left(\frac{\rho}{H}\right)^3
 \left(\frac{\kappa}{H}\right)
 \frac{W(\mu,c_\sigma)}
 {[1+\delta P_\gamma]^{3/2}}
 r_{\max}^{3/2}.
 \label{eq:conclusion_quadrupole_summary}
\end{equation}
In the vacuum-dominated limit $|\delta P_\gamma|\ll1$, the denominator may be set to unity. The exact result retains a phase-dependent boundary term, which can be important near the principal-series threshold and becomes less important at large $\mu$. If the power-spectrum criterion is the active upper bound on $\rho/H$, the largest allowed quadrupole has the same parametric dependence $W/|\widetilde P|^{3/2}$ as the corresponding squeezed nonlinearity.

Some aspects of these results follow directly from the analytical structure of the calculation. These include the nonanalytic $r^{3/2\pm i\mu}$ dependence, the leading helicity hierarchy, and the powers of $\rho$ and $\kappa$. By contrast, results at very small sound speed, particularly $c_\sigma\lesssim10^{-2}$, remain conditional on the cutoff staying above $H$ and on loop and higher-order corrections remaining controlled. The power-spectrum criterion used in the numerical analysis is a conservative restriction on the sourced tensor power and should not be identified with the fundamental weak-coupling boundary.

The amplitudes presented here are primordial quantities rather than observational forecasts. Planck uncertainty for the squeezed, scale-invariant tensor-tensor-tensor template is about $40\pm80$, while forecasts suggest roughly a factor of 3 improvement in $\sigma(f_{\rm NL}^{ttt,\rm sq})$ from a future T+E analysis, potentially increasing to as much as a factor of 100 when B-mode polarization is included~\cite{Philcox:2024wqx, Shiraishi:2019yux, Shiraishi:2013vha}. Therefore, future CMB observations such as LiteBIRD~\cite{LiteBIRD:2024twk} are hopeful to probe the gravitational wave collider signal and corresponding quadrupolar anisotropy predicted in this work, where $f_{\rm NL}$ can reach values as large as  $\mathcal{O}(1-10)$ (see Table~\ref{tab:1}). An explicit comparison  with CMB data would require radiation transfer, estimator normalization, instrumental noise, foregrounds, sky coverage, and delensing. The quadrupolar modulation provides an alternative way of probing the same squeezed coupling through anisotropic short-scale tensor power. In the minimal setup considered here, however, the tensor spectrum remains red or nearly scale invariant, so the signal is not automatically enhanced at interferometer frequencies. Reaching those scales would require an additional mechanism for amplifying the tensor spectrum while preserving the consistency conditions of the mixed sector.

Several extensions would be useful. A treatment of the full massive spin-two field should include the helicity-$0$ and helicity-$\pm1$ sectors and determine whether they introduce additional observables or constraints. The radiative stability of the interaction hierarchy and the strong-coupling scale at small $c_\sigma$ also deserve a dedicated analysis. A direct solution with time-dependent $c_\sigma(\tau)$ would test the horizon-crossing prescription used for the scale-dependent case. Parity-violating interactions would provide a separate extension in which the two tensor helicities need not behave identically. At the phenomenological level, an interesting direction would be to develop observational probes of primordial tensor non-Gaussianities at small scales \cite{Powell:2019kid,JimenezCruz:2025wqa,Ciprini:2026pvz}, including the quadrupolar anisotropies they can induce in the gravitational wave background accessible to interferometers \cite{Allen:1996gp}. Importantly, identifying the distinctive frequency-dependent signatures of the gravitational-wave collider will require dedicated estimators, extending existing approaches \cite{LISACosmologyWorkingGroup:2022kbp}.

The main result is that the tensor power spectrum, the nonanalytic squeezed bispectrum, and the induced quadrupolar anisotropy are controlled by the same underlying parameters. They therefore provide consistency checks on one another. A signal in the squeezed tensor bispectrum would predict an associated contribution to the tensor power and to the quadrupolar modulation, while limits on either of the latter restrict the allowed bispectrum amplitude. This correlation between the two- and three-point observables is the central phenomenological consequence of the model.

\section*{Acknowledgements}
SB and MF acknowledge the ``Consolidaci\'{o}n Investigadora'' grant CNS2022-135590. The work of SB and MF is partially supported by the Spanish Research Agency (Agencia Estatal de Investigaci\'{o}n)
through the Grant IFT Centro de Excelencia Severo Ochoa No CEX2020-001007-S, funded by
MCIN/AEI/10.13039/501100011033. AG acknowledges support
from the Royal Society, UK, funding reference: NIF\ R1\ 253963. GT is partially supported by the STFC grant [UKRI4243], by the Royal Society grant IES R3 43186, and by the
Leverhulme Trust grant RF-2026-166 9. Initial discussions on this project have been carried out during the 2025 ``The Dawn of Gravitational Wave Cosmology'' workshop, supported by the Fundaci\'{o}n Ram\'{o}n Areces and hosted by the ``Centro de Ciencias de Benasque Pedro Pascual''. We thank both the CCBPP and the Fundaci\'{o}n Areces for creating a stimulating and very productive environment for research.
Standard AI models  assisted us in developing the   results described in Appendix B, which we then  checked numerically and analytically. AG thanks Arko Bhaumik for discussion on the topic, and is grateful to Andrea Tesi for the slowness of the motorbike ride during a trip to Prato, so that a certain calculation could be finished.

\newpage

\appendix

\section{Propagators and mixed propagator}
\label{appendixA}

This appendix gives out the propagators used in the in-in calculation and gives the intermediate steps leading to the mixed propagator $\mathcal G_\pm$.  We use the Schwinger-Keldysh contour, so every field is doubled as $\varphi\to(\varphi_+,\varphi_-)$.  The path integral contains
\begin{equation}
    S[\varphi_+]-S[\varphi_-]
    =\int d\tau d^3x\,\big[\mathcal L(\varphi_+)-\mathcal L(\varphi_-)\big].
\end{equation}
The $+$ branch is time ordered and the $-$ branch is anti-time ordered.  This bookkeeping is essential because late-time expectation values are not ordinary in-out amplitudes.
\subsection{Bulk-to-bulk propagators}\label{appendA:B2Bulk}
For the graviton mode function in Eq.~\eqref{gammamode}, the contour-ordered propagator is
\begin{align}
G_{++}(k;\tau_1,\tau_2)
&=\gamma_k(\tau_1)\gamma_k^*(\tau_2)\Theta(\tau_1-\tau_2)
 +\gamma_k^*(\tau_1)\gamma_k(\tau_2)\Theta(\tau_2-\tau_1),\\
G_{--}(k;\tau_1,\tau_2)
&=\gamma_k^*(\tau_1)\gamma_k(\tau_2)\Theta(\tau_1-\tau_2)
 +\gamma_k(\tau_1)\gamma_k^*(\tau_2)\Theta(\tau_2-\tau_1),\\
G_{+-}(k;\tau_1,\tau_2)&=\gamma_k^*(\tau_1)\gamma_k(\tau_2),
\qquad
G_{-+}(k;\tau_1,\tau_2)=\gamma_k(\tau_1)\gamma_k^*(\tau_2).
\end{align}
For the helicity-two mode of the massive field, using Eq.~\eqref{sigmode},
\begin{align}
D_{++}(k;\tau_1,\tau_2)
&=\sigma_k(\tau_1)\sigma_k^*(\tau_2)\Theta(\tau_1-\tau_2)
 +\sigma_k^*(\tau_1)\sigma_k(\tau_2)\Theta(\tau_2-\tau_1),\\
D_{+-}(k;\tau_1,\tau_2)&=\sigma_k^*(\tau_1)\sigma_k(\tau_2),
\qquad
D_{-+}(k;\tau_1,\tau_2)=\sigma_k(\tau_1)\sigma_k^*(\tau_2).
\end{align}
Equivalently, in terms of Hankel functions,
\begin{align}
D_{++/--}(k;\tau_1,\tau_2)
&={\pi H^2\over4}e^{-\pi {\rm Im}\nu}(\tau_1\tau_2)^{3/2}
\Big[
H_\nu^{(1)}(-c_\sigma k\tau_1)H_{\nu^*}^{(2)}(-c_\sigma k\tau_2)
\Theta(\tau_{1/2}-\tau_{2/1})\\
&\hspace{2.2cm}
+H_{\nu^*}^{(2)}(-c_\sigma k\tau_1)H_\nu^{(1)}(-c_\sigma k\tau_2)
\Theta(\tau_{2/1}-\tau_{1/2})
\Big],\\
D_{+-/-+}(k;\tau_1,\tau_2)
&={\pi H^2\over4}e^{-\pi {\rm Im}\nu}(\tau_1\tau_2)^{3/2}
H_\nu^{(1)}(-c_\sigma k\tau_{2/1})
H_{\nu^*}^{(2)}(-c_\sigma k\tau_{1/2}).
\end{align}
The notation $++/--$ means that the upper/lower choice is used consistently for the corresponding branch ordering.

\subsection{Bulk-to-boundary propagators}\label{appendA:B2Boundary}
Since we are interested in using the mixed propagator technique to calculate late-time graviton correlation functions (tensor power spectrum and bispectrum) the external graviton is evaluated at late time, let us write explicitly the bulk-to-boundary graviton propagator setting $\tau_1\to0$,
\begin{align}
G_+(k,\tau)&\equiv G_{++}(k;0,\tau)
=\gamma_k(0)\gamma_k^*(\tau)
={2H^2\over M_{\rm Pl}^2k^3}(1-ik\tau)e^{ik\tau},\\
G_-(k,\tau)&\equiv G_{-+}(k;0,\tau)
=\gamma_k^*(0)\gamma_k(\tau)
={2H^2\over M_{\rm Pl}^2k^3}(1+ik\tau)e^{-ik\tau}.
\end{align}
These are the two graviton lines that attach a bulk vertex to a late-time external operator.
\subsection{The mixed propagator}\label{appenA:mixed}

The interaction $(a^3/2)\rho M_{\rm Pl}\sigma_{ij}\dot\gamma^{ij}$ generates $\mathcal G_\pm$ through one bilinear insertion. Accordingly, the full mixed propagator contains exactly one power of $\rho$:
\begin{align}
\label{eq:mixed_prop_appendix_start}
\mathcal G_\pm(k,\tau)
&={i\rho M_{\rm Pl}\over2}\int_{-\infty}^{0}{d\tau'\over(-H\tau')^3}
\left[
D_{\pm+}(k;\tau,\tau')\partial_{\tau'}G_+(k,\tau')
-D_{\pm-}(k;\tau,\tau')\partial_{\tau'}G_-(k,\tau')
\right].
\end{align}
The relative minus sign is the Schwinger-Keldysh sign difference between the two branches.

Changing variable to $z=-k\tau$ and specializing to the massive case $\nu=i\mu$, Eq.~\eqref{eq:mixed_prop_appendix_start} can be written as
\begin{align}
\label{eq:mixed_prop_Ipm}
\mathcal G_\pm(k,\tau)&=\frac{\pi\rho H}{4M_{\rm Pl}k^3}I_\pm(\mu,c_\sigma;z),
\qquad z=-k\tau,\\
\label{eq:Ipm_definition_appendix}
I_\pm(\mu,c_\sigma;z)
&\equiv
z^{3/2}e^{-\pi\mu}
\Bigg[
2\,{\rm Im}\left(
H_{i\mu}^{(1)}(c_\sigma z)
\int_0^\infty {dz'\over\sqrt{z'}}H_{-i\mu}^{(2)}(c_\sigma z')e^{-iz'}
\right)\nonumber\\
\hspace{1.3cm}
&+iH_{i\mu}^{(1)}(c_\sigma z)
\int_0^z {dz'\over\sqrt{z'}}H_{-i\mu}^{(2)}(c_\sigma z')e^{\mp iz'}
-iH_{-i\mu}^{(2)}(c_\sigma z)
\int_0^z {dz'\over\sqrt{z'}}H_{i\mu}^{(1)}(c_\sigma z')e^{\mp iz'}
\Bigg].
\end{align}
The three terms have a simple interpretation.  The integral extending to infinity incorporates the Bunch--Davies initial condition and integrates over the evolution of the massive mode; the two finite integrals encode the branch-dependent time ordering between the mixing vertex and the bulk point $\tau$.

The mixed propagator enters calculations of the tensor power spectrum and bispectrum. We show in the subsequent Appendices that for both the cases the correlators can be analyticlaly estimated to a great extent. 

\section{Analytical estimation of corrections to the tensor power spectrum}
\label{appendixB}

This appendix derives the analytic formulae for the two contributions $\mathcal C_1$ and $\mathcal C_2$ in the tensor power correction in Eq.~\eqref{eq:delp0} using the form of $I_{\pm}(\mu,\cs;z)$ in Eq.~\eqref{eq:Ipm_definition_appendix}.  The two terms $\mathcal C_1$ and $\mathcal C_2$ have different origins correspoinding to the first and the second line of Eq.~\eqref{eq:Ipm_definition_appendix} respectively.  $\mathcal C_1$ is the square of a non-time-ordered integral and is relatively direct.  $\mathcal C_2$ contains a nested time-ordered integral and is most efficiently evaluated by expanding one Bessel function in a series.

We first rewrite
\begin{equation}
H_{i\mu}^{(1)}(c_\sigma x)
=J_{i\mu}(c_\sigma x)+iY_{i\mu}(c_\sigma x)
={e^{\pi\mu}\over\sinh(\pi\mu)}J_{i\mu}(c_\sigma x)
-{1\over\sinh(\pi\mu)}J_{-i\mu}(c_\sigma x).\label{eq:HasJ}
\end{equation}
This identity converts all integrals into standard Bessel integrals with complex order.

\subsection{Integral without time ordering}
\label{sec:integralwoto}
The integral entering $\mathcal C_1$ is
\begin{equation}
    \mathcal J_0(\mu,c_\sigma)
    \equiv
    \int_0^\infty dx\,x^{-1/2}H_{i\mu}^{(1)}(c_\sigma x)e^{ix}.
\end{equation}
The oscillatory factor is regulated by the standard $i\epsilon$ prescription, $e^{ix}\to e^{ix-\epsilon x}$, and the result follows by analytic continuation of the Laplace transform of $x^{\lambda-1}J_\nu(\beta x)$.  One obtains
\begin{equation}
\mathcal J_0(\mu,c_\sigma)
={e^{\pi\mu/2-i\pi/4}\over\cosh(\pi\mu)}
\sqrt{2\pi\over c_\sigma}
{_2}F_1\!\left({1\over2}-i\mu,{1\over2}+i\mu;1;{c_\sigma-1\over2c_\sigma}\right).
\end{equation}
Substituting this expression into Eq.~\eqref{eq:def_c1} gives Eq.~\eqref{eq:c1expr}.  This same integral also appears in $C_\mu(c_\sigma)$ in Appendix~\ref{appendixA}, which provides a useful normalization check.

The closed form for $\mathcal J_0$ gives two direct numerical checks.  At $c_\sigma=1$ the hypergeometric argument vanishes and ${}_2F_1=1$, so Eq.~\eqref{eq:c1expr} reduces immediately to $\mathcal C_1=\pi^2/\cosh^2(\pi\mu)$.  For $c_\sigma\ll1$, the leading hypergeometric behavior is proportional to $\sqrt{c_\sigma}$, canceling the explicit $c_\sigma^{-1/2}$ in $\mathcal J_0$. 
%\suki{ A numerical implementation should recover both limits continuously and should reproduce the logarithmic oscillations generated by the conjugate powers $c_\sigma^{\pm i\mu}$.}

\subsection{Integral with time ordering}
\label{appen:powtimeordered}
The nested integral in $\mathcal C_2$ is
\begin{equation}
    \int_0^\infty {dx\over\sqrt x}H_{i\mu}^{(1)}(c_\sigma x)e^{-ix}
    \int_x^\infty {dy\over\sqrt y}H_{-i\mu}^{(2)}(c_\sigma y)e^{-iy}=\int_0^\infty {dx\over\sqrt x}H_{-i\mu}^{(2)}(c_\sigma x)e^{-ix}
    \int_0^x {dy\over\sqrt y}H_{i\mu}^{(1)}(c_\sigma y)e^{-iy} .
\end{equation}

The inner integral can be written in terms of Bessel functions using Eq.~\eqref{eq:HasJ}.  Expanding the Bessel functions and exponentials gives
\begin{align}
 y^{-1/2}J_{\pm i\mu}(c_\sigma y)e^{-iy}
 &=\sum_{n=0}^\infty a_n^\pm y^{n-1/2\pm i\mu},\\
 a_n^\pm
&\equiv
\left({c_\sigma\over2}\right)^{\pm i\mu}
{(-i)^n\over n!}
{ {}_2F_1\!\left(-{n\over2},{1-n\over2};1\pm i\mu;c_\sigma^2\right)
\over\Gamma(1\pm i\mu)} .\label{eq:anpmdef}
\end{align}
Therefore, 
\begin{eqnarray}
\mathcal{C}_2&=&-\pi{\rm Re}\sum_{n=0}^{\infty} \bigg[ \frac{e^{\pi\mu}}{\sinh(\pi\mu)}\mathcal{I}_n^+(\mu,\cs)-\frac{e^{-\pi\mu}}{\sinh(\pi\mu)}\mathcal{I}_n^-(\mu,\cs)\bigg],\label{eq:c2asIn}\\
\mathcal I_n^\pm
&=&
{a_n^\pm\over n+1/2\pm i\mu}
\int_0^\infty dx\,x^{n\pm i\mu}H_{-i\mu}^{(2)}(c_\sigma x)e^{-ix}.\label{eq:Inpm}
\end{eqnarray}

The remaining integral is evaluated by rotating the contour $x\to -iy$ and using standard hypergeometric representations of Bessel transforms.  The result is
\begin{align}
\mathcal I_n^+
&=
{2a_n^+e^{\pi\mu}(-i)^n\over\sqrt\pi(n+1/2+i\mu)}
{(2c_\sigma)^{i\mu}\over(c_\sigma+1)^{n+1+2i\mu}}
{\Gamma(n+1+2i\mu)\Gamma(n+1)\over\Gamma(n+3/2+i\mu)}\\
&\quad\times
_2F_1\!\left(n+1+2i\mu,{1\over2}+i\mu;n+{3\over2}+i\mu;{1-c_\sigma\over1+c_\sigma}\right),\\
\mathcal I_n^-
&=
{2a_n^-(-i)^n\over\sqrt\pi(n+1/2-i\mu)}
{(2c_\sigma)^{i\mu}\over(c_\sigma+1)^{n+1}}
{\Gamma(n+1-2i\mu)\Gamma(n+1)\over\Gamma(n+3/2-i\mu)}\\
&\quad\times
_2F_1\!\left(n+1,{1\over2}-i\mu;n+{3\over2}-i\mu;{1-c_\sigma\over1+c_\sigma}\right).
\label{eq:Inpm_closed}
\end{align}
The resummation over $n$ can be done writing the following Euler representations 
\begin{align}
&\frac{1}{n+1/2\pm i\mu}=\int _0^1 ds~ s^{n-1/2\pm i\mu}\nonumber,\\
G_n^{+}&=\frac{\Gamma(n+3/2+i\mu)}{\Gamma(1/2+i\mu)\Gamma(n+1)}\int _0^1 dt~ t^{-1/2+i\mu}(1-t)^n(1-ut)^{-(n+1+2i\mu)},\nonumber \\
G_n^{-}&=\frac{\Gamma(n+3/2-i\mu)}{\Gamma(1/2-i\mu)\Gamma(n+1)}\int _0^1 dt~ (1-t)^{-1/2-i\mu}t^n(1-ut)^{-(1/2+i\mu)},\nonumber
\end{align}
where $u\equiv (1-\cs)/(1+\cs)$, and $G_n^{\pm}$ are the hypergeometric functions appearing in $\mathcal{I}_n^{\pm}$ respectively. Recognising the resulting $n$-summations as binomial series, $\mathcal{C}_2$ can be written in the form of expression~\eqref{eq:c2expr}, in terms of convergent integrals 
\begin{align}
\mathcal{J}_{+}(\mu,\cs)&\equiv \int_0^1\int_0^1ds~ dt ~(st)^{-1/2+i\mu}\bigg((1-ut+s(1-t))(1-ut+us(1-t))\bigg)^{-(1/2+i\mu)} \\
\mathcal{J}_{-}(\mu,\cs)&\equiv \int_0^1\int_0^1ds~ dt\bigg(s(1-t)(1-ut)\bigg)^{-(1/2+i\mu)}\bigg((1+st)(1+ust)\bigg)^{-1/2+i\mu}.\label{eq:Jpm}
\end{align}
Substituting Eq.~\eqref{eq:Jpm} into Eq.~\eqref{eq:c2expr} gives the expression for $\mathcal C_2(\mu,c_\sigma)$.  
\subsection*{Limit $\cs \ll 1$}\label{subsec:lim1}
In the limit $\cs \ll 1$, $D_+ = D_- = 1 - t + s(1-t) = (1-t)(1+s)$, and therefore, 
\begin{align}
\mathcal{J}_+(\mu, \cs \ll 1) &= \int_0^1 ds \, s^{-\frac{1}{2}+i\mu} (1+s)^{-(1+2i\mu)} \int_0^1 dt \, t^{-\frac{1}{2}+i\mu} (1-t)^{-(1+2i\mu)}\nonumber \\
&= \frac{\left[\Gamma\left(\frac{1}{2}+i\mu\right) \Gamma(-i\mu)\right]^2}{\sqrt{\pi} \, \Gamma\left(\frac{1}{2}-i\mu\right)}\\
\mathcal{J}_-(\mu, \cs \ll 1) &= \int_0^1 dt \, t^{-\frac{1}{2}+i\mu} (1-t)^{-1-2i\mu} \int_0^t d(st) \, (st)^{-\frac{1}{2}-i\mu} (1+st)^{-1+2i\mu}\nonumber \\
&= \frac{\Gamma\left(\frac{1}{2}-i\mu\right) \Gamma(i\mu) \Gamma(-i\mu)}{\sqrt{\pi}}.
\label{limits:smallcsJpm}
\end{align}
\subsection*{Limit $\cs \rightarrow 1$}\label{subsec:lim2}
On the other extreme, with $\cs\rightarrow 1$, setting $u \simeq 0$ ($c_\sigma = 1$) yields:
\begin{equation}
\mathcal{J}_+(\mu, 1) = \int_{0}^{1}\int_{0}^{1} ds ~dt~ (st)^{b-1} \left[ 1 + s(1-t) \right]^{-b}, \quad \text{where } b = \frac{1}{2} + i\mu.
\end{equation}  
Performing the $t$-integral via the Gauss hypergeometric representation gives:
\begin{equation}
\int_{0}^{1} dt~ t^{b-1} \left( 1 - \frac{s}{1+s} t \right)^{-b} = \frac{1}{b} ~2F_1\left[b, b; b+1; \frac{s}{1+s}\right].
\end{equation}
Substituting this back into the $s$-integral, changing variables to $w = \frac{s}{1+s}$ leads to:
\begin{align}
\mathcal{J}_+(\mu, 1) &= \int_{0}^{1/2} dw~ w^{b-1} (1-w)^{-b} \ln\left( \frac{1-w}{w} \right).
= \sum_{j=0}^{\infty} \frac{(-1)^j}{(j+b)^2} \nonumber \\
&= \frac{1}{4} \left[ \psi^{(1)}\left( \frac{1}{4} + \frac{i\mu}{2} \right) - \psi^{(1)}\left( \frac{3}{4} + \frac{i\mu}{2} \right) \right].
\end{align}    
Similarly, 
\begin{equation}
\mathcal{J}_-(\mu, 1) = \int_{0}^{1}\int_{0}^{1} ds~ dt~ \left[ s(1-t) \right]^{-\frac{1}{2}-i\mu} (1 + st)^{-\frac{1}{2}+i\mu}.
\end{equation}  
Substituting $t \to 1 - t$ maps the integrand directly into the complex conjugate of $J_+(\mu, 1)$:
\begin{equation}
\mathcal{J}_-(\mu, 1) = \left[ \mathcal{J}_+(\mu, 1) \right]^* = \Phi\left( -1, 2, \frac{1}{2} - i\mu \right) = \frac{1}{4} \left[ \psi^{(1)}\left( \frac{1}{4} - \frac{i\mu}{2} \right) - \psi^{(1)}\left( \frac{3}{4} - \frac{i\mu}{2} \right) \right].
\end{equation}
Therefore, 
\begin{equation}
\mathcal{C}_2(\mu, 1) = -2 \mathrm{Re}[\Phi\left(-1,2,\frac{1}{2}+i\mu\right) ]= -\frac{1}{2} \mathrm{Re}\left[ \psi^{(1)}\left(\frac{1}{4} + \frac{i\mu}{2}\right) - \psi^{(1)}\left(\frac{3}{4} + \frac{i\mu}{2}\right) \right].
\end{equation}
In the limit $c_\sigma =1$, the integrals $\mathcal{J}_{\pm}$ were previously found in the calculations~\cite{Chen:2017ryl}.  This limit therefore is an important check on both the normalization and the branch choices in the nested integral.

\subsection{Asymptotes of $\mathcal{C}_{1,2}$ at different limits}
\label{subsec:C1C2limits}
In the large-mass limit, with $\cs <1$, the non-time-ordered contribution $\mathcal C_1$ can be estimated. Realizing that the hypergeometric function in Eq.~\eqref{eq:c1expr} can be written in terms of the Legendre polynomial $P_{i\mu-1/2}(1/\cs)$, its large $\mu$ asymptote is 
\begin{eqnarray}
\lim _{\mu \rightarrow \infty} {}_2F_1[1/2-i\mu,1/2+i\mu,1,\frac{\cs-1}{2\cs}]&\simeq &\sqrt{\frac{2\cs}{\pi \mu \sqrt{1-\cs ^2}}}\cos (\mu \theta -\pi/4)\nonumber \\
{\rm with}~ ~ ~ \theta \equiv \cosh^{-1}(1/\cs)&=&\ln \bigg[\frac{1+\sqrt{1-\cs^2}}{\cs}\bigg].\nonumber \\
{\rm so~ that,}
\lim _{\mu \rightarrow \infty}\mathcal{C}_1&\simeq & \frac{8\pi}{\mu} e^{-2\pi\mu}\frac{\cos ^2 (\mu \theta -\pi/4)}{\sqrt{1-\cs^2}}.
\end{eqnarray}
The signature of the off-shell process is evident from this limit, since $\mathcal{C}_1$ decays as $e^{-2\pi\mu}$. On the other hand, the first term dominates for large $\mu$ in $\mathcal{C}_2$, leading to
\begin{equation}
\lim _{\mu \rightarrow \infty}\mathcal{C}_2\simeq \frac{-4}{1+\cs} {\rm Re}\bigg[(\lim _{\mu \rightarrow \infty}\mathcal{J}_+) \bigg(\frac{\cs}{1+\cs}\bigg)^{2i\mu}\bigg].
\end{equation}
For general subluminal sound speeds ($\cs < 1$), applying the stationary phase approximation to the double integral yields a power-law decay $\mathcal{J}_+(\mu, \cs) \underset{\mu \gg 1}{\sim} \frac{i}{2\mu} + \mathcal{O}\left(\frac{1}{\mu^2}\right)$. When combined with the oscillatory phase factor $\left(\frac{2\cs}{1+\cs}\right)^{2i\mu}$, the leading real part decays as:
\begin{equation}\lim _{\mu \rightarrow \infty}\mathcal{C}_2\simeq \frac{2}{\mu(1+\cs)} \sin\left[ 2\mu \ln\left(\frac{2\cs}{1+\cs}\right) \right] + \mathcal{O}\left(\frac{1}{\mu^2}\right)
\end{equation}
This analytical estimate, together with the numerical evaluation of $\mathcal J_+$, shows that $\mathcal{C}_2$ decays as a power law at large $\mu$, and therefore, it is the main contributor to the total correction to the tensor power $\tilde {P}$ in the large-mass limit. 

For $\cs \ll 1$,
\begin{equation}
\lim _{\cs \ll 1} {}_2F_1[1/2-i\mu,1/2+i\mu,1,\frac{\cs-1}{2\cs}]\simeq \sqrt{2\cs}\bigg[\frac{\Gamma (2i\mu)}{\Gamma ^2({1\over 2}+i\mu)}(2\cs)^{-i\mu}+\frac{\Gamma (-2i\mu)}{\Gamma ^2({1\over 2}-i\mu)}(2\cs)^{i\mu}\bigg],
\end{equation}
so that the contribution from $\mathcal{C}_1$ is completely oscillatory. On the other hand, for $\cs\rightarrow 1$, 
\begin{equation}
\lim _{\cs \rightarrow 1} {}_2F_1[1/2-i\mu,1/2+i\mu,1,\frac{\cs-1}{2\cs}]\simeq 1+\bigg({1\over 8} +{\mu ^2 \over 2}\bigg)(\cs-1),
\end{equation}
leading to mild increase with $\cs$ near unity. For $\cs =1$, 
\begin{equation}
\mathcal{C}_1(\mu,\cs =1)=\frac{\pi ^2}{\cs\cosh ^2(\pi\mu)},
\end{equation}
matching results derived in previous literature~\cite{Chen:2012ge,Chen:2017ryl} (modulo the additional factor 4 for the tensor power spectrum in this work).

The limits for $\mathcal{C}_2$ for small $\cs$ are straightforward, since $\mathcal{J}_{\pm}(\mu,\cs \ll1)$ are $\mu$-dependent constants (see Eqs.~\eqref{limits:smallcsJpm}). Therefore, $\mathcal{C}_2(\mu, \cs \ll1)$ contains oscillations $e^{\pi\mu}\cos[2\mu\log(2\cs)]$, which dominates over the constant part coming from $\mathcal{J}_-$. For $\cs=1$, $\mathcal{J}_{\pm}$ lead to polygamma functions so that,
\begin{eqnarray}
\mathcal{C}_{2}(\mu, \cs=1)&=&-\frac{1}{2}\mathrm{Re}\left[\psi^{(1)}\left(\frac{1}{4}+\frac{i\mu}{2}\right)-\psi^{(1)}\left(\frac{3}{4}+\frac{i\mu}{2}\right)\right],
\end{eqnarray}
matching the limits from previous literature~\cite{Chen:2012ge,Chen:2017ryl}.

It is interesting to note that near a cancellation between $\mathcal C_1$ and $\mathcal C_2$, the plotted total is sensitive to numerical precision. For the sum $\mathcal{C}_1 + \mathcal{C}_2$ to become negative, two criteria must hold simultaneously:\\
(a) Phase Criterion: The sine term in $\mathcal{C}_2$ must be negative to pull the overall sum down. At its maximum negative trough, $\sin\left[2\mu\ln\left(\frac{2c_\sigma}{1+c_\sigma}\right)\right] = -1$.
\\
(b)Envelope Dominance Criterion: At this negative peak, the magnitude of the negative term $|\mathcal{C}_2|_{\text{max}}$ must strictly exceed the maximum envelope of $\mathcal{C}_1$ so that,
$|\mathcal{C}_2|_{\text{max}} > \mathcal{C}_1|_{\text{max}}$.

Substituting the upper envelope of $\mathcal{C}_1$  and the maximum trough magnitude of $\mathcal{C}_2$ into the dominance condition:
\begin{equation}
    \frac{2}{\mu(1 + c_\sigma)} > \frac{8\pi}{\mu \sqrt{1 - c_\sigma^2}} e^{-2\pi\mu}
\end{equation}
Within the leading large-$\mu$ envelopes used above, this gives the asymptotic estimate:
\begin{equation}
    \mu > \mu_{\text{threshold}}(c_\sigma) \equiv \frac{1}{2\pi} \ln\left( 4\pi \sqrt{\frac{1 + c_\sigma}{1 - c_\sigma}} \right)=\frac{\ln(4\pi)}{2\pi} + \frac{1}{2\pi} \tanh^{-1}(c_\sigma).
\end{equation}
Even as $c_\sigma \ll 1$, there is an intrinsic lower mass threshold $\mu_{\text{min}} = \frac{\ln(4\pi)}{2\pi} \approx 0.402$ imposed by the $4\pi$ prefactor.
As $c_\sigma$ approaches 1, the threshold $\mu_{\text{threshold}} \to \infty$ due to the $1/\sqrt{1-c_\sigma^2}$ divergence in $\mathcal{C}_1$.

For arbitrary $\mu$ and $c_\sigma$, the sign must instead be determined from the full expressions in Eq.~\eqref{eq:c1expr}-~\eqref{eq:c2expr}. Writing $\mathcal{J}_+ = |\mathcal{J}_+| e^{i\Phi_+}$, and rewriting the total oscillatory phase as:
\begin{equation}
    \xi (\mu, c_\sigma) \equiv 2\mu \ln\left(\frac{2c_\sigma}{1+c_\sigma}\right) + \Phi_+(\mu, c_\sigma)
\end{equation}

$\mathcal{C}_2$ can be cast into an envelope-phase representation:
\begin{equation}
    \mathcal{C}_2(\mu, c_\sigma) = -A_{\rm gen}(\mu, c_\sigma) \cos\left[ \xi(\mu, c_\sigma) \right] + \mathcal{R}_-(\mu, c_\sigma)
\end{equation}

where the dominant envelope magnitude $A_{\rm gen}$ and residual sub-leading term $\mathcal{R}_-$ are:
\begin{align}
    A_{\rm gen}(\mu, c_\sigma) &\equiv \frac{2 e^{\pi\mu}}{(1+c_\sigma)\sinh(\pi\mu)} |\mathcal{J}_+(\mu, c_\sigma)| \\
    \mathcal{R}_-(\mu, c_\sigma) &\equiv \frac{2 e^{-\pi\mu}}{(1+c_\sigma)\sinh(\pi\mu)} \operatorname{Re}[\mathcal{J}_-(\mu, c_\sigma)]
\end{align}
For the total sum $\mathcal{C}_1 + \mathcal{C}_2$ to be strictly negative, $\mathcal{C}_1$:
\begin{equation}    \mathcal{C}_1(\mu, c_\sigma) + \mathcal{R}_-(\mu, c_\sigma) < A_{\rm gen}(\mu, c_\sigma) \cos\left[ \xi(\mu, c_\sigma) \right]
\end{equation}
A useful condition for possible negative values ($\cos[\xi] \approx 1$) is that the envelope ratio satisfies:
\begin{equation}
    \frac{\mathcal{C}_1(\mu, c_\sigma) + \mathcal{R}_-(\mu, c_\sigma)}{A_{\rm gen}(\mu, c_\sigma)} < 1.
\end{equation}
Substituting Eqs.~\eqref{eq:c1expr},~\eqref{eq:c2expr}, and the definitions of $A_{\rm gen}$ and $\mathcal{R}_-$ yields the following condition obtained from the full expressions:
\begin{equation}
    \frac{\pi^2 (1+c_\sigma) \tanh(\pi\mu)}{2 c_\sigma \cosh(\pi\mu)} \frac{\left| {}_2F_1\left( \frac{1}{2} - i\mu, \frac{1}{2} + i\mu, 1, \frac{c_\sigma - 1}{2c_\sigma} \right) \right|^2}{|\mathcal{J}_+(\mu, c_\sigma)|} + e^{-2\pi\mu} \frac{\operatorname{Re}[\mathcal{J}_-(\mu, c_\sigma)]}{|\mathcal{J}_+(\mu, c_\sigma)|} < 1
\end{equation}

For $\mu \gtrsim 0.5$, the second term involving $\mathcal{J}_-$ is suppressed by $e^{-2\pi\mu} \ll 1$ and can be safely neglected.
Negative values require the hypergeometric contribution in $\mathcal C_1$ to fall below the complete interference envelope after the residual term and phase are included. The actual boundary is obtained numerically from the full condition above, not from the leading large-$\mu$ estimate.
Numerically, we find that this boundary is only attained for $\mu \gg 1$.

\section{Elements of the bispectrum}\label{AppenC}
\subsection{Contribution from the helicities}\label{appendix:helicity}
The helicity contribution to the bispectrum in Eq.~\eqref{eq:bisp0} arises from (i) contracting 3 external $\gamma ^{\lambda _n}(\hat{\textbf{k}}_n)$-legs with 3 internal $\dot{\gamma}_{i_nj_n}(x_n)$ legs, leading to a polarization tensor $\epsilon _{i_nj_n}^{\lambda _n}(\hat{\textbf{k}}_n)$ for each leg $n$; (ii) contracting 3 remaining $\sigma$-operators at $x_n$ with the 3 $\sigma$-operators inside the cubic vertex $\sigma_{ab}(y) \sigma_{bc}(y) \sigma_{ca}(y)$ at $y$, leading to 3 combinations $\sum _{s_n}\epsilon _{i_nj_n}^{s _n}(\hat{\textbf{p}}_n)\epsilon _{kl}^{s _n *}(\hat{\textbf{p}}_n)$, where $kl=ab,bc,ca$ for vertices $x_1,x_2,x_3$ respectively. Along with the delta functions in momentum and helicity, and using the relations
\begin{align}
\epsilon ^{L/R}_{ij}(\hat{\textbf{k}})\epsilon ^{R/L}_{ij}(\hat{\textbf{k}})=1; ~ ~ ~ \epsilon ^{L/R}_{ij}(\hat{\textbf{k}})\epsilon ^{L/R}_{ij}(\hat{\textbf{k}})=0; ~ ~ ~ \epsilon ^{L/R~*}_{ij}(\hat{\textbf{k}})=\epsilon ^{L/R}_{ij}(-\hat{\textbf{k}})=\epsilon ^{R/L}_{ij}(\hat{\textbf{k}}),\label{eq:polnorm}
\end{align}
the combined contributions for the helicities reduces the 9 polarization tensors into 3, as 
\begin{equation}
\mathcal A^{\lambda_1\lambda_2\lambda_3}
\equiv
\epsilon^{\lambda_1}_{\alpha\beta}(-\hat k_1)
\epsilon^{\lambda_2}_{\beta\gamma}(-\hat k_2)
\epsilon^{\lambda_3}_{\gamma\alpha}(-\hat k_3).
\end{equation}
To evaluate this contribution for different shapes, we choose a reference momentum triangle in the $(x,y)$ plane, parameterizing the wave vectors in terms of their azimuthal angles with respect to $\hat{\textbf{k}}_1$, so that $\hat{\textbf{k}}_i=(\cos \varphi _i,\sin \varphi _i,0)$, with $\varphi _1=0$. The resulting polarization tensor is
\begin{equation}
\epsilon^{R/L}(\hat{\boldsymbol{k}}_i) = \frac{1}{2} 
\begin{pmatrix} 
-\sin^2\varphi_i & \cos\varphi_i \sin\varphi_i & \mp i \sin\varphi_i \\[6pt] 
\cos\varphi_i \sin\varphi_i & -\cos^2\varphi_i & \pm i \cos\varphi_i \\[6pt] 
\mp i \sin\varphi_i & \pm i \cos\varphi_i & 1 
\end{pmatrix}.
\end{equation}
For $r_2\equiv k_2/k_1$ and $r_3\equiv k_3/k_1$, the product of polarization tensors can be written completely in terms of these ratios as 
\begin{align}
\mathcal{A}^{\lambda_1 \lambda_2 \lambda_3} &=& -\frac{1}{64 r_2^2 r_3^2} \left[ 2 r_2 + (1 + r_2^2 - r_3^2) \lambda_1 \lambda_2 \right] \left[ -2 r_3 + (-1 + r_2^2 - r_3^2) \lambda_1 \lambda_3 \right]\nonumber \\
&&\times \left[ 2 r_2 r_3 + (-1 + r_2^2 + r_3^2) \lambda_2 \lambda_3 \right].\label{exp:Al1l2l3}
\end{align}
In the squeezed limit, $r_2=1$ and $r_3\ll1$, the helicity contractions reduce to $\mathcal A^{LLL/RRR}=\mathcal A^{LLR/RRL}=1/4$ and $\mathcal A^{LRR/RLL}=k_3^2/(16k_2^2)$.  The final channel is therefore suppressed by two additional powers of the squeezed ratio. 

\subsection{Analytical forms}\label{appendix:analytical}
Here we simplify the form of $I_{\pm}(\mu,\cs;z)$ in Eq.~\eqref{eq:Ipm_definition_appendix} in terms of functions of $z$, depending on the parameters $\mu$ and $\cs$. 
For the infinte integral in the first line of Eq.~\eqref{eq:Ipm_definition_appendix}, we follow the same approach as taken in Appendix~\ref{sec:integralwoto} to arrive at
\begin{align}
C_\mu(c_\sigma)&\equiv i\int_0^\infty {dz\over\sqrt z}e^{iz}H_{i\mu}^{(1)}(c_\sigma z)\\
&={e^{i\pi/4+\pi\mu/2}\over\cosh(\pi\mu)}
\sqrt{2\pi\over c_\sigma}
{}_2F_1\!\left({1\over2}-i\mu,{1\over2}+i\mu;1;{c_\sigma-1\over2c_\sigma}\right),\label{def:cmu0}.
\end{align} 
For the second line in Eq.~\eqref{eq:Ipm_definition_appendix}, it is useful to rewrite this part in terms of Bessel functions using Eq.~\eqref{eq:HasJ} and its complex conjugate. Thus, 
\begin{align}
\label{I2gpm}
&+iH_{i\mu}^{(1)}(c_\sigma z)
\int_0^z {dz'\over\sqrt{z'}}H_{-i\mu}^{(2)}(c_\sigma z')e^{\mp iz'}
-iH_{-i\mu}^{(2)}(c_\sigma z)
\int_0^z {dz'\over\sqrt{z'}}H_{i\mu}^{(1)}(c_\sigma z')e^{\mp iz'} \nonumber \\ &=\left(-{ie^{\pi\mu}\over\sinh\pi\mu}g^{\mp}_{+}
+{i\over\sinh\pi\mu}g^{\mp}_{-} \right)H_{-i\mu}^{(2)}(c_\sigma z)
+
\left({ie^{\pi\mu}\over\sinh\pi\mu}g^{\mp}_{-}
-{i\over\sinh\pi\mu}g^{\mp}_{+}\right)H_{i\mu}^{(1)}(c_\sigma z),\\
{\rm where, ~ ~}&\hspace{1.3cm} g^{s}_{\pm}(\mu,\cs;z) \equiv 
 \int_0^z {dz'\over\sqrt{z'}}J_{\pm i\mu}(c_\sigma z')e^{ isz'}.\label{def:fpm}
\end{align}
Note that there are two types of sign indices here, the upper ones are coming from the exponential, while the lower ones are for the different signs in the index of the Bessel function $J_{\pm i\mu}(c_\sigma z')$ .Using the series expansion of Bessel function and the exponential function in Eq.~\eqref{def:fpm} and using well-known identities of the Pocchammer symbols, 
we obtain the series expansion for $g^s_{\pm}$
\begin{align}
g^{s}_{\pm}(\mu,\cs;z)=\left({c_\sigma\over2}\right)^{\pm i\mu} \sum _n \frac{(si)^n}{n!} \frac{z^{n+1/2\pm i\mu}}{n+1/2\pm i\mu}
{ {}_2F_1\!\left(-{n\over2},{1-n\over2};1\pm i\mu;c_\sigma^2\right)
\over\Gamma(1\pm i\mu)}.\label{eq:gsseries}
\end{align}
This series expansion is consistent with the definition of coefficients $a_n^{\pm}$ in Eq.~\eqref{eq:anpmdef} in Appendix~\ref{appen:powtimeordered}.
 A more convenient form of $g^s_{\pm}$ can be achieved with the Euler integral representation of the Bessel functions:
 \begin{align}
 J_{\pm i\mu}(c_\sigma x)&=\left({\cs x \over 2}\right)^{\pm i\mu} {1\over {\sqrt{\pi}\Gamma (1/2\pm i\mu)}} \int _{-1}^{1} dt (1-t^2)^{\pm i\mu -1/2}e^{i\cs xt}\nonumber \\
 {\rm so ~that ~}\nonumber \\
 g^{s}_{\pm}(\mu,\cs;z)&= \left({\cs\over2}\right)^{\pm i\mu}\frac{e^{\mp \pi\mu/2+i\pi/4}}{\sqrt{\pi}\Gamma(1/2\pm i\mu)}\int_{-1}^{1} dt~ \frac{(1-t^2)^{-1/2\pm i\mu}}{(s+\cs t)^{1/2\pm i\mu}}\blueedit{\gamma[1/2\pm i\mu,-iz(s+\cs t)]}.\label{eq:gpm_Eurler}
 \end{align}
To calculate the bispectrum, we will use the following form of $I_\pm(\mu,c_\sigma;z)$ that contains (semi-)analytical forms for the integrals in its Hankel function form in Eq.~\eqref{eq:Ipm_definition_appendix}. Using Eq.~\eqref{def:cmu0}, and~\eqref{I2gpm},
\begin{align}
\label{eq:Ipmdefexp}
I_\pm(\mu,c_\sigma;z)&=
z^{3/2}e^{-\pi\mu}
\Bigg[
\left(C_\mu-{ie^{\pi\mu}\over\sinh\pi\mu}g^{\mp}_{+}
+{i\over\sinh\pi\mu}g^{\mp}_{-} \right)H_{-i\mu}^{(2)}(c_\sigma z)\nonumber \\
&+
\left(C_\mu^*+{ie^{\pi\mu}\over\sinh\pi\mu}g^{\mp}_{-}
-{i\over\sinh\pi\mu}g^{\mp}_{+}\right)H_{i\mu}^{(1)}(c_\sigma z)
\Bigg].
\end{align}
where $\gamma[a,z]=\int_0^z t^{a-1}e^{-t}$ is the lower incomplete gamma function. We use this expression to evaluate $I_{\pm}(\mu,\cs;z)$ for general $z$. 

When computing diagrams, conjugating a diagram exchanges $+$ and $-$ vertices.  This provides a useful check: the sum over all Schwinger-Keldysh assignments must give a real late-time correlator.  It is also the reason why many final expressions can be written using the imaginary or real part of a single representative branch integral.

\subsection{Calculation of the coefficients $w_{\pm}(\mu,\cs)$}\label{AppendixA2}
\begin{figure}[!]
\centering
\includegraphics[width=0.49\textwidth]{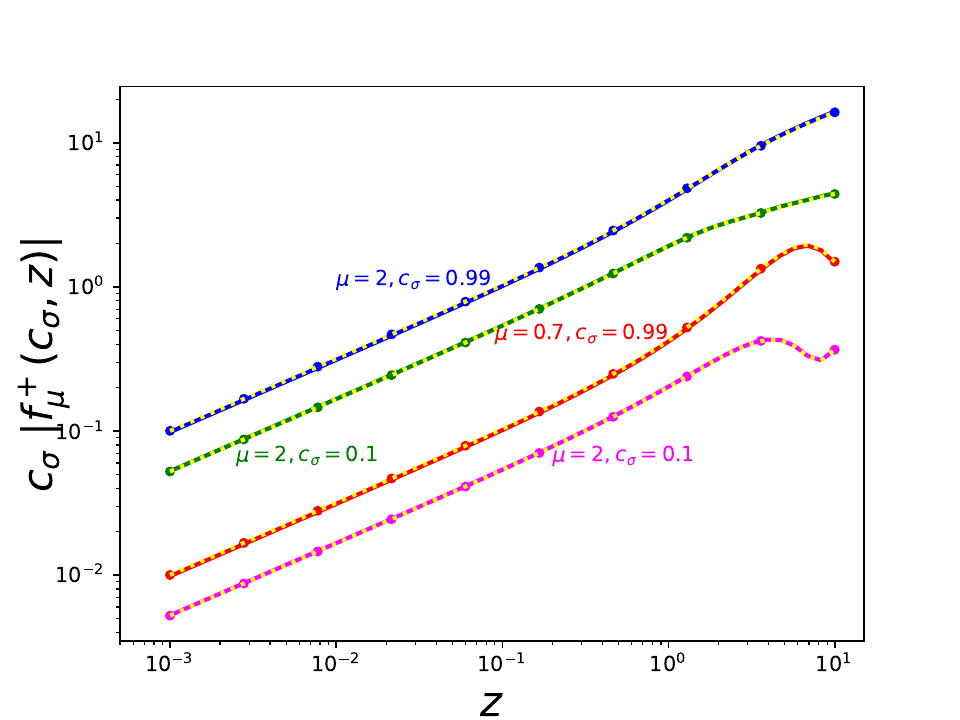}
\includegraphics[width=0.49\textwidth]{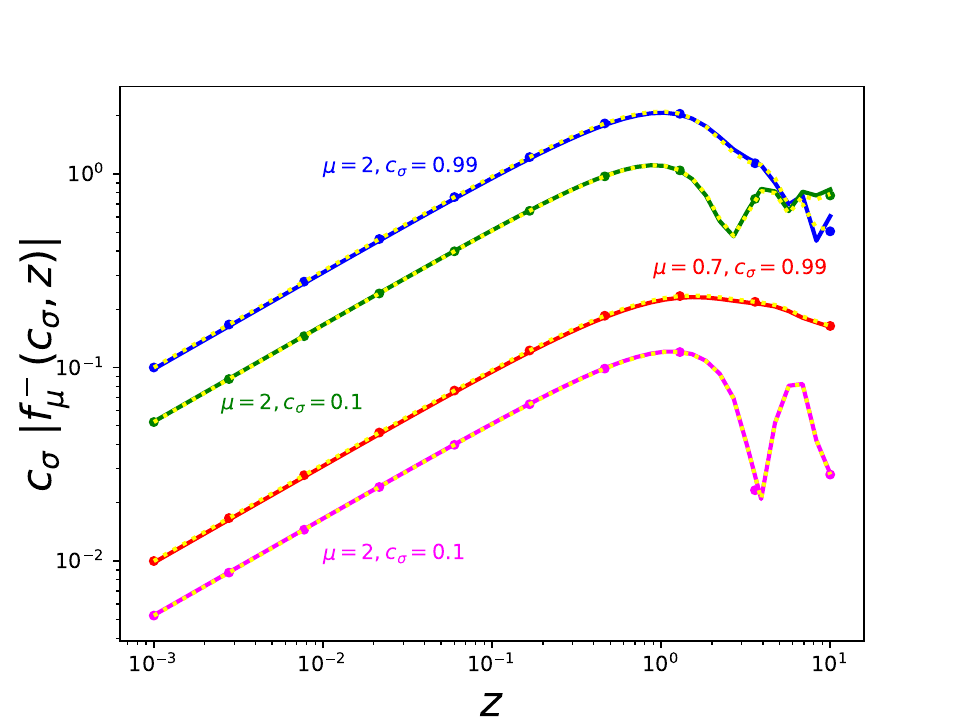}
\caption{\small $\vert f^{\pm}_{\mu}\vert$ as functions of $z$ for few different combinations of $(\mu,\cs)$. The solid blue, green, red and magenta solid lines denote the semi-analytical form of Eq.~\eqref{eq:fpm}; scattered dots in corresponding colors are for numerical integration $\int_0^z {dz'\over\sqrt{z'}}J_{\pm i\mu}(c_\sigma z')e^{ -iz'}$, and dashed yellow lines in each case are for the analytical series expansion of Eq.~\eqref{eq:fpmseries}. }
\label{fig:fpfmcomp}
\end{figure}

For analytical estimation of the squeezed bispectrum in Sec.~\ref{subsec:constantcs}, we utilize limits of the analytical form of $I_\pm(\mu,c_\sigma;z)$ derived in the previous Appendix. Noting that in Eq.~\eqref{def:calI}, we calculate the bispectrum only in terms of the `+' fields, corresponding to $I_+(\mu,c_\sigma;z)$, and since the squeezed coefficients $w_{\pm}$ in Eq.~\eqref{eq:wpm} only pertain to $I_+(\mu,c_\sigma;z)$, we define $f_{\mu}^\pm(c_\sigma,z)\equiv g^{-}_{\pm}(\mu,\cs;z)$ in 
Eq.~\eqref{def:fpm}, so that, using Eq.~\eqref{eq:gpm_Eurler},
\begin{align}
f_{\mu}^\pm(c_\sigma,z)&=\int_0^z {dz'\over\sqrt{z'}}J_{\pm i\mu}(c_\sigma z')e^{ -iz'}\nonumber \\
&=\left({\cs\over2}\right)^{\pm i\mu}\frac{e^{\pm \pi\mu/2-i\pi/4}}{\sqrt{\pi}\Gamma(1/2\pm i\mu)}\int_{-1}^{1} du~ \frac{(1-u^2)^{-1/2\pm i\mu}}{(1-\cs u)^{1/2\pm i\mu}}\gamma[1/2\pm i\mu,iz(1-\cs u)].
\label{eq:fpm}
\end{align}
On the other hand, the series expansion for $f_{\mu}^\pm(c_\sigma,z)$ using Eq.~\eqref{eq:gsseries} is
\begin{align}
f_{\mu}^\pm(c_\sigma,z)=\left({c_\sigma\over2}\right)^{\pm i\mu} \sum _n \frac{(-i)^n}{n!} \frac{z^{n+1/2\pm i\mu}}{n+1/2\pm i\mu}
{ {}_2F_1\!\left(-{n\over2},{1-n\over2};1\pm i\mu;c_\sigma^2\right)
\over\Gamma(1\pm i\mu)},\label{eq:fpmseries}
\end{align}
which makes the small-$z$ behavior explicit: $f_{\mu}^\pm \sim z^{1/2\pm i\mu}$.  Substitution into the direct Hankel representation gives
\begin{align}\label{eq:I_small_z_audited}
 I_+(\mu,c_\sigma;z\ll1)
 &\simeq z^{3/2}\left[
 s_+(\mu,c_\sigma)\left(\frac{c_\sigma}{2}\right)^{i\mu}z^{i\mu}
 +s_-(\mu,c_\sigma)\left(\frac{c_\sigma}{2}\right)^{-i\mu}z^{-i\mu}
 \right]+I_{+,\rm analytic}(z),
 \\
 s_{\pm}&\equiv \frac{2}{\sinh (\pi\mu)}\sqrt{\frac{\pi}{\cs}}\frac{{}_2F_1\!\left({1\over2}-i\mu,{1\over2}+i\mu;1;{c_\sigma-1\over2c_\sigma}\right)}{\Gamma(1\pm i\mu)[\sinh(\pi\mu /2)\pm i\cosh(\pi\mu /2)]}.\label{eq:spm}
\end{align}
where $I_{+,\rm analytic}$ collects terms analytic in the squeezed momentum ratio.
%The branch coefficients $s_\pm$ must be evaluated from the direct Hankel formula with a common $i\epsilon$ prescription; their previously quoted closed form should be checked against that definition before being used for precision normalization.
To evaluate the integral in Eq.~\eqref{eq:wpm} 
\begin{equation}
\mathcal{K}_{\pm}\equiv \int_0^{\infty} dz~z^{-5/2\pm i\mu} I_+^2(\mu,\cs)
\end{equation}
a displaced contour $z=z_0\pm iy$ is used to control the oscillatory ultraviolet tail, and the integral is divided into low- and high-$z$ parts at the matching point $z_0$.  The low-$z$ part is evaluated numerically, while the high-$z$ contribution is treated with the large-argument Hankel expansion. The ultraviolet tail is evaluated by applying the same contour deformation directly to Eq.~\eqref{eq:Ipm_definition_appendix}.
Therefore,
\begin{align}
&\mathcal{K}_{\rm UV,\pm}(\mu,\cs,z\gg1)\simeq e^{-\pi\mu}[J_0^{\pm}+J_a^{\pm}+J_b^{\pm}], {\rm where, }\nonumber \\
J_a^{\pm}&\equiv \frac{-2i}{\pi\cs}e^{i\pi/2-2i\cs z_0}\bigg(C_\mu -\frac{ie^{\pi\mu}}{\sinh(\pi\mu)}f_+^{\infty}+\frac{i}{\sinh(\pi\mu)}f_-^{\infty}\bigg)^2\nonumber \\
&\hspace{1.3cm}\times\int_0^{\infty} dy~e^{-2\cs y}(z_0-iy)^{-1/2\pm i\mu},\nonumber \\
J_b^{\pm}&\equiv \frac{2i}{\pi\cs}e^{-i\pi/2+2i\cs z_0}\bigg(C_\mu ^{*} +\frac{ie^{\pi\mu}}{\sinh(\pi\mu)}f_-^{\infty}-\frac{i}{\sinh(\pi\mu)}f_+^{\infty}\bigg)^2\nonumber \\
&\hspace{1.3cm}\times \int_0^{\infty} dy~e^{-2\cs y}(z_0+iy)^{-1/2\pm i\mu},\nonumber \\
{\rm and~ }J_0^{\pm}&\equiv -\frac{z_0^{1/2\pm i\mu}}{1/2\pm i\mu},\\
{\rm using}\nonumber \\
f_{\pm}^{\infty}\equiv f_{\mu}^\pm(c_\sigma,z\gg 1)
&\simeq\left({\cs\over2}\right)^{\pm i\mu}\frac{\Gamma({1/2} \pm i\mu)}{(\cs^2-1)^{{1\over 4} \pm {i\mu\over 2}}}
{}_2F_1\!\left( {1\over 4} \pm {i\mu\over 2},{1\over 4} \pm {i\mu\over 2}, 1\pm i\mu, \frac{\cs^2}{\cs^2-1} \right),\label{eq:fpminf}
\end{align}
The separate $c_\sigma=1$ limit must be treated from its regular exact expression because asymptotic formulae containing powers of $c_\sigma^2-1$ are nonuniform at unity. At $\cs=1$, the series expansion of $f^{\pm}_\mu$ using Eq.~\eqref{eq:fpmseries} is sufficient, which, after using Gamma-function representation of the Hypergeometric series at $\cs=1$, and Hypergeometric identities, gives
\begin{equation}
f_{\mu}^{\pm}(\mu,\cs=1;z)
=
\frac{z^{i\mu+\frac12}}
{2^{i\mu}\left(i\mu+\frac12\right)\Gamma(1+i\mu)^2}
\,{}_2F_2\left(
\frac12+i\mu,\frac12+i\mu;
\frac32+i\mu,1+2i\mu;
\mp 2iz
\right).\label{eq:fpmcs1}
\end{equation}
On the other hand, $\cs=1$ in Eq.~\eqref{def:cmu0} leads to 
\begin{equation}
C_\mu(c_\sigma =1)={e^{i\pi/4+\pi\mu/2}\over\cosh(\pi\mu)}
\sqrt{2\pi\over c_\sigma}
.\label{eq:cmucs1}
\end{equation}
Expressions~\eqref{eq:fpmcs1} and~\eqref{eq:cmucs1} match exactly with the previous derivations for $\cs=1$ in Ref.~\cite{Chen:2017ryl}.
Figure~\ref{fig:fpfmcomp} shows the comparisons of $\vert f^{\pm}_{\mu}\vert$ derived with numerical integration (scattered dots), semi-analytical form (solid lines) of Eq.~\eqref{eq:fpm}, and analytical series expansion of Eq.~\eqref{eq:fpmseries} (dotted lines) for few different combinations of $(\mu,\cs)$.

At large $\mu$, $w_+\sim e^{-\pi\mu}$, while $w_-\sim e^{-2\pi\mu}$, explicitly shown in Fig.~\ref{fig:wpm}, so the $+$-coefficient dominates. For the effective ansatz $c_\sigma(k)=c_{\rm ini}(k/k_*)^{s_2}$, the long and short legs evaluate the mixed propagator at different effective sound speeds.  To leading order in the slow variation one may keep the same mixed-propagator kernel but replace the sound speed by its horizon-crossing value for each leg.

The ultraviolet and infrared scalings of the coefficient integral can be read from the mixed kernel.  The late-time form $I_+\sim z^{3/2\pm i\mu}$ gives
\begin{equation}
 z^{-5/2\pm i\mu}I_+^2=\mathcal O(z^{1/2+i\alpha\mu}),
 \label{eq:w_integral_ir_scaling}
\end{equation}
where $\alpha$ is a real branch-dependent coefficient, so the integral is locally convergent at $z=0$.  At large $z$, convergence is oscillatory and depends on the Bunch-Davies contour deformation rather than absolute falloff on the real axis.  This distinction explains why infrared truncation is comparatively benign, whereas the ultraviolet tail and contour choice control the numerical error budget for $w_\pm$.

\bibliography{ref_cc} 
\bibliographystyle{JHEP}

\end{document}